\documentclass[11pt]{article}
\usepackage{geometry}
\usepackage[english]{babel}
\usepackage[utf8x]{inputenc}
\usepackage{amsmath}
\usepackage{amsthm}
\usepackage{ragged2e}
\usepackage{amssymb}
\usepackage{graphicx}
\usepackage{booktabs}
\usepackage{enumerate}
\usepackage{setspace}

\usepackage{microtype}
\usepackage{bm}
\usepackage{color}
\usepackage[colorlinks=true]{hyperref}
\usepackage{natbib}

\def\name{Brice Romuald Gueyap Kounga}
\def\dept{Department of Economics}
\def\univ{Western University}

\hypersetup{
        urlcolor = blue,
        pdfauthor = {\name},
        pdfkeywords = {Nonparametric, Estimation, Dyadic data, Nonseparable, Networks, Inference},
        pdftitle = {Nonseparable Dyadic Regression},
        pdfsubject = {Econometric Theory},
        pdfpagemode = UseNone,
        citecolor = blue
}

\makeatletter
\renewcommand*{\@fnsymbol}[1]{\ensuremath{\ifcase#1\or \dagger\or \dagger\or \ddagger\or \mathsection\else \mathparagraph\fi}}
\makeatother
\newtheorem{assumption}{Assumption}
\newtheorem{definition}{Definition}
\newtheorem{lemma}{Lemma}
\newtheorem{example}{Example}
\newtheorem{theorem}{Theorem}
\newtheorem{corollary}{Corollary}
\newtheorem{proposition}{Proposition}

\newcommand{\E}{\mathbb{E}}
\newcommand{\R}{\mathbb{R}}
\newcommand{\PP}{\mathbb{P}}
\newcommand{\Var}{\mathrm{Var}}
\newcommand{\Cov}{\mathrm{Cov}}
\newcommand{\X}{\mathcal{X}}
\newcommand{\F}{\mathcal{F}}
\newcommand{\M}{\mathcal{M}}
\newcommand{\G}{\mathcal{G}}

\newcommand{\EE}{\mathcal{E}}
\newcommand{\ind}{\mathbf{1}}
\DeclareMathOperator*{\argmin}{arg\,min}
\newcommand{\pto}{\stackrel{p}{\longrightarrow}}
\newcommand{\dto}{\stackrel{d}{\longrightarrow}}

\title{\textbf{Nonseparable Dyadic Regression}}

\author{\name\footnote{\dept, \univ, London, Ontario, Canada. E-mail: \href{mailto:bgueyapk@uwo.ca}{bgueyapk@uwo.ca}. This paper previously circulated under the title ``Nonparametric Regression with Dyadic Data''; the present version supersedes it.\\
Special thanks to Nail Kashaev and Roy Allen for their comments and suggestions in the preparation of this paper. I thank Nirav Mehta, my classmates in the 2022--2023 Communication and Professional Development Workshop, and the participants of the AMM Reading Group at \univ\ for helpful comments and suggestions.}}
\date{September 14, 2026}

\begin{document}
\onehalfspacing
\maketitle

\begin{abstract}
\noindent This paper studies a nonseparable model for dyadic outcomes, such as bilateral trade flows, in which the outcome depends on both agents' characteristics and on a scalar unobservable through an unknown function increasing in that unobservable. I establish identification of a normalized structural function and the error distribution, propose kernel plug-in estimators, and derive a two-regime central limit theory under dyadic dependence in which a shared-agent variance component generically dominates. An agent-level bootstrap is proved consistent in that regime. Simulations show independence-based intervals undercover severely while the bootstrap substantially improves coverage. In bilateral trade data, conditional dispersion falls by more than half between small and large exporters.
\end{abstract}

\vspace{0.2cm}
\noindent\textbf{Keywords:} Dyadic data; nonseparable models; nonparametric identification; exchangeable arrays; bootstrap.

\vspace{0.1cm}
\noindent\textbf{JEL classification:} C12, C14, C31.

\section{Introduction}

Many important economic outcomes are defined for \emph{dyads}: trade flows between pairs of countries \citep{tinbergen1962shaping, anderson2003gravity}, liabilities across banks, R\&D partnerships across firms \citep{konig2019r}, supply-chain linkages \citep{atalay2011network}, risk sharing across households \citep{de2004risk, fafchamps2007formation}, friendships between individuals \citep{christakis2020empirical}, and correlated voting among seating neighbors in parliaments \citep{harmon2019peer}, among many others. Empirical work on such data overwhelmingly relies on linear (gravity-style) regressions, in which observables and unobservables enter additively. Yet economic theory rarely delivers additivity: in a gravity equation, for instance, unobserved bilateral frictions interact multiplicatively with country characteristics. This paper studies the nonparametric analysis of dyadic outcomes when the unobservable is allowed to interact with observables in an unrestricted way.

Specifically, I consider the model
\begin{equation}
Y_{ij}=g(X_i,X_j,e_{ij}), \qquad i\neq j, \label{eq:model}
\end{equation}
where $X_i\in\X\subset\R^{p}$ are observed agent characteristics, $e_{ij}$ is a scalar unobservable, and $g$ is unknown, continuous, and strictly increasing in $e_{ij}$. Both $g$ and the distribution $F_e$ of $e_{ij}$ are treated nonparametrically. The model nests additive dyadic regression, random-coefficient specifications, and multiplicative ``random gravity'' models (Example \ref{ex:gravity} below), and it delivers objects that additive models cannot: quantile-structural functions in which the effect of the composite dyad-level shock varies with the pair's characteristics, and counterfactual outcome distributions whose shape, not only location, responds to the covariates.

In an additive specification $Y_{ij}=f(X_i,X_j)+e_{ij}$, covariate changes shift the conditional distribution without changing its shape, and the effect of the shock does not vary with the pair's characteristics. Model \eqref{eq:model} lets the same reduction in frictions raise trade modestly for a pair of small economies and substantially for a pair of large ones, and lets the researcher ask how the lower tail of trade responds to country characteristics, or what a given dyad's outcome would have been had its shock sat at a different quantile. The price is that the objects of interest are functionals of conditional \emph{distributions}, so the estimation theory cannot be borrowed from the dyadic conditional-mean literature, and dyadic dependence changes it in a first-order way.

The paper makes three contributions. \emph{First}, I give transparent conditions under which a normalized representative of the pair $(g,F_e)$ is point identified. The identification argument is a population-level statement about the conditional distribution of a single dyad's outcome, and it adapts the classical analysis of \cite{rosa2003} for nonadditive models with i.i.d.\ data; I state precisely the support and normalization conditions this adaptation requires, which include a choice of location--scale normalization $(\bar{x},\bar{e},\alpha)$ or, alternatively, homogeneity of degree one along rays of the covariate space. Identification here is deliberately modest: it is estimation and inference where dyadic data change the analysis fundamentally.

\emph{Second}, and this is the core of the paper, I develop the asymptotic theory of kernel plug-in estimators of $F_e$ and $g$ under dyadic dependence. Outcomes on dyads sharing an agent covary: exports from China to France covary with exports from China to the United States through Chinese fundamentals. Formally, the array $\{(Y_{ij},X_i,X_j)\}$ is jointly exchangeable and dissociated, so that by the Aldous--Hoover representation \citep{aldous1981} the composite error can be written $e_{ij}=\tau(U_i,U_j,V_{ij})$ for i.i.d.\ agent components $U_i$ and pair shocks $V_{ij}$. The variance of the dyadic kernel distribution estimator has \emph{two} contributions: a shared-agent contribution of order $(Nh^{d})^{-1}$, obtained from its H\'ajek projection onto agent-level components, and a pair-level contribution of order $(nh^{2d})^{-1}$, where $N$ is the number of agents, $n=N(N-1)$ the number of directed dyads, and $d$ the dimension of the conditioning subvector. Consistency requires $Nh^{d}\to\infty$, so the shared-agent term dominates unless it is exactly zero, and variance formulas that treat the $n$ dyads as an i.i.d.\ sample are too small by a factor that grows without bound. The correct limit theory is a two-regime CLT: rate $\sqrt{Nh^{d}}$ with projection variance $\Omega_1$ in the generic case, and, under the stronger condition (S) of Section \ref{consistency}, rate $\sqrt{nh^{2d}}$ with variance $\Omega_2$ plus a reverse-dyad contribution when the conditioning coordinates coincide; zero projection variance alone does not deliver the second limit. This is the structural-estimation analog of the degeneracy phenomenon in \cite{graham2022kernel} and \cite{cattaneo2024}; see \cite{NBERw28548} for the corresponding minimax theory. Because the estimator smooths only over the $2d$ conditioning coordinates, standard second-order positive kernels satisfy all rate conditions in the generic regime, so the plug-in quantile inversion remains well defined.

The mechanism is simple and applies to any plug-in estimator built from dyadic data. The sample contains $n=N(N-1)$ dyads but only $N$ independent draws of agent-level randomness, since agent $i$'s type enters all $2(N-1)$ dyads involving $i$; averaging over dyads drives down pair-level noise at the rate of the number of dyads but cannot average away agent-level variation faster than the number of agents allows. At a conditioning point roughly $Nh^{d}$ agents contribute, so agent-level variation imposes a variance floor of order $(Nh^{d})^{-1}$, of which the pair-level term $(nh^{2d})^{-1}$ is essentially the square; consistency forces $Nh^{d}\to\infty$, so the floor dominates whenever present. It is absent exactly when $\Omega_1=0$, which can occur through cancellation even with agent effects, and the faster Gaussian limit is established under condition (S).

\emph{Third}, because $\Omega_1$ depends on conditional moments of unobservables and is unpleasant to estimate directly, I propose an agent-level (``pigeonhole'') bootstrap that resamples agents with replacement and rebuilds the induced dyads. The scheme replicates the shared-agent dependence automatically \citep{davezies2021,menzel2021}, and Theorem \ref{theo:boot} proves its consistency in the nondegenerate regime, which obtains whenever the estimator is consistent and $\Omega_1>0$; in the degenerate designs examined it is conservative, covering at $0.985$ or above. In simulations it raises coverage to $0.90$ to $0.94$ against a nominal $0.95$, for both the error distribution and the structural function, exactly where i.i.d.-formula intervals fail.

The Monte Carlo evidence of Section \ref{simulation} quantifies each element of this story with the estimands held fixed across designs, to the accuracy reported in Appendix A.14. With independent errors, independence-formula intervals cover near the nominal 95\% level; a common agent component inducing a pairwise error correlation of one third collapses coverage to $0.42$ at $N=50$ and $0.29$ at $N=200$, coverage \emph{falling} as the network grows because the neglected variance component grows relative to the term the formula captures. Computed analytically for the design, $\Omega_1$ predicts the Monte Carlo variance to within two percent at every sample size, and the agent bootstrap raises coverage to between $0.90$ and $0.94$. A third experiment shows that under dyadic dependence a naive leave-one-dyad-out criterion selects bandwidths less than half the oracle value and pays about $40$ percent in mean squared error, while the leave-pair-out criterion tracks the oracle. The prescriptions for applied work are concrete: leave out pairs when selecting bandwidths, undersmooth for inference, and resample agents, never dyads.

\paragraph{Related literature.} This paper sits at the intersection of two literatures that have developed rapidly. The first is nonparametric reduced-form analysis of dyadic data: \cite{graham2022kernel} study kernel density estimation for undirected dyadic data and uncover the degeneracy dichotomy central to this paper; \cite{NBERw28548} derive minimax rates for dyadic Nadaraya--Watson regression; \cite{cattaneo2024} and \cite{chiang2023} provide uniform distributional approximations and inference for dyadic kernel estimators and exchangeable arrays; \cite{davezies2021} develop the empirical-process theory for jointly exchangeable arrays on which my proofs draw; \cite{menzel2021} studies bootstrap inference under multiway cluster dependence, including the dyadic case and its degeneracies; see \cite{GRAHAM202023} and \cite{GRAHAM2020111} for surveys. \cite{shimizu2025} derives kernel regression asymptotics under \emph{cluster sampling} with growing cluster sizes and finds the same broad phenomenon, a within-group covariance term indexed by $\lambda=\lim n^{-1}\sum_g n_g^2h^{d}$ that i.i.d.\ formulas miss. Dyadic data lie outside that framework: each dyad belongs to \emph{two} overlapping agent-groups, so no partition into independent blocks exists, and if each agent is viewed as a cluster of size $2(N-1)$ the analog of $\lambda$ is of order $Nh^{d}\to\infty$, the divergent case his assumptions exclude. Where a finite $\lambda$ inflates the variance at an unchanged rate, the dyadic overlap changes the rate itself. Closest in the object studied is \cite{chen2025}, who develops dyadic-robust inference for \emph{linear} quantile regression, with a sandwich covariance estimator and $t$ and Wald statistics for undirected and directed networks, applied to a gravity specification on the \cite{silva2006} data. The overlap is real: under Assumption \ref{ass:structure}, $g(x_1,x_2,e_q)$ at $e_q=F_e^{-1}(q)$ \emph{is} the $q$-th conditional quantile of $Y_{ij}$ (Section \ref{sec:sep}). Three things separate the papers. His quantile is linear in a fixed covariate list and the estimand is a coefficient vector at a parametric rate; here $g$ is unrestricted in $(X_i,X_j)$ and the rate is nonparametric, which is what makes the degeneracy dichotomy arise. His analysis is at a fixed quantile level and does not identify $F_e$, so it delivers no counterfactual indexed by the shock; the normalization of Section \ref{sec:id} is what converts quantile curves into such an object. And his two asymptotic frameworks, adapted from \cite{tabordmeehan2019} and \cite{janson1988}, are indexed by the \emph{density} of the dyad set, whereas the two regimes here are indexed by degeneracy of the H\'ajek projection, a property of the error structure; the dichotomies do not correspond. His finding that dyadic-robust standard errors are about twice their independence-based counterparts on trade data corroborates the magnitude in Section \ref{empirical} for a quite different estimator. \cite{canen2024} show that the standard dyadic-robust variance estimator fails when network spillovers link dyads sharing no agent; dissociation is exactly the assumption their spillovers violate. Concurrently, \cite{hounyo2026} establish a uniform projection theorem for dyadic processes and a corrected Gaussian bootstrap valid in both regimes, for linear conditional means under a null of correct specification; their correction is the natural refinement of the resampling scheme of Section \ref{sec:boot}. Relative to this literature, the estimands here are \emph{structural}, $g$ and $F_e$ rather than conditional means or densities, and the plug-in map involves a quantile inversion whose asymptotic behavior under dyadic dependence appears not to have been characterized before.

The second literature concerns unobserved heterogeneity in network models. \cite{Graham} studies link formation with degree heterogeneity; \cite{auerbach2022} identifies a partially linear model with agent-level unobservables using network data; \cite{andrei} identifies and estimates network models in which agent unobservables interact through an unknown coupling function, with an additive idiosyncratic error. Closest in spirit, \cite{diegert2024} establish nonparametric identification of the conditional distribution of a dyadic outcome given two-way unobserved heterogeneity, and of the heterogeneity distribution itself, in a fully nonseparable model $Y_{ij}=H(U_i,U_j,V_{ij})$ without covariates. Their analysis is identification-only and deliberately agnostic about functional structure; it does not deliver a structural response function with respect to observed covariates, nor estimation or inference theory. The present paper is complementary: by imposing scalar monotonicity of the composite shock and independence from covariates, I obtain the structural function $g$ itself, and hence quantile counterfactuals in $(X_i,X_j)$, together with a complete estimation and inference theory. The monotone-scalar-unobservable tradition follows \cite{rosa2003} and \cite{matzkin2015estimation}; see also \cite{vuong2017counterfactual} for counterfactual mappings in nonseparable models.

Two scope limitations deserve emphasis at the outset. Because I assume $e_{ij}\perp(X_i,X_j)$, the model is of the \emph{random effects} type: through the Aldous--Hoover representation, the agent components $U_i$ entering $e_{ij}$ must be independent of observed characteristics. Correlated heterogeneity of the fixed-effects type studied by \cite{Graham}, \cite{auerbach2022}, and \cite{andrei} is ruled out. Second, because $e_{ij}$ is a scalar composite of $(U_i,U_j,V_{ij})$, counterfactuals are with respect to the composite dyad-level shock; the model does not separately identify the roles of the two agents' unobservables, and it cannot answer questions such as ``what if agent $i$'s unobservable were higher in all of its relationships.'' Section \ref{conclusion} discusses extensions in both directions.

The remainder of the paper proceeds as follows. Section \ref{sec:model} presents the model and the identification analysis. Section \ref{sec:estimation} defines the estimators. Section \ref{consistency} develops the two-regime asymptotic theory and the bootstrap. Section \ref{simulation} reports the Monte Carlo evidence. Section \ref{empirical} applies the methods to bilateral trade among the world's largest economies. Section \ref{conclusion} concludes. All proofs are in Appendix A; Appendix B extends the framework to observed dyad-level covariates.

\section{The Model}\label{sec:model}

\subsection{Setup and assumptions}

A random sample of $N$ agents is drawn from a large population; agent $i$ carries an observed characteristic vector $X_i\in\X\subset\R^{p}$. For each ordered pair $i\neq j$ the econometrician observes the directed outcome $Y_{ij}$ (with the convention $Y_{ii}=0$), so the sample contains $n=N(N-1)$ directed dyads. Outcomes are generated by model \eqref{eq:model}, where $g:\X\times\X\times\EE\to\R$ and the scalar unobservable $e_{ij}$ is supported on $\EE\subset\R$.

The first assumption specifies the dependence structure of the data.

\begin{assumption}[Sampling and dependence structure]\label{ass:sampling}
~\begin{enumerate}[(i)]
\item $\{(X_i,U_i)\}_{i=1}^N$ is i.i.d., where $U_i$ is an unobserved agent-level random element;
\item $\{(V_{ij},V_{ji})\}_{i<j}$ is i.i.d.\ across unordered pairs, with $(V_{ij},V_{ji})$ exchangeable, and independent of $\{(X_i,U_i)\}_{i=1}^N$;
\item $e_{ij}=\tau(U_i,U_j,V_{ij})$ for an unknown measurable function $\tau$;
\item $(U_i,V_{ij})_{i\ne j}$ is independent of $(X_1,\dots,X_N)$.
\end{enumerate}
\end{assumption}

Assumption \ref{ass:sampling}(i)--(iii) states that the array $\{(Y_{ij},X_i,X_j)\}$ is jointly exchangeable and dissociated; by the representation theorem of \cite{aldous1981} this is the canonical structure for dyadic data generated from a random sample of agents, and it delivers as a \emph{consequence} the familiar ``local dependence'' property: variables on dyads sharing at least one agent may be dependent, while variables on dyads sharing no agent are independent. For instance, exports from China to France may covary with exports from China to the United States (through $U_{\text{China}}$), but not with exports from Canada to Nigeria. Deriving local dependence from the exchangeable-array representation, rather than imposing it verbally, is what makes the structure usable in proofs. Part (iv) is the substantive restriction: it makes the model a \emph{random effects} model, ruling out correlation between agents' unobserved components and their observed characteristics. Special cases include $e_{ij}=U_i+U_j+V_{ij}$ \citep{graham2022kernel} and $e_{ij}=h(U_i,U_j)+V_{ij}$ \citep{andrei}.

\begin{assumption}[Structural function and error distribution]\label{ass:structure}
~\begin{enumerate}[(i)]
\item $g$ is continuous on $\X\times\X\times\EE$;
\item $g(x_1,x_2,\cdot)$ is strictly increasing on $\EE$ for every $(x_1,x_2)\in\X\times\X$;
\item $e_{ij}$ has a continuous, strictly increasing distribution function $F_e$ on $\EE$;
\item the joint density $f_X$ of $(X_i,X_j)$ satisfies $f_X(x_1,x_2)>0$ for all $(x_1,x_2)$ in the support $\X\times\X$.
\end{enumerate}
\end{assumption}

Under Assumption \ref{ass:sampling}(iv), $e_{ij}\perp(X_i,X_j)$ with common marginal $F_e$ across dyads. Assumption \ref{ass:structure}(i)--(ii) guarantees that $g(x_1,x_2,\cdot)$ has a continuous strictly increasing inverse in its last argument, denoted $m(x_1,x_2,\cdot)$, and part (iii) guarantees that conditional distribution functions of $Y_{ij}$ are continuous and strictly increasing where they matter, so that the quantile inversions below are well defined. For all $(x_1,x_2)\in\X^2$ and $y$,
\begin{align}
F_{Y|X}(y|x_1,x_2)&=\PP\left(g(X_i,X_j,e_{ij})\le y\,|\,X_i=x_1,X_j=x_2\right)
= F_e\!\left(m(x_1,x_2,y)\right),\label{eq:map}
\end{align}
so the observable conditional distribution is a known functional of the two structural unknowns. Note that strict monotonicity in a \emph{scalar} unobservable rules out discrete or censored outcomes; the weak-monotonicity extension, which requires generalized inverses and delivers only partial identification of $F_e$ on flat regions of $g$, is left to future work.

A word on direction. The model treats direction symmetrically: the same $g$ governs $Y_{ij}$ and $Y_{ji}$, with $(e_{ij},e_{ji})$ exchangeable by Assumption \ref{ass:sampling}. The estimators below pool ordered pairs, which is appropriate under this symmetry. For undirected data ($Y_{ij}=Y_{ji}$) all results hold with $n=N(N-1)/2$ and obvious modifications.

\subsection{Identification}\label{sec:id}

Let $\F$ be the set of continuous, strictly increasing distribution functions, $\G$ the set of functions satisfying Assumption \ref{ass:structure}(i)--(ii), and $\M$ the corresponding set of inverse functions. Identification follows \cite{rosa2003}: the observational-equivalence argument, the normalizations, and the quantile-matching structure below are hers, restated in dyadic notation. I add the support conditions the argument requires and the observation that it is indifferent to cross-dyad dependence.

\begin{definition}
$(g,F_e)$ is identified in $\G\times\F$ if for all $(g',F'_e)\in\G\times\F$, $F_{Y,X}(\cdot\,;g,F_e)=F_{Y,X}(\cdot\,;g',F'_e)$ implies $(g,F_e)=(g',F'_e)$, where $F_{Y,X}(\cdot\,;g',F'_e)$ denotes the joint distribution of $(Y_{ij},X_i,X_j)$ generated by $(g',F'_e)$.
\end{definition}

\begin{definition}
$g'\in\G$ is observationally equivalent to $g\in\G$ if there exist $F_e,F'_e\in\F$ with $F_{Y,X}(\cdot\,;g,F_e)=F_{Y,X}(\cdot\,;g',F'_e)$.
\end{definition}

\begin{lemma}[Observational equivalence]\label{lem:obseq}
$m$ and $m'$ are observationally equivalent if and only if there exists a continuous, strictly increasing $u:\EE\to\R$ such that $m'=u\circ m$.
\end{lemma}

Lemma \ref{lem:obseq} says that the data pin down $m$ (equivalently $g$) only up to a monotone relabeling of the unobservable: $(u\circ m,\,F_e\circ u^{-1})$ generates the same observables as $(m,F_e)$. Identification therefore requires selecting a representative from each equivalence class. The next lemma, an \emph{existence-of-representative} result rather than a property of the true $g$, shows that a convenient normalization is always available.

\begin{lemma}[Normalization]\label{lem:norm}
Fix $(\bar{x}_1,\bar{x}_2)\in\X\times\X$. For every $g\in\G$ there exists an observationally equivalent $g^\ast\in\G$ (with associated error distribution $F^\ast_e\in\F$) such that
\begin{equation}
g^\ast(\bar{x}_1,\bar{x}_2,e)=e \quad\text{for all } e. \label{eq:norm}
\end{equation}
\end{lemma}

Henceforth ``the'' identified objects are the representative $(g^\ast,F_e^\ast)$ satisfying \eqref{eq:norm}; I drop the asterisk. Partition $X_i=(X_i^0,X_i^1)$ and strengthen \eqref{eq:norm} to
\begin{equation}
g(x^0_1,\bar{x}^1_1,x^0_2,\bar{x}^1_2,e)=e \qquad\text{for all } x^0_1,x^0_2 \text{ and given } (\bar{x}^1_1,\bar{x}^1_2). \label{eq:norm2}
\end{equation}
Unlike \eqref{eq:norm}, this is \emph{not} a free normalization: Lemma \ref{lem:norm} delivers the identity at a single point through a single monotone relabeling $u$, and imposing it at every $x^0$ requires the same $u$ to invert $g(x^0_1,\bar x^1_1,x^0_2,\bar x^1_2,\cdot)$ for all $x^0$, which holds only if that map does not depend on $x^0$. Equation \eqref{eq:norm2} thus combines a normalization with the restriction that $g$ at the $X^1$-normalization point be invariant in $X^0$; the following example gives a leading class in which it holds.

\begin{example}
Restriction \eqref{eq:norm2} holds for the random-coefficient structure $g(x_1,x_2,e)=e\cdot s(x^1_1,x^1_2)+t(x_1,x_2)$ whenever $t$ vanishes and $s$ equals one at $x^1=\bar{x}^1$, since then $g$ at the normalization point is the identity in $e$ for every $x^0$.
\end{example}

\begin{assumption}\label{ass:condind}
$e_{ij}$ is independent of $(X^1_i,X^1_j)$ conditional on $(X^0_i,X^0_j)$, with the same partition $X=(X^0,X^1)$ as in \eqref{eq:norm2}. Moreover, for every $(x^0_1,x^0_2)$ in the support of $(X^0_i,X^0_j)$ the conditional distribution function $F_{e|X^0}(\cdot\mid x^0_1,x^0_2)$ is continuous and strictly increasing on $\EE$.
\end{assumption}

The second clause is not implied by the first, nor by continuity of the marginal $F_e$, and without it Theorem \ref{t1} is false. Take $U_i$ i.i.d.\ standard normal, $X^0_i=U_i$, $e_{ij}=U_i+U_j$, $g(x_1,x_2,e)=e$, $X^1_i$ an independent standard normal, and $\bar x^1_1=\bar x^1_2=0$. The marginal error law is $N(0,2)$, and conditional on $(X^0_i,X^0_j)$ the error is trivially independent of $(X^1_i,X^1_j)$, being degenerate at $x^0_1+x^0_2$. The conditional outcome distribution is then a step function, the inversion in \eqref{eq:idg} returns the atom rather than $g(x_1,x_2,e)$, and $g$ is unidentified off the conditional support: $g'(x_1,x_2,e)=e+A(x^1_1,x^1_2)\left(e-x^0_1-x^0_2\right)$ with $A=\left((x^1_1)^2+(x^1_2)^2\right)/\left(1+(x^1_1)^2+(x^1_2)^2\right)$ is continuous, strictly increasing in $e$, satisfies \eqref{eq:norm2}, and agrees with $g$ at every observed dyad yet differs elsewhere. Under Assumption \ref{ass:sampling}(iv) the clause is automatic, since $F_{e|X^0}=F_e$.

\begin{assumption}[Support]\label{ass:support}
$(x^0_1,\bar{x}^1_1,x^0_2,\bar{x}^1_2)\in\X\times\X$ for all $(x^0_1,x^0_2)$ in the support of $(X^0_i,X^0_j)$.
\end{assumption}

Assumption \ref{ass:support} is substantive: the normalization point $\bar{x}^1$ must be attainable together with every value of the free coordinates.

\begin{theorem}[Identification]\label{t1}
Let Assumptions \ref{ass:sampling}(i)--(iii), \ref{ass:structure}, \ref{ass:condind} and \ref{ass:support} and normalization \eqref{eq:norm2} hold. Then for all $e\in\EE$ and $x_1=(x^0_1,x^1_1),x_2=(x^0_2,x^1_2)\in\X$:
\begin{align}
F_{e|X^0}(e|x^0_1,x^0_2) &= F_{Y|X}\!\left(e\,\big|\,(x^0_1,\bar{x}^1_1),(x^0_2,\bar{x}^1_2)\right),\label{eq:idF}\\
g(x_1,x_2,e) &= F_{Y|X}^{-1}\!\left(F_{Y|X}\!\left(e\,\big|\,(x^0_1,\bar{x}^1_1),(x^0_2,\bar{x}^1_2)\right)\,\Big|\,x_1,x_2\right).\label{eq:idg}
\end{align}
If in addition $e_{ij}\perp(X_i,X_j)$ (Assumption \ref{ass:sampling}(iv)) and $X^0$ is not an argument of $g$, then $F_e(e)=F_{Y|X^1}(e|\bar{x}^1_1,\bar{x}^1_2)$ and $g(x_1,x_2,e)=F^{-1}_{Y|X^1}\big(F_{Y|X^1}(e|\bar{x}^1_1,\bar{x}^1_2)\,\big|\,x^1_1,x^1_2\big)$.
\end{theorem}

Equation \eqref{eq:idg} has the familiar quantile-matching interpretation: $g(x_1,x_2,e)$ is the same quantile of $Y_{ij}$ given $(x_1,x_2)$ that $e$ is of $e_{ij}$.

Theorem \ref{t1} runs entirely through the conditional distribution of a single dyad's outcome, so dependence of the errors across dyads plays no role in identification: all that is used is the conditional law of $e_{ij}$ given its own pair's covariates and the fact that this law is common across dyads, both delivered by Assumption \ref{ass:sampling} however strongly dyads sharing an agent covary. Cross-dyad dependence affects precision, the subject of Section \ref{consistency}, not what is identified. By the same token the identified object is the marginal $F_e$, not the joint law of the shocks, so counterfactuals that move a single dyad's shock are within reach while those that move one agent's component in all its relationships are not. The boundary for identification is Assumption \ref{ass:condind}: correlation between $e_{ij}$ and $X^0$ is not fatal, since the argument conditions on $X^0$, but correlation with $X^1$ given $X^0$, or a conditional error law with atoms or gaps in $\EE$, would break it. Under Assumption \ref{ass:sampling}(iv) both concerns disappear.

An alternative normalization exploits homogeneity restrictions implied by economic theory.

\begin{assumption}[Homogeneity of degree one]\label{ass:hd1}
The error support $\EE$ lies strictly on one side of zero and $\bar e\in\EE$, so that $e/\bar e>0$ for every $e\in\EE$. The value $g(x^0_1,\bar{x}^1_1,x^0_2,\bar{x}^1_2,\bar{e})$ does not depend on $(x^0_1,x^0_2)$; write $\alpha$ for it. And for all $x^0$ and all $\lambda>0$ with $\lambda\bar{x}^1\in\X^1$ and $\lambda\bar{e}\in\EE$,
$g(x^0_1,\lambda\bar{x}^1_1,x^0_2,\lambda\bar{x}^1_2,\lambda\bar{e})=\lambda\alpha$.
\end{assumption}

\begin{example}[Random gravity model]\label{ex:gravity}
$Y_{ij}=X_i^{\,a}X_j^{\,b}e_{ij}^{\,1-a-b}$ is homogeneous of degree one in $(X_i,X_j,e_{ij})$.
\end{example}

\begin{assumption}[Ray support]\label{ass:ray}
$(x^0_1,(e/\bar{e})\bar{x}^1_1,x^0_2,(e/\bar{e})\bar{x}^1_2)\in\X\times\X$ for all $e\in\EE$ and all relevant $(x^0_1,x^0_2)$.
\end{assumption}

Assumption \ref{ass:ray} is a large-support condition along the ray $\{\lambda\bar{x}^1:\lambda\in\EE/\bar{e}\}$; it restricts how far the identification travels with bounded covariates. The point $(\bar{x}^1,\bar{e})$ is chosen by the econometrician and $\alpha$ fixes the scale of the error, so that together they select one representative from the observationally equivalent class; they are \emph{not} estimated. Under Assumptions \ref{ass:sampling}--\ref{ass:condind}, \ref{ass:hd1}, and \ref{ass:ray}, arguments identical to Theorem \ref{t1} give, for the case where $X^0$ is not an argument of $g$,
\begin{align}
F_e(e) &= F_{Y|X^1}\!\left(\tfrac{e}{\bar{e}}\,\alpha\,\Big|\,\tfrac{e}{\bar{e}}\bar{x}^1_1,\tfrac{e}{\bar{e}}\bar{x}^1_2\right),\label{eq:idFhd1}\\
g(x_1,x_2,e) &= F_{Y|X^1}^{-1}\!\left(F_{Y|X^1}\!\left(\tfrac{e}{\bar{e}}\,\alpha\,\Big|\,\tfrac{e}{\bar{e}}\bar{x}^1_1,\tfrac{e}{\bar{e}}\bar{x}^1_2\right)\Big|\,x^1_1,x^1_2\right).\label{eq:idghd1}
\end{align}
The general case with $X^0$ present is analogous and omitted.

\section{Estimation}\label{sec:estimation}

All identifying formulas are compositions of conditional distribution functions of $Y_{ij}$ given subvectors of $(X_i,X_j)$ and their inverses. Let $W_i$ denote a subvector of $X_i$ with $\dim(W_i)=d$, and let $w=(w_1,w_2)$ be an evaluation point. The plug-in estimator smooths \emph{only over the $2d$ conditioning coordinates and the outcome}:
\begin{equation}
\widehat{F}_{Y|W}(y|w_1,w_2)
=\frac{\sum_{1\le i\ne j\le N}\bar{k}\!\left(\frac{y-Y_{ij}}{h_y}\right)K\!\left(\frac{w_1-W_i}{h}\right)K\!\left(\frac{w_2-W_j}{h}\right)}
{\sum_{1\le i\ne j\le N}K\!\left(\frac{w_1-W_i}{h}\right)K\!\left(\frac{w_2-W_j}{h}\right)},\label{eq:Fhat}
\end{equation}
where $K:\R^d\to\R$ is a kernel, $\bar{k}(t)=\int_{-\infty}^{t}k(s)\,ds$ for a univariate kernel $k$, and $h,h_y\to0$ are bandwidths. Two features of \eqref{eq:Fhat} are worth noting. It does not require estimating the full $(2p+1)$-dimensional joint density, only the $2d$-dimensional localization actually needed, which weakens the smoothness and bandwidth requirements substantially; and it allows a separate outcome bandwidth $h_y$, matching how such estimators are implemented in practice. When $K$ and $k$ are everywhere-positive densities (e.g.\ Gaussian), $\widehat{F}_{Y|W}(\cdot|w_1,w_2)$ is continuous and strictly increasing, so its inverse
\[
\widehat{F}^{-1}_{Y|W}(s|w_1,w_2)=\inf\{y:\widehat{F}_{Y|W}(y|w_1,w_2)\ge s\}
\]
is a well-defined point. (With kernels that are not everywhere positive, monotone rearrangement \citep{chernozhukov2009} restores monotonicity without affecting first-order asymptotics, and the inverse is then read as a generalized inverse.)

The estimators of $F_e$ and $g$ replace population objects in \eqref{eq:idF}--\eqref{eq:idg} or \eqref{eq:idFhd1}--\eqref{eq:idghd1} by their kernel counterparts. For instance, under the homogeneity normalization with $X^0$ absent,
\begin{align}
\widehat{F}_e(e)&=\widehat{F}_{Y|X^1}\!\left(\tfrac{e}{\bar{e}}\alpha\,\Big|\,\tfrac{e}{\bar{e}}\bar{x}^1_1,\tfrac{e}{\bar{e}}\bar{x}^1_2\right),\nonumber\\
\widehat{g}(x_1,x_2,e)&=\widehat{F}^{-1}_{Y|X^1}\!\left(\widehat{F}_e(e)\,\Big|\,x^1_1,x^1_2\right).\label{eq:ghat}
\end{align}

Bandwidth choice can be disciplined by a cross-validation criterion that respects the dependence structure, in the spirit of the cluster-sampling cross-validation of \cite{shimizu2025}. Let $\psi$ be a measurable transform of the outcome with $\E[\psi(Y_{ij})^2\mid W_i,W_j]$ bounded, and $m_\psi(x_1,x_2)=\E[\psi(Y_{ij})\mid W_i=x_1,W_j=x_2]$; the choices $\psi(y)=y$ and $\psi(y)=\ind(y\le t)$ cover the conditional-mean pilot and the conditional distribution function at any threshold. For each ordered dyad $(i,j)$ let $\widehat m_{\psi,-(ij)}(\cdot\,;h)$ denote the kernel estimator computed from the subsample of dyads sharing \emph{no} index with $(i,j)$, with the convention that a ratio with vanishing denominator equals zero, and define the \emph{leave-pair-out} criterion
\[
\mathrm{CV}_\psi(h)=\frac{1}{n}\sum_{i\ne j}\omega(W_i,W_j)\left[\psi(Y_{ij})-\widehat m_{\psi,-(ij)}(W_i,W_j;h)\right]^{2},
\]
for a bounded nonnegative weight $\omega$. Deleting \emph{both} agents of the held-out dyad is essential: dyads sharing even one agent with $(i,j)$ are dependent with it, and a criterion that keeps them in the training sample systematically understates out-of-sample error and selects too small a bandwidth.

\begin{theorem}[Dyadic cross-validation]\label{theo:cv}
Under Assumption \ref{ass:sampling}, for bounded $\psi$ (or, more generally, whenever $\E[\psi(Y_{ij})^2\mid W_i,W_j]$ is bounded),
\begin{multline*}
\E\left[\mathrm{CV}_\psi(h)\right]=\E\left[\omega(W_i,W_j)\,\sigma^2_\psi(W_i,W_j)\right]\\
+\E\left[\omega(W_i,W_j)\left\{m_\psi(W_i,W_j)-\widehat m_{\psi,-(ij)}(W_i,W_j;h)\right\}^{2}\right],
\end{multline*}
where $\sigma^2_\psi(x_1,x_2)=\Var(\psi(Y_{ij})\mid W_i=x_1,W_j=x_2)$.
\end{theorem}

The first term does not involve $h$, so minimizing the expected criterion is exactly equivalent to minimizing $R_{N-2}(h)\equiv\E[\omega\{m_\psi-\widehat m_{\psi,-(ij)}\}^2]$, the weighted risk of the estimator \eqref{eq:Fhat} computed from $N-2$ agents. The decomposition is finite-sample and exact, the dyadic counterpart of the cluster-sampling result of \cite{shimizu2025}; the proof, in Appendix A.11, rests only on the fact that a dyad and the subsample sharing no agent with it are independent under Assumption \ref{ass:sampling}.

Two further properties of the criterion are worth recording, though neither delivers a selection theorem. The first is that $\mathrm{CV}_\psi$ concentrates around its expectation on a fixed-size grid; the second is that discarding two of $N$ agents shifts the risk-optimal bandwidth by a negligible amount. The discussion after the corollary explains why concentration in levels does \emph{not} imply that the minimizer of the criterion agrees with the minimizer of the risk, and that claim is not made.

\begin{corollary}[Concentration and the effect of deleting two agents]\label{cor:cv}
Let $\psi$ be bounded, which for part (i) is stronger than the condition imposed in Theorem \ref{theo:cv} and covers the choice $\psi(y)=\ind(y\le t)$ used throughout but not the unbounded pilot $\psi(y)=y$; let $\omega$ be bounded, nonnegative, and supported in a compact set $\mathcal S$ contained in the interior of the support of $(W_i,W_j)$ and on which $f_W$ is bounded away from zero. Let $\mathcal H_N=\{h_{1,N},\dots,h_{J,N}\}$ be a grid of $J$ bandwidth \emph{sequences}, $J$ fixed and not depending on $N$, each satisfying Assumption \ref{ass:bw}, and write $\varepsilon_N=\max_{h\in\mathcal H_N}(Nh^{2d})^{-1/2}$. Let $\widehat h=\argmin_{h\in\mathcal H_N}\mathrm{CV}_\psi(h)$.
\begin{enumerate}[(i)]
\item $\max_{h\in\mathcal H_N}\left|\mathrm{CV}_\psi(h)-\E[\mathrm{CV}_\psi(h)]\right|=O_p(\varepsilon_N)$, and in particular $\pto0$.
\item The minimizers over $h>0$ of the leading risk $\underline R_M(h)=C_Bh^{4}+C_V/(Mh^{d})$, with $C_B,C_V>0$ not depending on $M$, satisfy $h^{\ast}_{N-2}/h^{\ast}_{N}=\left(N/(N-2)\right)^{1/(4+d)}=1+O(N^{-1})$.
\end{enumerate}
\end{corollary}

Part (i) is proved in Appendix A.12 by an Efron--Stein argument: replacing one agent moves $\mathrm{CV}_\psi$ by $O(1/(Nh^{d}))$, because that agent carries $O(Nh^{d})$ of the $O(N^2h^{2d})$ effective kernel weight at any evaluation point in $\mathcal S$, and replacing one pair shock moves it by $O(1/(N^2h^{2d}))$; summing the squared perturbations gives $\Var(\mathrm{CV}_\psi(h))=O(1/(Nh^{2d}))$ under Assumption \ref{ass:bw}. The trimming weight is what makes these bounds available uniformly over the random held-out evaluation points rather than only at a fixed interior point, and the proof pays for it with a concentration step for the leave-pair-out denominators.

What this does \emph{not} deliver is consistent selection of the risk-minimizing bandwidth, and the obstacle is instructive. Let $h^{\circ}_N$ minimize $R_{N-2}$ on the grid and $\Delta_N$ be the gap to the second-best risk; since $\E[\mathrm{CV}_\psi(h)]=\text{const}+R_{N-2}(h)$, an error uniformly below $\Delta_N/2$ would force $\widehat h=h^{\circ}_N$. But Assumption \ref{ass:bw} requires $Nh^{d+4}\to0$, so the risk is variance-dominated, $R_{N-2}(h)\asymp C_V/(Nh^{d})$, and with $h_{\min}$ the smallest grid bandwidth $\Delta_N\le\max_{\mathcal H_N}R_{N-2}=O\big((Nh_{\min}^{d})^{-1}\big)$ while $\varepsilon_N=(Nh_{\min}^{2d})^{-1/2}$. Hence
\[
\frac{\Delta_N}{\varepsilon_N}\;\le\;C\,\frac{(Nh_{\min}^{d})^{-1}}{(Nh_{\min}^{2d})^{-1/2}}\;=\;C\,N^{-1/2}\;\longrightarrow\;0,
\]
the reverse of what the argument needs, by a factor of $\sqrt N$, and coarsening the grid does not help. The obstruction is that part (i) bounds the fluctuation of $\mathrm{CV}_\psi$ in \emph{levels}, which is dominated by the irreducible term $\E[\omega\sigma^2_\psi]$ of Theorem \ref{theo:cv}; that term does not depend on $h$ and cancels from every comparison the selector makes. A selection result must bound the fluctuation of the \emph{differences} $\mathrm{CV}_\psi(h)-\mathrm{CV}_\psi(h')$, which are of smaller order, and establishing that under dyadic dependence is not attempted here. Theorem \ref{theo:cv} is unaffected, part (ii) concerns the leading term $\underline R_M$ only, and the case for the criterion rests on the exact decomposition together with the Monte Carlo evidence of Section \ref{simulation}, not on a selection theorem. Selection and undersmoothing are also separate steps: nothing forces the grid to satisfy the undersmoothing part of Assumption \ref{ass:bw}, and in Section \ref{empirical} all exhibits are computed at $h_{\mathrm{CV}}$ with the undersmoothed $h_{\mathrm{CV}}N^{-1/5}$ reported as a sensitivity check.

Three practical remarks. The normalization point matters for precision, since $\widehat{F}_e$ uses only data local to $(\bar{x}^1_1,\bar{x}^1_2)$, so $\bar{x}^1$ should sit in a high-density region. The limit theory requires undersmoothing $h$ relative to rule-of-thumb choices (Assumption \ref{ass:bw}) for a centered limit, and the Monte Carlo quantifies the cost of ignoring this. And boundary bias afflicts \eqref{eq:Fhat} near the edges of the support of $W$, so evaluation points should be interior.

\paragraph{Implementation.} The complete procedure, as used in Sections \ref{simulation} and \ref{empirical}, is the following.
\begin{enumerate}[1.]
\item \emph{Inputs.} The directed dyads $\{(Y_{ij},X_i,X_j):i\ne j\}$, the conditioning subvector $W$, the normalization point $\bar x^1$ (a high-density point, for instance the agent-level median of $W$) and, under the homogeneity normalization, $(\bar e,\alpha)$; Gaussian $K$ and $k$, so that $\bar k=\Phi$.
\item \emph{Bandwidths.} Choose $h$ by minimizing $\mathrm{CV}_\psi(h)$ over a grid, with $\widehat m_{\psi,-(ij)}$ computed from the dyads sharing no agent with $(i,j)$; set $h_y=\widehat\sigma_Y n^{-0.3}$. For inference under Assumption \ref{ass:bw}, replace $h$ by $hN^{-1/5}$.
\item \emph{Conditional distribution.} For each evaluation point compute $\widehat F_{Y|W}(y\mid w_1,w_2)$ from \eqref{eq:Fhat} on a grid of $y$ values, and its inverse by linear interpolation of the grid values.
\item \emph{Structural objects.} Evaluate the identifying formulas \eqref{eq:idF}--\eqref{eq:idg}, or \eqref{eq:idFhd1}--\eqref{eq:idghd1}, at the kernel estimates of step 3, which gives $\widehat F_e$ and $\widehat g$ as in \eqref{eq:ghat}; shock quantiles are $\widehat e_q=\widehat F_e^{-1}(q)$.
\item \emph{Inference.} Draw $I_1,\dots,I_N$ i.i.d.\ uniform on $\{1,\dots,N\}$, keep the dyads $(I_a,I_b)$ with $a\ne b$ and $I_a\ne I_b$, and repeat steps 3 and 4 on this sample with the bandwidths of step 2 held fixed; after $B$ repetitions form percentile intervals or bootstrap standard errors (Section \ref{sec:boot}).
\end{enumerate}

\section{Asymptotic Theory}\label{consistency}

This section develops the distribution theory for $\widehat{F}_{Y|W}$, $\widehat{F}_e$, and $\widehat{g}$ under the dyadic dependence implied by Assumption \ref{ass:sampling}. The natural route is to treat the $n$ dyads as an i.i.d.\ sample and invoke functional delta-method results of the kind in \cite{newey1994kernel}. That route is valid only in the degenerate regime $\Omega_1=0$, and then only under condition (S) of Theorem \ref{theo:cdf}(ii); under genuine dyadic dependence it misses the dominant variance term. I first explain the mechanism, then state the corrected results.

\subsection{The H\'ajek projection and the two variance regimes}\label{sec:hajek}

Fix an interior evaluation point $(w_1,w_2)$ with $f_W(w_1),f_W(w_2)>0$, where $f_W$ is the marginal density of $W_i$ (by random sampling the joint density of $(W_i,W_j)$ factors: $f(w_1,w_2)=f_W(w_1)f_W(w_2)$). Write the centered numerator kernel as
\[
\psi_{ij}(y)=K\!\left(\tfrac{w_1-W_i}{h}\right)K\!\left(\tfrac{w_2-W_j}{h}\right)
\left[\bar{k}\!\left(\tfrac{y-Y_{ij}}{h_y}\right)-F_{Y|W}(y|w_1,w_2)\right].
\]
Dyads sharing no agent contribute no covariance (dissociation). The variance of $\sum_{i\ne j}\psi_{ij}$ therefore has two nontrivial parts:
\begin{enumerate}[(a)]
\item \emph{Diagonal terms} ($O(N^2)$ of them, each $O(h^{2d})$): these produce, after normalization, the familiar term $\Omega_2/(nh^{2d})$ with
\[
\Omega_2(y;w)=\frac{F_{Y|W}(y|w)\left[1-F_{Y|W}(y|w)\right]}{f_W(w_1)f_W(w_2)}\left(\textstyle\int K^2\right)^{2}.
\]
This is exactly the i.i.d.-style variance that an independence calculation delivers.
\item \emph{Shared-agent terms} ($O(N^3)$ of them, each $O(h^{3d})$): for dyads $(i,j)$ and $(i,k)$, conditional on $W_i\approx w_1$, $W_j\approx w_2$, $W_k\approx w_2$, the indicators $\ind(Y_{ij}\le y)$ and $\ind(Y_{ik}\le y)$ covary through the shared type $\zeta_i=(X_i,U_i)$. Define the \emph{agent-deviation functions}
\begin{align*}
\delta_s(\zeta;y,w)&=\E\left[\PP\left(Y_{ij}\le y\mid \zeta_i,\zeta_j\right)\,\middle|\,\zeta_i=\zeta,\,W_j=w_2\right]-F_{Y|W}(y|w_1,w_2),\\
\delta_r(\zeta;y,w)&=\E\left[\PP\left(Y_{ji}\le y\mid \zeta_i,\zeta_j\right)\,\middle|\,\zeta_i=\zeta,\,W_j=w_1\right]-F_{Y|W}(y|w_1,w_2),
\end{align*}
the deviations of the outcome distribution induced by an agent's \emph{entire} type $\zeta_i=(X_i,U_i)$, observed and unobserved components alike, when it occupies the sender ($W_i$ near $w_1$) or receiver ($W_i$ near $w_2$) slot; both have conditional mean zero given $W_i$ by construction. Conditioning on the full type is essential: when $W_i$ is a strict subvector of $X_i$, components of $X_i$ excluded from $W_i$ shift the conditional outcome distribution and contribute to the shared-agent variance even if the errors are independent across dyads. When $W_i=X_i$ (the leading case), $\delta_s$ reduces to a function of $(U_i;\,X_i=w_1)$. The shared-agent covariances aggregate to a term $\Omega_1/(Nh^{d})$ with
\begin{equation}
\begin{split}
\Omega_1(y;w)=\left(\textstyle\int K^2\right)
\left\{\frac{\E\!\left[\delta_s(\zeta_i;y,w)^2\,\middle|\,W_i=w_1\right]}{f_W(w_1)}
+\frac{\E\!\left[\delta_r(\zeta_i;y,w)^2\,\middle|\,W_i=w_2\right]}{f_W(w_2)}\right\}\\
+\ind(w_1=w_2)\,C(y;w),
\end{split}\label{eq:Om1}
\end{equation}
where $C(y;w)=2\left(\int K^2\right)\E\!\left[\delta_s(\zeta_i;y,w)\,\delta_r(\zeta_i;y,w)\mid W_i=w_0\right]/f_W(w_0)$ is the sender--receiver cross-moment that arises when the two conditioning points coincide at $w_0$.
\end{enumerate}
The two normalized variance terms are $\Omega_1/(Nh^d)$ and $\Omega_2/(nh^{2d})$, and their ratio is of order $Nh^{d}$. Consistency of \eqref{eq:Fhat} requires $nh^{2d}\to\infty$, i.e.\ $Nh^{d}\to\infty$. Hence \emph{whenever the estimator is consistent, the shared-agent term dominates unless $\Omega_1= 0$}. This is the exact analog, for conditional distribution functions, of the degeneracy dichotomy in \cite{graham2022kernel} and \cite{cattaneo2024}, and it drives everything that follows.

Degeneracy is defined here by $\Omega_1=0$. At distinct conditioning points this asks that $\E[\delta_s^2\mid W_i=w_1]$ and $\E[\delta_r^2\mid W_i=w_2]$ both vanish; at a coincident point $w_1=w_2=w_0$ it asks only that $\E[(\delta_s+\delta_r)^2\mid W=w_0]=0$ by \eqref{eq:Om1}, so a sender effect exactly offset by a receiver effect suffices. Three implications, and their failing converses, are what the results below rely on. Condition (S) of Theorem \ref{theo:cdf}(ii), which requires the conditional law of $(Y_{ij},Y_{ji})$ given \emph{both} full types to depend on them only through $(W_i,W_j)$, implies $\Omega_1=0$. Dyad-independent errors, $e_{ij}=\tau(V_{ij})$, imply $\Omega_1=0$ when $W_i=X_i$, but not when $W_i$ is a strict subvector of $X_i$, because excluded covariates then contribute to the shared-agent variance. Neither converse holds: the Rademacher design below has $W_i=X_i$, a genuine agent component, and $\Omega_1=0$, and it violates (S). Generic network dependence makes $\Omega_1>0$, so degeneracy is the exception, but it is not equivalent to the absence of agent effects.

That $\Omega_1=0$ does not imply dyad-independent errors is easy to see. Let $U_i$ be i.i.d.\ Rademacher, $V_{ij}$ i.i.d.\ $N(0,\sigma^2)$, and $e_{ij}=U_iU_j+V_{ij}$, with covariates independent of both. Conditional on either sign of $U_i$ the law of $e_{ij}$ is the same mixture $\tfrac12N(1,\sigma^2)+\tfrac12N(-1,\sigma^2)$, so $\delta_s\equiv\delta_r\equiv0$ and $\Omega_1=0$; indeed $e_{ij}$ and $e_{ik}$ are pairwise \emph{independent}, their joint law being an even mixture of product measures with identical margins. What fails is higher-order independence: around a triangle $(U_iU_j)(U_jU_k)(U_kU_i)=1$ identically, and conditioning on \emph{both} types reveals $U_iU_j$, so condition (S) fails. Theorem \ref{theo:cdf}(ii) does not cover this configuration, because the surviving variance component is a completely degenerate $U$-statistic with a Gaussian-chaos rather than normal limit (Appendix A.10). Thus $\Omega_1=0$ locates the rate but does not by itself deliver a normal limit; condition (S) does.

Two features of this characterization are worth recording. It is local to the evaluation point rather than global, so an array can be degenerate at one point and not at another. And at \emph{distinct} conditioning points $w_1\ne w_2$ the two deviations enter through separate terms of \eqref{eq:Om1}, so $\Omega_1=0$ there does require $\delta_s$ and $\delta_r$ to vanish separately; at a coincident point they enter only through $\delta_s+\delta_r$, and exact cancellation suffices. The antisymmetric design following Proposition \ref{prop:regimeZ} in Appendix B is an instance in which that cancellation occurs with both deviations far from zero. Generic network dependence, and also partial conditioning ($d<p$) with covariate-dependent outcomes, implies $\Omega_1>0$; degeneracy is the exception rather than the rule, but it is not equivalent to the absence of agent effects, and the two examples above show two ways it can arise despite them.

\subsection{Main results}

\begin{assumption}[Smoothness]\label{ass:smooth}
Let $\zeta_i=(X_i,U_i)$ denote an agent's full type. At each evaluation point $(y_\circ,w)$ there are a compact neighborhood $\mathcal N_y$ of $y_\circ$ and neighborhoods of $w_1,w_2$ on which:
\begin{enumerate}[(i)]
\item $f_W$ is twice continuously differentiable, bounded, and bounded away from zero;
\item $(y,w')\mapsto F_{Y|W}(y|w')$ is twice continuously differentiable with bounded derivatives, and $f_{Y|W}(\cdot|w)$ is continuous, bounded, and bounded away from zero on $\mathcal N_y$;
\item uniformly in $(\zeta,\zeta')$: $y\mapsto\PP(Y_{ij}\le y\mid\zeta_i=\zeta,\zeta_j=\zeta')$ admits a density bounded by $\bar C$ with a bounded continuous derivative on $\mathcal N_y$;
\item uniformly in $(\zeta,y)\in\text{(types)}\times\mathcal N_y$: $w'\mapsto\E[\PP(Y_{ij}\le y|\zeta_i=\zeta,\zeta_j)\mid W_j=w']f_W(w')$ is twice continuously differentiable with bounded derivatives near $w_2$, and symmetrically for the receiver slot near $w_1$;
\item the maps $w'\mapsto\E[\delta_s(\zeta_i;y_\circ,w)^2\mid W_i=w']f_W(w')$ and its $\delta_r$ and $\delta_s\delta_r$ analogs are continuous at the relevant points.
\end{enumerate}
\end{assumption}

\begin{assumption}[Kernels]\label{ass:kernel}
$K$ is a bounded, \emph{nonnegative}, symmetric, second-order kernel with $\int K=1$ and $\int\|u\|^{2}K(u)\,du<\infty$; $k$ is a bounded symmetric density; either $K,k$ have compact support or are Gaussian. Nonnegativity of $K$ is used throughout and not only for convenience: it makes $\widehat F_{Y|W}$ a weighted average of indicators and hence a genuine distribution function, which is what allows the plug-in inversion of Section \ref{sec:estimation} to be well defined, and it makes the denominators in Appendices A.11 and A.12 nonnegative. Where a higher-order kernel is required, as in the degenerate regime with $d\ge2$, monotone rearrangement \citep{chernozhukov2009} restores monotonicity of $\widehat F$ but the inverse must then be read as a generalized inverse.
\end{assumption}

\begin{assumption}[Bandwidths]\label{ass:bw}
As $N\to\infty$: $h\to0$, $h_y\to0$, $Nh^{2d}/\log N\to\infty$, and the undersmoothing conditions $Nh^{d+4}\to0$ and $Nh^{d}h_y^{4}\to0$ hold.
\end{assumption}

Assumption \ref{ass:bw} is satisfiable by $h\propto N^{-a}$ with $a\in\left(\frac{1}{d+4},\frac{1}{2d}\right)$, a nonempty interval for all $d\le 3$; in particular \emph{standard second-order positive kernels suffice}, where a formulation that localized in all $p$ coordinates of each agent's characteristics would demand kernel orders incompatible with a positive kernel whenever $d<p$. The simplification arises because \eqref{eq:Fhat} smooths only over $2d$ coordinates.

\begin{theorem}[Two-regime CLT for the conditional distribution function]\label{theo:cdf}
Let Assumptions \ref{ass:sampling}, \ref{ass:structure}, \ref{ass:smooth}--\ref{ass:bw} hold at an interior point $(y,w)$.
\begin{enumerate}[(i)]
\item (Nondegenerate regime.) If $\Omega_1(y;w)>0$, then
\[
\sqrt{Nh^{d}}\left(\widehat{F}_{Y|W}(y|w)-F_{Y|W}(y|w)\right)\dto N\!\left(0,\Omega_1(y;w)\right).
\]
\item (Degenerate regime.) Suppose condition (S) holds: \emph{the conditional distribution of $(Y_{ij},Y_{ji})$ given $(\zeta_i,\zeta_j)$ depends on the agents' types only through $(W_i,W_j)$.} Condition (S) implies $\E[\delta_s^2\mid W_i=w_1]=\E[\delta_r^2\mid W_i=w_2]=0$ and hence $\Omega_1(y;w)=0$; it holds in the leading case where $W_i=X_i$ and the errors are i.i.d.\ across dyads. If additionally $Nh^{d+2}\to0$ and $Nh^{d}h_y^{2}\to0$ (for Gaussian kernels, $Nh^{d+2}\log N\to0$), then
\[
\sqrt{nh^{2d}}\left(\widehat{F}_{Y|W}(y|w)-F_{Y|W}(y|w)\right)\dto N\!\left(0,\Omega_2(y;w)+\ind(w_1=w_2)\,\Omega_2'(y;w)\right),
\]
where
\[
\Omega_2'(y;w)=\Cov\!\left(\ind(Y_{ij}\le y),\ind(Y_{ji}\le y)\mid W_i=W_j=w_0\right)\left(\textstyle\int K^2\right)^2\big/f_W(w_0)^{2}
\]
is the same-pair reverse-dyad term (zero when $e_{ij}\perp e_{ji}$). If $\Omega_1=0$ but (S) fails, a third, interaction variance component of the same order arises; see the discussion at the end of Appendix A.10.
\end{enumerate}
Moreover $\sup_{y}\left|\widehat{F}_{Y|W}(y|w)-F_{Y|W}(y|w)\right|\pto0$ in both regimes.
\end{theorem}

Since $\widehat{F}_e(e)$ is $\widehat{F}_{Y|W}$ evaluated at a (known) point determined by the normalization, Theorem \ref{theo:cdf} applies to it directly. The next result treats the structural function; recall $\widehat{g}$ composes an estimated quantile at one conditioning point with an estimated distribution at another.

Because $\widehat g$ combines estimated conditional distributions at \emph{two} evaluation points, its asymptotic variance involves their covariance. The two H\'ajek projections are built from kernels localized at the four points $w_1,w_2,\tilde w_1,\tilde w_2$, and a cross term survives whenever two of these \emph{coincide}; it is then of the same order $(Nh^{d})^{-1}$ as the variance terms themselves. Define accordingly
\begin{multline}
\Omega_{12}(y,w;\tilde y,\tilde w)=\left(\textstyle\int K^2\right)\Bigg[
\ind(w_1=\tilde w_1)\,\frac{C_{ss}(w_1)}{f_W(w_1)}
+\ind(w_1=\tilde w_2)\,\frac{C_{sr}(w_1)}{f_W(w_1)}\\
+\ind(w_2=\tilde w_1)\,\frac{C_{rs}(w_2)}{f_W(w_2)}
+\ind(w_2=\tilde w_2)\,\frac{C_{rr}(w_2)}{f_W(w_2)}\Bigg],\label{eq:Om12}
\end{multline}
where, for $a,b\in\{s,r\}$, $C_{ab}(v)=\E\left[\delta_a(\zeta_i;y,w)\,\delta_b(\zeta_i;\tilde y,\tilde w)\mid W_i=v\right]$. This reduces to $\Omega_1(y;w)$ when the two points coincide and vanishes when the four localization points are pairwise distinct.

The degenerate regime needs a cross-point term of its own, and for a different reason. There the variance is carried by single dyads rather than by shared agents, so a covariance between the two evaluation points survives only if one dyad can be localized at both. For distinct ordered points that happens in exactly one configuration: the \emph{reversed} point $\tilde w=(w_2,w_1)$, at which the dyad $(j,i)$ is localized precisely when $(i,j)$ is localized at $w$. Since Assumption \ref{ass:sampling} permits $Y_{ij}$ and $Y_{ji}$ to be dependent, the resulting covariance is not negligible, and it is of the same order as the variance terms themselves. Define
\begin{equation}
\Omega_2^{\mathrm{rev}}(y,w;\tilde y)=\frac{\Cov\!\left(\ind(Y_{ij}\le y),\,\ind(Y_{ji}\le\tilde y)\mid W_i=w_1,W_j=w_2\right)}{f_W(w_1)f_W(w_2)}\left(\textstyle\int K^2\right)^{2},\label{eq:Om2rev}
\end{equation}
which vanishes when $e_{ij}\perp e_{ji}$ and which coincides with $\Omega_2$ when $Y_{ij}=Y_{ji}$ almost surely.

\begin{theorem}[CLT for the structural function]\label{theo:g}
Let $\widehat{g}(w,e)=\widehat{F}^{-1}_{Y|W}\big(\widehat{F}_{Y|\widetilde{W}}(\tilde{y}|\tilde{w})\,\big|\,w\big)$ with population counterpart $g(w,e)$, where $(\tilde{y},\tilde{w})$ is the normalization-determined evaluation point with $\dim(\widetilde{W}_i)=\tilde{d}$, and suppose $(w_1,w_2)\ne(\tilde{w}_1,\tilde{w}_2)$ or $d\neq\tilde d$. Let Assumptions \ref{ass:sampling}, \ref{ass:structure}, \ref{ass:smooth}--\ref{ass:bw} hold at both points and set $d^\ast=\max(d,\tilde{d})$. In the nondegenerate regime,
\[
\sqrt{Nh^{d^\ast}}\left(\widehat{g}(w,e)-g(w,e)\right)\dto N\!\left(0,\;\Sigma_g\right),
\]
where, writing $y_0=g(w,e)$,
\begin{multline*}
\Sigma_g=\Big[\ind(d^\ast=\tilde{d})\,\Omega_1(\tilde{y};\tilde{w})+\ind(d^\ast=d)\,\Omega_1(y_0;w)\\
-2\,\ind(d=\tilde d)\,\Omega_{12}(\tilde y,\tilde w;y_0,w)\Big]\Big/ f_{Y|W}(y_0\mid w)^{2}.
\end{multline*}
When the four localization points are pairwise distinct, $\Omega_{12}=0$ and the variance is the sum of the two separate contributions.
In the degenerate regime (condition (S) of Theorem \ref{theo:cdf}(ii) at both evaluation points, with that part's bandwidth conditions) the same statement holds with rate $\sqrt{nh^{2d^\ast}}$ and
\begin{multline*}
\Sigma_g^{\mathrm{deg}}=\Big[\ind(d^\ast=\tilde d)\left\{\Omega_2(\tilde y;\tilde w)+\ind(\tilde w_1=\tilde w_2)\Omega_2'(\tilde y;\tilde w)\right\}\\
+\ind(d^\ast=d)\left\{\Omega_2(y_0;w)+\ind(w_1=w_2)\Omega_2'(y_0;w)\right\}\\
-2\,\ind(d=\tilde d)\,\ind\!\left((\tilde w_1,\tilde w_2)=(w_2,w_1)\right)\Omega_2^{\mathrm{rev}}(y_0,w;\tilde y)\Big]\Big/f_{Y|W}(y_0\mid w)^{2}.
\end{multline*}
The reversed-point term is a necessity, and the formula is sharp: for an undirected array, $Y_{ij}=Y_{ji}$, with $\tilde w=(w_2,w_1)$, exchanging summation indices in \eqref{eq:Fhat} shows that $\widehat F_{Y|W}(y|w_1,w_2)$ and $\widehat F_{Y|W}(y|w_2,w_1)$ are \emph{identically equal} for every sample, so $\widehat g(w,e)=e$ exactly with zero sampling variance. Then $\Omega_2^{\mathrm{rev}}=\Omega_2$ and the bracket is $\Omega_2+\Omega_2-2\Omega_2=0$, as it must be; without the term the formula would predict positive variance for an estimator that has none. Setting it aside, the familiar independence-based formulas are recovered after accounting for the reduced smoothing dimension.
\end{theorem}

Regarding feasibility of the bandwidth conditions across the two regimes: the nondegenerate results require only Assumption \ref{ass:bw}, which second-order kernels satisfy for all $d\le3$. The degenerate regime is more demanding: the conditions $Nh^{d+2}\to0$ and $Nh^{2d}/\log N\to\infty$ of Theorem \ref{theo:cdf}(ii) are jointly satisfiable with a second-order kernel only when $d=1$; for $d\ge2$ the degenerate-regime CLT requires a kernel of order exceeding $2$ (with monotone rearrangement restoring invertibility of $\widehat F$), reflecting an intrinsic bias-versus-degenerate-variance tradeoff. Since degeneracy requires $\Omega_1=0$, and dyad-independent errors with $W_i=X_i$ are the leading case in which that holds, this restriction is of limited practical consequence, but it should be borne in mind when the errors are suspected to be (nearly) independent across dyads.

These results make precise, and quantitative, why i.i.d.\ formulas fail. Confidence intervals built on $\Omega_2/(nh^{2d})$ have width of order $(Nh^{d})^{-1}$ while the true sampling standard deviation is of order $(Nh^{d})^{-1/2}$. Their ratio $\left(Nh^{d}\right)^{-1/2}\to0$: nominal 95\% intervals cover with probability tending to \emph{zero}. The Monte Carlo below shows the collapse is quantitatively severe already at $N=100$.

\subsection{Inference via the agent-level bootstrap}\label{sec:boot}

The projection variance $\Omega_1$ involves the conditional moments $\E[\delta_s^2|\cdot]$, $\E[\delta_r^2|\cdot]$ of deviations indexed by \emph{unobservables}; estimating it directly requires estimating agent-level mixed moments and is unattractive. I instead recommend the \emph{agent-level (pigeonhole) bootstrap}: draw $I_1,\dots,I_N$ i.i.d.\ uniform on $\{1,\dots,N\}$, and form the bootstrap sample $\{(Y_{I_aI_b},X_{I_a},X_{I_b}):a\ne b,\ I_a\ne I_b\}$; compute $\widehat{F}^{\,\ast}_e$, $\widehat{g}^{\,\ast}$ on the bootstrap sample; repeat $B$ times; form percentile intervals. Because resampling is at the agent level, dyads sharing a resampled agent remain dependent in the bootstrap world, so the scheme replicates the shared-agent variance automatically.

\begin{theorem}[Bootstrap validity]\label{theo:boot}
Under the conditions of Theorem \ref{theo:cdf}(i), conditional on the data,
\begin{multline*}
\sup_{t}\Big|\PP^{\ast}\!\left(\sqrt{Nh^{d}}\left(\widehat{F}^{\,\ast}_{Y|W}(y|w)-\widehat{F}_{Y|W}(y|w)\right)\le t\right)\\
-\PP\!\left(\sqrt{Nh^{d}}\left(\widehat{F}_{Y|W}(y|w)-F_{Y|W}(y|w)\right)\le t\right)\Big|\pto0,
\end{multline*}
and the corresponding statement holds for $\widehat{g}$ under the conditions of Theorem \ref{theo:g}.
\end{theorem}

The proof (Appendix A.9) rests on an \emph{exact} multinomial-weight representation of the bootstrap statistic, in which the dropped duplicate-agent draws need no separate treatment and the conditional variance of the linear part is $\sum_lt_l^2$ exactly, combined with a Lindeberg--Feller argument for the resulting sum of conditionally i.i.d.\ terms; see \cite{praestgaard1993} for the general theory of exchangeably weighted sums and \cite{davezies2021} for resampling exchangeable arrays. In the degenerate case the plain pigeonhole bootstrap is known to be conservative \citep{menzel2021}; the corrected Gaussian multiplier scheme of \cite{hounyo2026}, valid in both regimes for dyadic regression, could be adapted to the plug-in quantile map here at the cost of estimating the degeneracy index. The bootstrap intervals are what the Monte Carlo below evaluates.

\section{Monte Carlo}\label{simulation}

Outcomes are generated by the homogeneous-of-degree-one design
\[
Y_{ij}=0.3\,X_i^{2}X_j^{2}(-e_{ij})^{-3},\qquad X_i\sim N(6,1)\ \text{i.i.d.},
\]
with $e_{ij}$ distributed as $N(-6,1)$ \emph{truncated to} $\EE=(-\infty,-0.5]$. The truncation is not cosmetic. The map $e\mapsto(-e)^{-3}$ has a pole at $e=0$: on the whole real line it is neither continuous nor monotone, and $\E[(-e)^{-3}]$ does not exist, because the singularity is not integrable against a density that is positive at the origin. Restricting the support to $\EE$ makes $g(x_1,x_2,e)=0.3x_1^{2}x_2^{2}(-e)^{-3}$ continuous, strictly increasing and bounded on $\EE$, with all moments finite, so that Assumptions \ref{ass:structure} and \ref{ass:hd1} hold as stated. The sampling algorithm matters, since truncation and the agent-effects representation interact: agent effects $U_i$ are drawn first, and conditional on them each $e_{ij}$ is drawn from its normal conditional law and redrawn if it exceeds $-0.5$. Under this law the marginal of $e_{ij}$ is not exactly $N(-6,1)$ and $\mathrm{Corr}(e_{ij},e_{ik})$ is not exactly $\lambda$. Appendix A.14 computes its moments directly by quadrature (Table \ref{tab:trunc}): the mean and variance deviate from $-6$ and $1$ by less than $10^{-6}$, the correlation from $\lambda$ by at most $3.8\times10^{-8}$, and $F_e(-6)$ lies between $0.5000000000$ and $0.5000000095$. The experiments use $\Phi(0)/\Phi(5.5)=0.5000000095$ as the estimand throughout; all discrepancies are negligible beside the $3.2$ percent Monte Carlo standard error on a variance at $R=2000$, but the untruncated quantities below are approximate benchmarks rather than identities. Across the $235{,}400{,}000$ error draws generated for Tables \ref{tab:coverage}--\ref{tab:sweep} the truncation bound was reached, and the draw replaced, $3$ times; Figure \ref{fig:curves} and Table \ref{tab:cv} use separate runs. Results are reported on the evaluation interval $[-8,-4]$, which is not the support of $e_{ij}$. The normalization constants are $\bar{x}^1_1=\bar{x}^1_2=6$, $\bar{e}=-6$, and $\alpha=g(6,6,-6)=1.8$, so the conditioning ray $(e/\bar{e})\bar{x}^1$ spans $[4,8]$, comfortably inside the effective support of $X$.

A single one-parameter family isolates the role of dyadic dependence, holding the estimands fixed to the accuracy reported in Table \ref{tab:trunc}. Let
\[
e_{ij}=-6+\sqrt{\lambda}\,(U_i+U_j)+\sqrt{1-2\lambda}\;V_{ij},\qquad U_i,V_{ij}\ \text{i.i.d.}\ N(0,1),
\]
again truncated to $\EE$. Before truncation the marginal law of $e_{ij}$ is $N(-6,1)$ for every $\lambda\in[0,1/2)$ and $\mathrm{Corr}(e_{ij},e_{ik})=\lambda$, so that, to the accuracy recorded below, the estimands do not move as $\lambda$ varies while $\lambda$ indexes the shared-agent dependence; Table \ref{tab:trunc} reports how far the truncated law departs from these values, the largest being $5.9\times10^{-7}$ in the variance. Two values are reported in full and three intermediate ones are used to trace the transition:
\begin{itemize}
\item $\lambda=0$ (\textbf{degenerate}), referred to below as DGP 1: $e_{ij}$ i.i.d.\ across dyads. Here $\Omega_1=0$.
\item $\lambda=1/3$ (\textbf{dyadic dependence}), referred to below as DGP 2: equivalent to $e_{ij}=-6+(U_i+U_j+V_{ij})/\sqrt3$, so $\Omega_1>0$.
\end{itemize}
The intermediate values $\lambda\in\{0.02,0.05,0.10\}$ are used in Table \ref{tab:sweep} to trace how quickly the failure of the independence formula sets in as the shared-agent component is switched on.
Estimation uses Gaussian kernels throughout, so the inversion in \eqref{eq:ghat} is exact.

\subsection{Estimation of $g$ and $F_e$}

\begin{figure}[!htbp]
    \centering
    \caption{Estimates of $g$ and $F_e$: mean over 100 replications}
    \includegraphics[width=\textwidth]{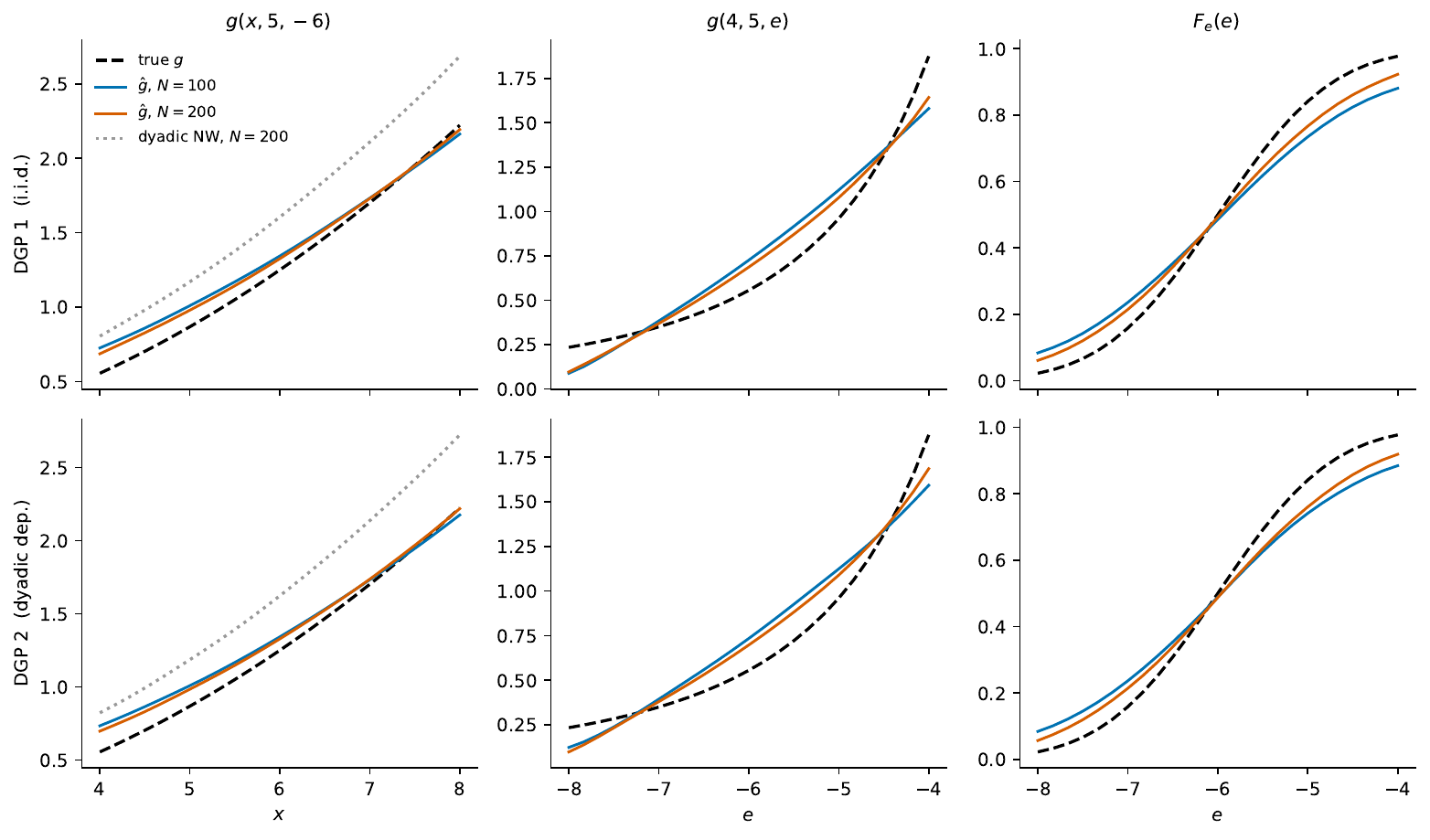}
    \label{fig:curves}
    {\footnotesize\singlespacing Notes: top row DGP 1 (i.i.d.\ errors), bottom row DGP 2 (dyadic dependence). Left: $g(x,5,-6)$; middle: $g(4,5,e)$; right: $F_e(e)$. Dashed black: truth; blue: $\widehat g$ (or $\widehat F_e$), $N=100$; red: $N=200$; dotted gray (left panels): dyadic Nadaraya--Watson estimate of the conditional mean, $N=200$, converging to its own estimand $\E[Y|x,5]\ne g(x,5,-6)$.}
\end{figure}

Figure \ref{fig:curves} reports, for $N\in\{100,200\}$ (hence $n\in\{9{,}900,\,39{,}800\}$ directed dyads) and $R=100$ replications, the mean of $\widehat{g}(x,5,-6)$ over $x\in[4,8]$, of $\widehat{g}(4,5,e)$ and $\widehat{F}_e(e)$ over $e\in[-8,-4]$, using the rule-of-thumb bandwidths $h=1.06N^{-1/5}$ and $h_y=1.06\widehat{\sigma}_Y n^{-1/5}$. Two lessons emerge. The estimators track the true functions and improve with $N$ under both DGPs: dyadic dependence does not compromise consistency, as Theorem \ref{theo:cdf} predicts. And the dyadic Nadaraya--Watson estimator of $\E[Y_{ij}|X_i=x,X_j=5]$ \citep{NBERw28548}, also shown, converges to a \emph{different} function: $e\mapsto(-e)^{-3}$ is strictly convex on $\EE$ and, by the truncation, integrable there, so by Jensen's inequality the conditional mean lies strictly above $g(x,5,-6)$. This is a gap between two estimands, each consistently estimated, not evidence that either procedure performs better. The figure plots replication means and says nothing about dispersion, which Table \ref{tab:coverage} reports directly.

\subsection{Inference: the collapse of i.i.d.-formula intervals}

Table \ref{tab:coverage} is the paper's core evidence. For $N\in\{50,100,200\}$ I construct nominal 95\% intervals for $F_e(-6)=0.5$ and for $g(5,5,-6)\approx0.868$ in two ways: from the \emph{i.i.d.-formula} variance $\Omega_2/(nh^{2d})$, the formula the natural i.i.d.\ intuition delivers, and from the agent-level bootstrap of Section \ref{sec:boot}. Bandwidths are undersmoothed, $h=1.06N^{-2/5}$, satisfying Assumption \ref{ass:bw}, so that bias is negligible and coverage isolates the variance effect. Because a coverage figure is itself an estimate, Monte Carlo standard errors are reported throughout; the i.i.d.\ columns use $R=2000$ replications and the bootstrap columns, which cost $B=199$ resamples each, use $R=200$, so their standard errors are near $0.02$.

Under DGP 1 (degenerate) i.i.d.-formula coverage is close to nominal at all sample sizes, though this reflects partial cancellation rather than exact validation: the degenerate-regime approximation additionally requires $Nh^{d+2}\to0$, whereas the designs have $Nh^{3}$ between $0.41$ and $0.54$, so the projection term has not yet died out and the analytic $\Omega_2$ understates the sampling variance. What DGP 1 establishes robustly is the contrast with DGP 2, where coverage collapses and \emph{deteriorates} as $N$ grows, from $0.421$ at $N=50$ to $0.292$ at $N=200$ for $F_e$ and from $0.534$ to $0.329$ for $g$, the signature of a variance formula of the wrong asymptotic order. The DGP 2 sampling variance exceeds its DGP 1 counterpart by factors of $8.6$, $14.6$ and $25.0$, growing with $Nh^{d}$ as the theory requires, and the bootstrap intervals for $\widehat F_e$ are correspondingly $3.1$ to $5.1$ times longer than the independence intervals ($2.5$ to $4.3$ for $\widehat g$).

\begin{table}[!htbp]
\centering
\caption{Coverage and length of nominal 95\% intervals}
\label{tab:coverage}
{\footnotesize
\begin{tabular}{lcccccccc}
\toprule
 & \multicolumn{4}{c}{$\widehat F_e(-6)$} & \multicolumn{4}{c}{$\widehat g(5,5,-6)$} \\
\cmidrule(lr){2-5}\cmidrule(lr){6-9}
 & \multicolumn{2}{c}{coverage} & \multicolumn{2}{c}{mean length} & \multicolumn{2}{c}{coverage} & \multicolumn{2}{c}{mean length} \\
\cmidrule(lr){2-3}\cmidrule(lr){4-5}\cmidrule(lr){6-7}\cmidrule(lr){8-9}
$N$ & i.i.d. & boot. & i.i.d. & boot. & i.i.d. & boot. & i.i.d. & boot. \\
\midrule
\multicolumn{9}{l}{\emph{$\lambda=0$: dyad-independent errors (DGP 1)}} \\
50 & 0.916 & 0.990 & 0.138 & 0.206 & 0.932 & 0.990 & 0.365 & 0.487 \\
 & (0.006) & (0.007) & & & (0.006) & (0.007) & & \\
100 & 0.925 & 0.985 & 0.087 & 0.137 & 0.932 & 0.990 & 0.211 & 0.307 \\
 & (0.006) & (0.009) & & & (0.006) & (0.007) & & \\
200 & 0.925 & 1.000 & 0.056 & 0.093 & 0.921 & 0.995 & 0.129 & 0.201 \\
 & (0.006) & (0.000) & & & (0.006) & (0.005) & & \\
\multicolumn{9}{l}{\emph{$\lambda=1/3$: dyadic dependence (DGP 2)}} \\
50 & 0.421 & 0.915 & 0.135 & 0.421 & 0.534 & 0.935 & 0.400 & 0.982 \\
 & (0.011) & (0.020) & & & (0.011) & (0.017) & & \\
100 & 0.338 & 0.940 & 0.087 & 0.352 & 0.432 & 0.915 & 0.206 & 0.704 \\
 & (0.011) & (0.017) & & & (0.011) & (0.020) & & \\
200 & 0.292 & 0.900 & 0.057 & 0.289 & 0.329 & 0.910 & 0.127 & 0.545 \\
 & (0.010) & (0.021) & & & (0.011) & (0.020) & & \\
\bottomrule
\end{tabular}}
\par\vspace{2mm}
{\footnotesize\singlespacing Notes: empirical coverage of nominal $95\%$ intervals and mean interval length, at the undersmoothed bandwidth $h=1.06N^{-2/5}$. Monte Carlo standard errors, $\sqrt{\widehat p(1-\widehat p)/R}$, are in parentheses. The i.i.d.\ columns use $R=2000$ replications; the bootstrap columns use $R=200$ replications with $B=199$ draws each, which is what the computation affords. ``i.i.d.'' intervals use $\Omega_2/(nh^{2d})$, the formula valid only in the degenerate regime $\Omega_1=0$ under condition (S); ``boot.'' intervals are agent-level percentile intervals. The error design is $e_{ij}=-6+\sqrt\lambda\,(U_i+U_j)+\sqrt{1-2\lambda}\,V_{ij}$ truncated to $\EE$, so the marginal law of $e_{ij}$, and hence both estimands, agree across rows to nine decimal places (Table \ref{tab:trunc}).}

\end{table}

The corrected variance formula \eqref{eq:Om1} can be checked against the table directly. At the $F_e(-6)$ evaluation point, $(w_1,w_2)=(6,6)$, $y_\circ=\alpha=1.8$, and under DGP 2 the deviation functions are $\delta_s(\zeta)=\delta_r(\zeta)=\Phi(-u/\sqrt2)-\tfrac12$ for $\zeta=(6,u)$, so that $\E[\delta_s^2\mid W=6]=\E[\delta_s\delta_r\mid W=6]=\arcsin(1/3)/(2\pi)\approx0.0541$. With Gaussian kernels ($\int K^2=1/(2\sqrt\pi)$) and $f_W(6)=\phi(0)$, equation \eqref{eq:Om1} gives $\Omega_1=(\int K^2)\cdot4\cdot0.0541/\phi(0)\approx0.153$, half of it contributed by the sender--receiver cross term $C(y;w)$. Table \ref{tab:omega1} compares the implied leading-term prediction $\Omega_1/(Nh)$ with the Monte Carlo sampling variance over $R=2000$ replications. The ratios are $1.019$, $1.001$ and $1.007$, and none differs from one by as much as one Monte Carlo standard error. There are no free parameters in the comparison: $\Omega_1$ is computed in closed form from the design. The theory does not merely rationalize the coverage failure, it predicts the magnitude of the extra variance.

\begin{table}[!htbp]
\centering
\caption{The projection variance $\Omega_1$ against the Monte Carlo variance, $\lambda=1/3$}
\label{tab:omega1}
{\footnotesize
\begin{tabular}{lcccc}
\toprule
$N$ & $\Omega_1/(Nh)$ predicted & Monte Carlo variance & (s.e.) & ratio \\
\midrule
50 & 13.802 & 14.068 & (0.445) & 1.019 \\
100 & 9.106 & 9.111 & (0.288) & 1.001 \\
200 & 6.008 & 6.047 & (0.191) & 1.007 \\
\bottomrule
\end{tabular}}
\par\vspace{2mm}
{\footnotesize\singlespacing Notes: the leading-term prediction $\Omega_1/(Nh)$ of Theorem \ref{theo:cdf}(i) against the Monte Carlo sampling variance of $\widehat F_e(-6)$ under $\lambda=1/3$, both $\times10^{3}$, over $R=2000$ replications. $\Omega_1=0.1530$ is computed in closed form from \eqref{eq:Om1} as described in the text, with no free parameters. Monte Carlo standard errors for a variance are $\widehat V\sqrt{2/(R-1)}$. The ratios are within two percent of one at every sample size and none differs from one by as much as one Monte Carlo standard error.}

\end{table}

In contrast, the agent-level bootstrap of Section \ref{sec:boot} raises coverage under dyadic dependence to $0.915$, $0.940$ and $0.900$ for $\widehat F_e(-6)$ at $N=50,100,200$, and $0.935$, $0.915$ and $0.910$ for $\widehat g$, against Monte Carlo standard errors of about $0.02$. Two qualifications belong with those numbers. Coverage is close to nominal but not nominal: at $N=200$ the shortfall for $\widehat F_e$ is $0.050$, which is $3.2$ Monte Carlo standard errors at the nominal $0.95$ and $2.4$ at the observed $0.900$, and it does not shrink with $N$ over the range examined. And in the degenerate design the bootstrap is conservative, covering at $0.985$ or above, the behavior known for pigeonhole-type resampling \citep{menzel2021} and the reason Theorem \ref{theo:boot} is stated for the nondegenerate regime; \cite{hounyo2026} is the route to removing the conservativeness. The $g$-interval coverage also deteriorates under DGP 2, the projection variances at the two conditioning points adding rather than offsetting because the four localization points are pairwise distinct, exactly as Theorem \ref{theo:g} predicts.

Table \ref{tab:sweep} asks how weak the dependence has to be before the independence formula becomes usable. The answer is: weaker than most applications will be. At $\lambda=0.05$, a correlation between dyads sharing an agent that few researchers would describe as substantial, i.i.d.-formula coverage is already down to $0.695$, some $23$ percentage points below the $0.925$ attained at $\lambda=0$ and $25$ below nominal; the agent bootstrap holds at or above $0.94$ throughout.

\begin{table}[!htbp]
\centering
\caption{How fast does the independence formula fail? Coverage as $\lambda\to0$}
\label{tab:sweep}
{\footnotesize
\begin{tabular}{lccccc}
\toprule
$\lambda=\mathrm{Corr}(e_{ij},e_{ik})$ & 0.00 & 0.02 & 0.05 & 0.10 & 1/3 \\
\midrule
Coverage, i.i.d.\ formula & 0.925 & 0.795 & 0.695 & 0.520 & 0.338 \\
Coverage, agent bootstrap & 0.985 & 0.965 & 0.985 & 0.950 & 0.940 \\
Var$(\widehat F_e(-6))\times10^{3}$ & 0.62 & 1.26 & 1.65 & 3.64 & 9.11 \\
\bottomrule
\end{tabular}}
\par\vspace{2mm}
{\footnotesize\singlespacing Notes: $N=100$, nominal $95\%$ intervals for $F_e(-6)$. The two end columns use the replication counts of Table \ref{tab:coverage}; the three interior columns use $R=200$. As $\lambda\to0$ the shared-agent variance component vanishes and the independence formula becomes correct, but the approach is slow: a correlation of $0.05$ between dyads sharing an agent, which few researchers would regard as substantial, is already enough to cost some $23$ percentage points of coverage relative to the independent case. The estimands do not vary with $\lambda$.}

\end{table}

\subsection{Cross-validation}

The third experiment tests Theorem \ref{theo:cv}. I take $\psi(y)=\ind(y\le1)$, so that the target $m_\psi$ is the conditional distribution function itself, evaluated at a threshold near the middle of the outcome distribution; because $e_{ij}$ has the same $N(-6,1)$ marginal in both designs, the target is the same known function
$m_\psi(x_1,x_2)=\Phi\big(6-(0.3x_1^2x_2^2)^{1/3}\big)$
under DGP 1 and DGP 2, so the two blocks of rows in Table \ref{tab:cv} differ only through the dependence structure. Over a grid of bandwidths I compare three selectors: the leave-pair-out criterion of Theorem \ref{theo:cv}; a naive leave-one-\emph{dyad}-out criterion, which deletes only the held-out dyad and leaves the dyads sharing an agent with it in the training sample; and the infeasible oracle that minimizes the realized average squared error in each replication. The leave-pair-out criterion is computed in closed form: by inclusion--exclusion the deleted contributions can be expressed through three matrix products, so the entire criterion costs $O(N^3)$ per bandwidth rather than requiring an explicit refit for every dyad.

\begin{table}[!htbp]
\centering
\caption{Bandwidth selection: leave-pair-out versus naive cross-validation}
\label{tab:cv}
\begin{tabular}{lc cc cc cc}
\toprule
 & & \multicolumn{2}{c}{Naive (leave-dyad-out)} & \multicolumn{2}{c}{Leave-pair-out} & \multicolumn{2}{c}{Oracle} \\
\cmidrule(lr){3-4}\cmidrule(lr){5-6}\cmidrule(lr){7-8}
DGP & $N$ & $\bar h$ & ASE & $\bar h$ & ASE & $\bar h$ & ASE \\
\midrule
1 (independent) & 50 & 0.293 & 0.0015 & 0.300 & 0.0015 & 0.293 & 0.0014 \\
1 (independent) & 100 & 0.233 & 0.0007 & 0.239 & 0.0007 & 0.236 & 0.0007 \\
2 (dyadic dep.) & 50 & 0.183 & 0.0086 & 0.381 & 0.0060 & 0.382 & 0.0046 \\
2 (dyadic dep.) & 100 & 0.151 & 0.0048 & 0.316 & 0.0034 & 0.323 & 0.0027 \\
\bottomrule
\end{tabular}
\par\vspace{2mm}
{\footnotesize\singlespacing Notes: 100 replications; $\psi(y)=\ind(y\le1)$, so the target is the conditional distribution function, identical across the two designs. $\bar h$ is the mean selected bandwidth and ASE the mean average squared error of the full-sample estimator at the selected bandwidth. ``Oracle'' minimizes the ASE in each replication and is infeasible. The naive criterion keeps dyads sharing an agent with the held-out dyad in the training sample.}

\end{table}

Table \ref{tab:cv} shows a clean dichotomy. Under DGP 1 the two feasible criteria coincide with each other and with the oracle, since the extra dyads were independent of the held-out dyad to begin with. Under DGP 2 the naive criterion selects bandwidths less than half the oracle value ($0.183$ against $0.382$ at $N=50$, $0.151$ against $0.323$ at $N=100$), because dyads sharing an agent with the held-out dyad carry information about its error and the criterion rewards undersmoothing; its average squared error is roughly $40$ percent higher. The leave-pair-out criterion tracks the oracle almost exactly ($0.381$ against $0.382$, $0.316$ against $0.323$), the residual gap reflecting only that the oracle re-optimizes in every replication.

\subsection{Bandwidth sensitivity}

The coverage experiment uses undersmoothed bandwidths because rule-of-thumb choices ($h\propto N^{-1/5}$) violate the undersmoothing conditions of Assumption \ref{ass:bw}: with $d=1$, $\sqrt{nh^{2}}\,h^{2}\propto N^{2/5}\to\infty$, so smoothing bias dominates the standard error and \emph{even under DGP 1} i.i.d.-formula intervals undercover badly (coverage of roughly $0.5$--$0.6$ in unreported experiments with $h\propto N^{-1/5}$). Practitioners should undersmooth for inference even when point estimation uses larger bandwidths.

\section{Empirical Application: Distributional Gravity}\label{empirical}

I apply the methods to bilateral merchandise trade, the setting for which the random gravity specification of Example \ref{ex:gravity} was designed and the leading source of dyadic data in economics.

\subsection{Data and implementation}\label{sec:data}

Data come from the CEPII Gravity database (version 202211), which merges reconciled bilateral trade flows from BACI with country-level covariates \citep{conte2022}. I use the 2019 cross-section, the last year unaffected by the pandemic, and retain the 120 largest economies by GDP. The outcome is $Y_{ij}=\log(\text{exports from }i\text{ to }j)$ and the covariate is $X_i=\log(\text{GDP}_i)$, a scalar, so $d=1$ and the bandwidth conditions of Assumption \ref{ass:bw} hold with Gaussian kernels.

The sample is built as follows. The $120$ largest economies give $120\times119=14{,}280$ ordered pairs. Of these, $1{,}059$ ($7.42$ percent) have a zero or unrecorded flow and are dropped. One economy retains no flow in either direction and leaves the sample, so $N=119$ and $n=13{,}221$. Every remaining economy appears both as an exporter and as an importer, with out-degrees between $74$ and $118$ and in-degrees between $68$ and $118$ against a maximum of $118$; $34$ economies have a complete row and column.

This is where the application departs from the theory. Assumption \ref{ass:sampling} observes \emph{all} ordered pairs among the sampled agents, whereas here $821$ of the $14{,}042$ ordered pairs among the retained economies, $5.85$ percent, are absent, and not at random: a flow is dropped precisely when it is zero or unrecorded, an event about the outcome. Deleting outcomes on the basis of their own value can change the conditional distribution of $e_{ij}$ among retained dyads and the effective covariate distribution, and hence the variance formulas of Section \ref{consistency} and the bootstrap that mimics them; nothing here establishes that it does not, and the weak-monotonicity extension discussed in the conclusion is the right treatment of the zeros. Dropping zero flows is standard in this literature, and \cite{chen2025} does the same, but standard is not innocuous. \emph{The results below are therefore an illustration of what the estimator and the inference procedure do on real dyadic data, not estimates whose sampling properties are covered by the theorems of this paper.} What the application can support is a comparison between specifications computed on identical data under identical conventions, which is what Section \ref{sec:sep} uses it for; what it cannot support is a claim that the reported intervals have nominal coverage.

Table \ref{tab:desc} reports descriptive statistics. Log exports span a wide range, and log GDP is the single scalar covariate; log distance is shown for reference; it is used in the validation exercise of Section \ref{sec:validate} and as the dyad-level covariate of Appendix B, but never in the estimates of this section.

\begin{table}[!htbp]
\centering
\caption{Descriptive statistics, 2019 cross-section}
\label{tab:desc}
{\footnotesize
\begin{tabular}{lccccccc}
\toprule
& Mean & S.D. & Min & p25 & Median & p75 & Max \\
\midrule
Log exports, $Y_{ij}$ & 10.01 & 3.72 & -5.81 & 7.80 & 10.46 & 12.61 & 19.87 \\
Log GDP, exporter $X_i$ & 18.97 & 1.56 & 16.66 & 17.76 & 18.85 & 19.85 & 23.79 \\
Log GDP, importer $X_j$ & 18.96 & 1.56 & 16.66 & 17.76 & 18.85 & 19.85 & 23.79 \\
Log distance (not used in estimation) & 8.61 & 0.85 & 4.01 & 8.15 & 8.81 & 9.23 & 9.90 \\
\bottomrule
\end{tabular}}
\par\vspace{2mm}
{\footnotesize\singlespacing Notes: 2019 cross-section. Of the $14{,}280$ ordered pairs among the $120$ largest economies, $1{,}059$ ($7.42$ percent) have zero or unrecorded flows and are dropped; one economy retains no flows at all, leaving $N=119$ economies and $n=13{,}221$ directed dyads, against $N(N-1)=14{,}042$ for a complete array. Exports and GDP are both in \emph{thousands} of current USD (BACI and World Bank, via CEPII); distance is in kilometres and is used only in Sections \ref{sec:validate} and \ref{sec:Z}. The medians reported here are dyad-level, each economy entering once per dyad; the normalization point $\bar x$ is the \emph{agent-level} median log GDP, $18.60$ (about \$120 billion), which is the relevant high-density point for the estimator.}

\end{table}

The identified pair is the normalized representative with $g(\bar x,\bar x,e)=e$ at $\bar x$ equal to the median log GDP (about $\$120$ billion), a high-density point as the guidance of Section \ref{sec:estimation} prescribes. The unobservable $e_{ij}$ is the composite pair-level shock, absorbing bilateral frictions such as distance, trade agreements, and common language, together with the interaction of the two countries' unobserved fundamentals; the random-effects Assumption \ref{ass:sampling}(iv) requires these to be independent of the two GDPs, a caveat discussed below. The bandwidth is chosen by the leave-pair-out cross-validation of Theorem \ref{theo:cv}, minimized at an interior grid point, $h_{\mathrm{CV}}=0.30$; the naive leave-one-dyad-out criterion, which keeps every dyad sharing an agent with the held-out pair in the training sample, selects the smallest bandwidth on the grid, $0.15$, the reward for undersmoothing that the discussion after Theorem \ref{theo:cv} predicts and Table \ref{tab:cv} shows in simulation. \emph{Every number and figure in this section is computed at $h=0.30$, $h_y=\widehat\sigma_Yn^{-0.3}=0.216$, with agent-bootstrap intervals ($B=399$) at the same bandwidth.} The undersmoothed $h_{\mathrm{us}}=h_{\mathrm{CV}}N^{-1/5}=0.115$ that Assumption \ref{ass:bw} requires for a centered limit appears only in Table \ref{tab:hsens}; the price of reporting at $h_{\mathrm{CV}}$ is that the intervals are not bias-corrected, and that table shows the conclusions do not turn on the choice.

\subsection{Results}

\begin{figure}[!htbp]
    \centering
    \caption{Distributional gravity: structural estimates, 2019}
    \includegraphics[width=\textwidth]{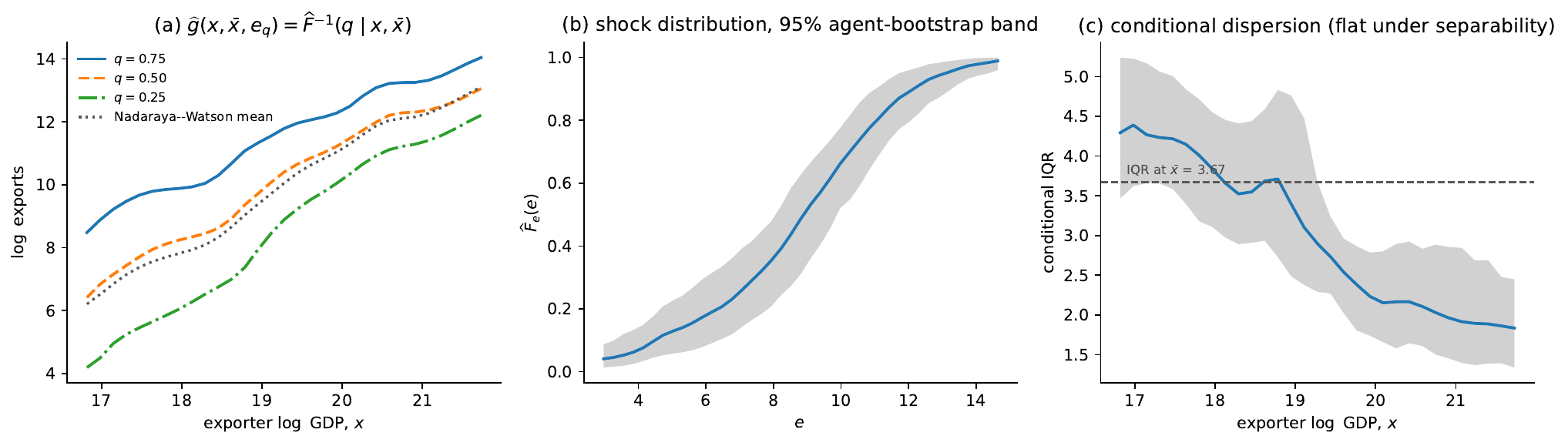}
    \label{fig:empirical}
    {\footnotesize\singlespacing Notes: Panel (a): $\widehat g(x,\bar x,e_q)$ for $q\in\{0.25,0.50,0.75\}$ over exporter log GDP $x$, importer fixed at $\bar x$, with the dyadic Nadaraya--Watson conditional mean for comparison; bandwidth $h_{\mathrm{CV}}=0.30$. Panel (b): $\widehat F_e$ with pointwise 95\% agent-bootstrap bands. Panel (c): the conditional interquartile range $\widehat g(x,\bar x,e_{0.75})-\widehat g(x,\bar x,e_{0.25})$, which additive separability requires to be constant in $x$, with a pointwise 95\% agent-bootstrap band; the dashed line marks the value $3.67$ attained at the normalization point $\bar x$. All three panels are computed at $h=h_{\mathrm{CV}}=0.30$ and $h_y=0.216$ from a single set of estimates. The bands are pointwise, not simultaneous: uniform inference is left to future work, so the band should not be read as a test of flatness over the whole range; Table \ref{tab:sep} reports the departure at three exporter sizes with agent-bootstrap standard errors, which is a measurement rather than a test.}
\end{figure}

Figure \ref{fig:empirical}(a) shows the structural function at the quartiles of the estimated shock distribution as a function of exporter size. The curves are far from parallel: the interquartile spread of $g$ in the shock argument \emph{compresses} markedly as exporter size rises, so the composite shock moves the trade of a small economy by an enormous factor while barely denting that of a very large one, an interaction an additive specification cannot express; Section \ref{sec:sep} quantifies it. The dyadic Nadaraya--Watson conditional mean lies about $0.23$ log points below the median structural curve, the empirical counterpart of the estimand gap in the simulations and not evidence that either estimator is biased. Panel (b) reports $\widehat F_e$: the interquartile range of the composite shock is $3.67$ log points with an agent-bootstrap standard error of $0.43$, and the bands are wide, reflecting how much of the information about the shock distribution is concentrated near the normalization point.

\begin{table}[!htbp]
\centering
\caption{Independence-based versus agent-bootstrap standard errors}
\label{tab:empirical}
{\footnotesize
\begin{tabular}{lcccc}
\toprule
Evaluation point $(x_1,x_2)$ & $\widehat F^{-1}_{Y|W}(0.5\mid x_1,x_2)$ & i.i.d.\ s.e. & Bootstrap s.e. & Ratio \\
\midrule
$(17.72,\,18.60)$ & 8.05 & 0.134 & 0.400 & 3.0 \\
$(17.72,\,21.09)$ & 10.87 & 0.123 & 0.378 & 3.1 \\
$(21.09,\,18.60)$ & 12.38 & 0.112 & 0.337 & 3.0 \\
$(18.60,\,21.09)$ & 11.71 & 0.161 & 0.457 & 2.8 \\
\bottomrule
\end{tabular}}
\par\vspace{2mm}
{\footnotesize\singlespacing Notes: the median-shock structural function, which by the identity of Section \ref{sec:sep} is the conditional median $\widehat F^{-1}_{Y|W}(0.5\mid x_1,x_2)$; $h=0.30$. ``i.i.d.\ s.e.'' uses $\Omega_2/(nh^{2d})$ divided by the squared conditional density, the formula valid only under dyad-independent errors; because the target is a fixed-quantile object the normalization stage cancels from its first-order expansion, so this column carries one covariate-density factor and no cross-point term $\Omega_{12}$. ``Bootstrap s.e.'' is the agent-level bootstrap standard deviation over $B=399$ draws, with the shock quantiles re-estimated inside each draw. The final column is their ratio: the independence formula understates the sampling standard deviation by this factor at every evaluation point.}

\end{table}

Table \ref{tab:empirical} contains the paper's central message in real data. Its target is $\widehat F^{-1}_{Y|W}(0.5\mid x_1,x_2)$, that is $\widehat g$ at the \emph{estimated} median shock with the quantile re-estimated inside every bootstrap draw; by the identity of Section \ref{sec:sep} this is the fixed-$q$ conditional quantile, for which the normalization stage cancels exactly: expanding as in Appendix A.8 with $\widehat s=\widehat F_e(\widehat e_q)=q$,
\[
\widehat g(w,\widehat e_q)-Q_w(q)=-\frac{\widehat F_{Y|W}(Q_w(q)\mid w)-q}{f_{Y|W}(Q_w(q)\mid w)}+o_p(r_N),
\]
so the leading variance is that of a \emph{single} conditional distribution function divided by the squared conditional density, and $\Omega_{12}$ of \eqref{eq:Om12} does not appear; it would if a numerical value of $e$ were held fixed, the object Theorem \ref{theo:g} describes. The independence-formula column is computed for the fixed-quantile functional reported here and carries one density factor; holding a numerical shock value fixed would define a different functional and require the corresponding cross-point covariance term.

At four evaluation points spanning small and large exporter--importer configurations, the agent-bootstrap standard errors exceed the independence-formula standard errors by factors between $2.8$ and $3.1$: i.i.d.-style inference on this entirely standard gravity dataset would report intervals about a third as wide as the bootstrap suggests, because dyads sharing a country are numerous and the shared-country variance component dominates. The comparison is indicative rather than a verified coverage statement, for the reasons of Section \ref{sec:data} and because the bootstrap standard error is itself an estimate. The magnitude is not peculiar to this estimator or sample: \cite{chen2025}, with a \emph{linear} quantile gravity regression on different trade data and an analytic dyadic-robust covariance matrix, reports standard errors about twice those ignoring dyadic dependence. The factors differ, but both say the correction is a multiple rather than a few percent.

Finally, the quantile counterfactual. For a median-size economy exporting to a large one (ninetieth percentile), moving the pair's composite shock from its lower to its upper quartile multiplies predicted exports by $e^{2.84}\approx17$, with a $95$ percent bootstrap interval of $[8,\,34]$ in levels. The point estimate is large because bilateral shocks are large. The interval is computed at $h_{\mathrm{CV}}$ with the shock quantiles re-estimated within each bootstrap draw, which is the functional described in Section \ref{sec:sep}; holding them fixed across draws would resample a different functional, with no generally signed effect on the width. A quantile regression can report how the $25$th and $75$th conditional percentiles of trade differ at this covariate configuration, and \cite{chen2025} does exactly that on gravity data; what requires the present framework is the reading of that difference as the effect of moving one pair's own composite shock between two fixed points of its distribution, holding the two economies' sizes at their observed values.

\subsection{Is the nonseparable model necessary?}\label{sec:sep}

Panel (a) of Figure \ref{fig:empirical} suggests a strong interaction between size and shocks, but the claim deserves to be quantified directly, because it is exactly what an additive model denies. Under separability, $Y_{ij}=m(X_i,X_j)+\varepsilon_{ij}$ with $\varepsilon_{ij}$ independent of the covariates, the conditional quantile function is $Q_\tau(Y\mid x_1,x_2)=m(x_1,x_2)+Q_\tau(\varepsilon)$: the quantile curves are parallel vertical shifts, so the conditional interquartile range does not vary with $(x_1,x_2)$. In the notation of this paper the same object is $\widehat g(x,\bar x,e_{0.75})-\widehat g(x,\bar x,e_{0.25})$, so additive separability delivers a sharp, directly testable restriction on an object the estimator already produces. (The exporter points are taken away from $\bar x$, where the normalization $\widehat g(\bar x,\bar x,e)=e$ would make the quantity mechanical.)

\begin{table}[!htbp]
\centering
\caption{Conditional interquartile range of log exports, by exporter size}
\label{tab:sep}
{\footnotesize
\begin{tabular}{lccc}
\toprule
 & \multicolumn{3}{c}{Exporter size $x$} \\
\cmidrule(lr){2-4}
 & 10th pct. & 40th pct. & 90th pct. \\
\midrule
Conditional interquartile range of $\log$ exports & 4.37 & 3.74 & 1.91 \\
(agent-bootstrap s.e.) & (0.40) & (0.39) & (0.37) \\
\midrule
\multicolumn{4}{l}{Difference, 10th percentile minus 90th percentile: 2.46 (s.e. 0.54, 95\% CI [1.30, 3.39], $t=4.54$)} \\
\bottomrule
\end{tabular}}
\par\vspace{2mm}
{\footnotesize\singlespacing Notes: the conditional interquartile range is $\widehat g(x,\bar x,e_{0.75})-\widehat g(x,\bar x,e_{0.25})$, equivalently $\widehat F^{-1}(0.75\mid x,\bar x)-\widehat F^{-1}(0.25\mid x,\bar x)$, with the importer held at $\bar x$. The exporter points avoid $\bar x$ itself, where $\widehat g(\bar x,\bar x,e)=e$ holds by construction. Under a model additively separable in a covariate-independent error this quantity does not depend on $x$. All quantities are computed at $h=h_{\mathrm{CV}}=0.30$; standard errors and the interval are nominal agent-bootstrap quantities, whose coverage is not established for this observation process (Section \ref{sec:data}), computed with $B=399$ draws, the shock quantiles being re-estimated within each draw.}

\end{table}

An identity simplifies what these curves are. The shock quantiles are estimated as $\widehat e_q=\widehat F^{-1}_{Y|\bar x,\bar x}(q)$ with the same normalization distribution used in $\widehat g$, so $\widehat F_e(\widehat e_q)=q$ by construction and
\[
\widehat g(x_1,x_2,\widehat e_q)=\widehat F^{-1}_{Y|W}\!\left(\widehat F_e(\widehat e_q)\,\big|\,x_1,x_2\right)=\widehat F^{-1}_{Y|W}(q\mid x_1,x_2)
\]
exactly, up to interpolation. The quartile curves are therefore estimated \emph{conditional quantile} curves: they require no more than monotonicity, and it is their interpretation as $g$ at a fixed shock that uses the model. The agent bootstrap re-estimates $\widehat e_q$ inside every draw, so the resampled object is the functional reported. Because $\widehat F_e(\widehat e_q)=q$ holds identically, the normalization stage cancels from the first-order expansion and the uncertainty of a fixed-$q$ curve is that of a single conditional quantile. The term $\Omega_{12}$ of Theorem \ref{theo:g} belongs to the different functional in which a numerical shock value is held fixed, which is not what the exhibits report; holding an estimated value fixed across draws would resample that other functional, with no generally signed effect.

Table \ref{tab:sep} and panel (c) of Figure \ref{fig:empirical} show a large and systematic departure from constancy. The conditional interquartile range falls from $4.37$ log points at the tenth percentile of exporter size to $1.91$ at the ninetieth, a difference of $2.46$ with an agent-bootstrap standard error of $0.54$, a $95$ percent interval of $[1.30,3.39]$ and $t=4.54$. The statistic is a descriptive measure of the departure with a bootstrap assessment of its sampling variability, not a test with a calibrated level: the reference distribution is not supplied by the theory for these data, since the observation process lies outside Assumption \ref{ass:sampling} and the intervals are computed at $h_{\mathrm{CV}}$ without bias correction. Economically, bilateral shocks are worth roughly a factor of $80$ in trade for a small exporter and a factor of $7$ for a large one: large economies are insured against bilateral circumstance in a way small economies are not. The difference is between $2.2$ and $2.5$ across bandwidths from $0.20$ to $0.50$ and $1.73$ at the undersmoothed bandwidth (Table \ref{tab:hsens}), and a leave-one-economy-out recomputation gives values between $2.27$ and $2.67$, always of the same sign. Two qualifications on what is being compared. The restriction is that the conditional quantiles are parallel in $x$, which is what $Y_{ij}=m(X_i,X_j)+\varepsilon_{ij}$ implies \emph{when $\varepsilon_{ij}$ is independent of the covariates}; an additive model with a residual whose dispersion depends on $x$ is not distinguished by this statistic, and the target is the location-shift specification the applied gravity literature estimates. And the divergence from the Nadaraya--Watson curve in panel (a) is a difference of estimands rather than evidence of misspecification, since an additive model with a skewed error would also produce it; what separates the two specifications is the shape of the conditional distribution at fixed $x$.

A second comparison asks how much the restriction costs out of sample, using the cross-validation criterion of Theorem \ref{theo:cv} as a scoring rule for the conditional distribution rather than as a bandwidth selector. For each threshold $t$ I compare the leave-pair-out score of the nonseparable estimator against its separable counterpart $\widehat F_\varepsilon(t-\widehat m(x_1,x_2))$, where $\widehat m$ is the dyadic Nadaraya--Watson mean and $\widehat F_\varepsilon$ the residual distribution.

\begin{table}[!htbp]
\centering
\caption{Out-of-sample comparison of conditional distribution estimates}
\label{tab:cmp}
{\footnotesize
\begin{tabular}{lccccc}
\toprule
Threshold $t$ & Nonseparable & Separable & Improvement & Score difference & (s.e.) \\
\midrule
7.80 & 0.1249 & 0.1268 & 1.6\% & +0.0020 & (0.0009) \\
10.46 & 0.1419 & 0.1450 & 2.2\% & +0.0032 & (0.0015) \\
12.61 & 0.1003 & 0.1026 & 2.3\% & +0.0023 & (0.0008) \\
\bottomrule
\end{tabular}}
\par\vspace{2mm}
{\footnotesize\singlespacing Notes: leave-pair-out cross-validation scores for the conditional distribution function, $\psi(y)=\ind(y\le t)$, lower is better; thresholds are the quartiles of $Y$. ``Nonseparable'' is the estimator of Section 3. ``Separable'' is $\widehat F_\varepsilon(t-\widehat m(x_1,x_2))$ under $Y_{ij}=m(X_i,X_j)+\varepsilon_{ij}$ with $\varepsilon$ independent of the covariates. \emph{Both} components of the separable benchmark are computed leave-pair-out: for each held-out dyad, its conditional mean and the residual distribution $\widehat F_\varepsilon$ are estimated from the dyads sharing no agent with it, so neither procedure sees the held-out observation. The score difference is the per-dyad mean of the separable minus the nonseparable squared error, with a standard error from a leave-one-agent-out jackknife over the $119$ economies.}

\end{table}

Table \ref{tab:cmp} reports the result. Both components of the separable benchmark are computed leave-pair-out: for each held-out dyad its conditional mean and the residual distribution $\widehat F_\varepsilon$ are estimated from the dyads sharing no agent with it, so neither procedure sees the observation it is scored on. The nonseparable estimator wins at every threshold, by between $1.6$ and $2.3$ percent of mean squared prediction error, with per-dyad score differences of $2.1$, $2.1$ and $2.9$ agent-clustered standard errors. Read together, the two exercises say that additivity costs little for predicting whether a flow falls below a threshold while describing the shape of the conditional distribution quite differently, and it is the shape that structural questions turn on; neither is a test with a calibrated level, for the reasons of Section \ref{sec:data}.

\subsection{What is the composite shock? A validation using distance}\label{sec:validate}

The model absorbs all bilateral frictions into $e_{ij}$, so the recovered shocks should behave the way frictions do. Inverting the estimated structural function gives $\widehat e_{ij}=\widehat F^{-1}_{Y|\bar x,\bar x}\big(\widehat F_{Y|X_i,X_j}(Y_{ij})\big)$ for every dyad, constructed without using distance. Its correlation with log distance is $-0.48$, and a regression on log distance has slope $-1.63$: farther pairs draw lower composite shocks, at the magnitude gravity regressions report. This is a reassuring specification check rather than an identified elasticity, since $\widehat e_{ij}$ is a residual-like object into which every friction is folded, and part of the gradient could reflect smoothing.

The same exercise speaks weakly to the model's key restriction. Assumption \ref{ass:sampling}(iv) requires $e_{ij}$ to be independent of the two GDPs, and the correlation between log distance, the largest observable component of the shock, and log GDP is $0.038$. This bears little weight: a zero linear correlation is not independence, distance is one component of $e_{ij}$ among many, and the assumption concerns the whole composite shock. The figure is consistent with the assumption and would have been a warning had it been large. Distance is not simply used as a regressor because it is a dyad-level covariate, whereas the framework of Section \ref{sec:model} conditions on agent characteristics; the extension to $g(X_i,X_j,Z_{ij},e_{ij})$ is developed in Appendix B and applied to distance there.

\begin{table}[!htbp]
\centering
\caption{Sensitivity of the headline quantities to the bandwidth}
\label{tab:hsens}
{\footnotesize
\begin{tabular}{lcc}
\toprule
Bandwidth $h$ & IQR of the shock at $\bar x$ & IQR$(p_{10})-$IQR$(p_{90})$ \\
\midrule
0.150 & 3.51 & 2.16 \\
0.200 & 3.64 & 2.40 \\
0.250 & 3.67 & 2.47 \\
0.300 (CV) & 3.67 & 2.46 \\
0.400 & 3.72 & 2.36 \\
0.500 & 3.79 & 2.23 \\
0.115 (undersmoothed) & 3.29 & 1.73 \\
\bottomrule
\end{tabular}}
\par\vspace{2mm}
{\footnotesize\singlespacing Notes: sensitivity of the two headline quantities to the smoothing bandwidth. $h_{\mathrm{CV}}=0.30$ is selected by the leave-pair-out criterion of Theorem \ref{theo:cv}; $h_{\mathrm{us}}=h_{\mathrm{CV}}N^{-1/5}=0.115$ is the undersmoothed bandwidth that Assumption \ref{ass:bw} requires for the centering of the limiting distribution. The sign and rough magnitude of the separability violation are unaffected across the range.}

\end{table}

\begin{table}[!htbp]
\centering
\caption{Effective local sample size at the evaluation points}
\label{tab:eff}
{\footnotesize
\begin{tabular}{lccc}
\toprule
Bandwidth & Exporter point & Effective dyads & Effective sender agents \\
\midrule
$h_{\mathrm{CV}}=0.30$ & $p_{10}$ & 483 & 25.3 \\
$h_{\mathrm{CV}}=0.30$ & $p_{40}$ & 654 & 32.8 \\
$h_{\mathrm{CV}}=0.30$ & $p_{90}$ & 240 & 10.8 \\
$h_{\mathrm{us}}=0.115$ & $p_{10}$ & 60 & 11.3 \\
$h_{\mathrm{us}}=0.115$ & $p_{40}$ & 93 & 16.1 \\
$h_{\mathrm{us}}=0.115$ & $p_{90}$ & 35 & 5.5 \\
\bottomrule
\end{tabular}}
\par\vspace{2mm}
{\footnotesize\singlespacing Notes: Kish effective sample sizes $(\sum w)^2/\sum w^2$ for the kernel weights at the evaluation point $(x,\bar x)$, and for the sender-slot agent weights alone. The nominal sample is $n=13{,}221$ dyads and $N=119$ economies; the table reports the local information actually available to the estimator, which is what the rate $\sqrt{Nh^{d}}$ of Theorem \ref{theo:cdf}(i) refers to.}

\end{table}

\subsection{Scope of the application}\label{sec:caveats}

Four limits bind what this exercise establishes. The observation process of Section \ref{sec:data} is the binding one: the array is $5.85$ percent short of complete, selected on the outcome, and no part of the theory covers that selection. The second is local information: the rate $\sqrt{Nh^{d}}$ makes the relevant count the number of \emph{agents} within a bandwidth, not the $13{,}221$ dyads, and Table \ref{tab:eff} reports Kish effective sample sizes of $11$ to $33$ sender economies at $h_{\mathrm{CV}}$ and $5$ to $16$ at the undersmoothed bandwidth; asymptotic approximations are being applied at effective agent counts in the low tens. The third is the random-effects restriction, which multilateral-resistance components correlated with GDP would violate; the correlated-heterogeneity extension discussed in the conclusion is the remedy. The fourth is that a single cross-section of a single network says nothing about dynamics, and a second network would be needed before any of these magnitudes could be called typical.

\section{Conclusion}\label{conclusion}

This paper studies a nonseparable dyadic regression model with a scalar monotone unobservable, providing identification of a normalized representative of the structural function and error distribution, kernel plug-in estimators, and, centrally, a corrected asymptotic theory under dyadic dependence. The main message is methodological: for structural nonparametric objects estimated from dyadic data, the shared-agent variance component generically dominates, i.i.d.-based inference collapses (with coverage deteriorating as the network grows), and agent-level resampling substantially restores it, with consistency proved in the nondegenerate regime for the model without a dyad-level covariate and conservative behavior in the degenerate case found in simulation. These conclusions extend beyond the specific model: any plug-in estimator built from dyadic kernel distribution estimators inherits the two-regime structure derived here.

Several extensions are underway or left for future work. First, correlated heterogeneity: replacing Assumption \ref{ass:sampling}(iv) with a control-function or fixed-effects-robust structure would connect this framework to \cite{Graham}, \cite{auerbach2022}, \cite{andrei}, and \cite{diegert2024}, whose identification of two-way heterogeneity distributions suggests a route to relaxing scalar monotonicity as well. Second, endogenous link formation: when $Y_{ij}$ is observed only for linked pairs, the model acquires a selection equation, and identification will require exclusion restrictions or a parametric selection structure. Third, discrete and censored outcomes under weak monotonicity, with partial identification of $F_e$ on flat regions of $g$. Fourth, uniform-in-$(x,e)$ inference bands for $\widehat g$ via the high-dimensional exchangeable-array approximations of \cite{chiang2023} and \cite{cattaneo2024}. Finally, extending the bilateral-trade application of Section \ref{empirical} to a panel of cross-sections would permit the shock distribution, and the counterfactuals built on it, to vary over time.

\section*{Data availability statement}
The data that support the findings of this study are derived from the CEPII Gravity database, version 202211 \citep{conte2022}, which is openly available at \url{http://www.cepii.fr/CEPII/en/bdd_modele/bdd_modele_item.asp?id=8}. The estimation sample, all code for the simulations and the empirical application, and instructions for reproducing every table and figure are provided in the replication package accompanying this article. 



\bibliographystyle{apalike}
\bibliography{mybib2026}

\clearpage
\appendix
\section*{Appendix: Proofs}\label{appendix}
\renewcommand{\thesection}{A}
\setcounter{equation}{0}
\renewcommand{\theequation}{A.\arabic{equation}}

\subsection*{A.1 \ Proof of Lemma \ref{lem:obseq}}
Using \eqref{eq:map} and Assumption \ref{ass:structure}(iv), $F_{Y,X}(\cdot;g,F_e)=F_{Y,X}(\cdot;g',F'_e)$ holds if and only if
\[
F'_e(m'(x_1,x_2,y))=F_e(m(x_1,x_2,y))\quad\text{for all }(x_1,x_2,y),
\]
that is, if and only if $m'=(F'_e)^{-1}\circ F_e\circ m\equiv u\circ m$, with $u=(F'_e)^{-1}\circ F_e$ continuous and strictly increasing by Assumption \ref{ass:structure}(iii). Conversely, given such $u$, set $F'_e=F_e\circ u^{-1}$; then $F'_e(m'(x_1,x_2,y))=F_e(m(x_1,x_2,y))$, so the two pairs generate the same observables. $\blacksquare$

\subsection*{A.2 \ Proof of Lemma \ref{lem:norm}}
Let $(g,F_e)$ be given with inverse $m$. Define $u(\cdot)=g(\bar{x}_1,\bar{x}_2,\cdot)$, which is continuous and strictly increasing by Assumption \ref{ass:structure}(i)--(ii). Set $m^\ast=u\circ m$ and let $g^\ast$ be the inverse of $m^\ast$ in its last argument, with $F^\ast_e=F_e\circ u^{-1}$. By Lemma \ref{lem:obseq}, $(g^\ast,F^\ast_e)$ is observationally equivalent to $(g,F_e)$. For any $y$: $m^\ast(\bar{x}_1,\bar{x}_2,y)=g(\bar{x}_1,\bar{x}_2,m(\bar{x}_1,\bar{x}_2,y))=y$, since $m(\bar x_1,\bar x_2,\cdot)$ and $g(\bar x_1,\bar x_2,\cdot)$ are inverse to each other. Hence $g^\ast(\bar{x}_1,\bar{x}_2,e)=e$ for all $e$. $\blacksquare$

\subsection*{A.3 \ Proof of Theorem \ref{t1}}
For $e\in\EE$ and $x_1=(x^0_1,x^1_1)$, $x_2=(x^0_2,x^1_2)$,
\begin{align}
F_{e|X^0}(e|x^0_1,x^0_2)
&=\PP\left(e_{ij}\le e\mid X_i=(x^0_1,x^1_1),X_j=(x^0_2,x^1_2)\right)\label{eq:pfa}\\
&=\PP\left(g(X_i,X_j,e_{ij})\le g(X_i,X_j,e)\mid X_i=x_1,X_j=x_2\right)\notag\\
&=F_{Y|X}\!\left(g(x_1,x_2,e)\mid x_1,x_2\right),\label{eq:key}
\end{align}
where \eqref{eq:pfa} uses Assumption \ref{ass:condind} and the equality in \eqref{eq:key} uses strict monotonicity. Setting $x^1=\bar{x}^1$ in \eqref{eq:key} and invoking normalization \eqref{eq:norm2} (well defined on the support by Assumption \ref{ass:support}) gives \eqref{eq:idF}. By the second clause of Assumption \ref{ass:condind} the conditional error distribution is continuous and strictly increasing on $\EE$, and with the continuity and strict monotonicity of $g$ in Assumption \ref{ass:structure} this makes $F_{Y|X}(\cdot|x_1,x_2)$ invertible on the corresponding outcome range. Combining \eqref{eq:key} with \eqref{eq:idF} and applying that inverse yields \eqref{eq:idg}. The simplifications when $X^0$ is absent are immediate. The homogeneity case follows the same steps with \eqref{eq:norm2} replaced by Assumption \ref{ass:hd1} evaluated at $\lambda=e/\bar{e}$ (well defined by Assumption \ref{ass:ray}). $\blacksquare$

\subsection*{A.4 \ Notation, conventions, and standing facts}

Fix the interior evaluation point $(y_\circ,w)$, $w=(w_1,w_2)$, and abbreviate $F(y)=F_{Y|W}(y|w_1,w_2)$, $F=F(y_\circ)$, $f_1=f_W(w_1)$, $f_2=f_W(w_2)$. Write $\zeta_i=(X_i,U_i)$ for agent $i$'s \emph{full type} and $W_i=W(X_i)$; by Assumption \ref{ass:sampling}(i) the $\zeta_i$ are i.i.d. Conditional on $\bm\zeta=(\zeta_1,\dots,\zeta_N)$, the pair $(Y_{ij},Y_{ji})$ is a function of $(V_{ij},V_{ji})$ alone, so the unordered-pair blocks $\{(Y_{ij},Y_{ji})\}_{i<j}$ are conditionally independent across pairs given $\bm\zeta$ (Assumption \ref{ass:sampling}(ii)); this conditional-independence structure is used repeatedly and is the only consequence of Assumption \ref{ass:sampling} the proofs require.

Set $K_{1i}=K((w_1-W_i)/h)$, $K_{2j}=K((w_2-W_j)/h)$, $\bar k_{ij}(y)=\bar k((y-Y_{ij})/h_y)$,
\[
p(\zeta,\zeta';y)=\PP\left(Y_{ij}\le y\mid \zeta_i=\zeta,\zeta_j=\zeta'\right),\qquad
\mu(\zeta,\zeta';y)=\E\left[\bar k_{ij}(y)\mid \zeta_i=\zeta,\zeta_j=\zeta'\right],
\]
\[
\bar p_s(\zeta;y)=\E\left[p(\zeta,\zeta_j;y)\mid W_j=w_2\right],\qquad
\bar p_r(\zeta;y)=\E\left[p(\zeta_i,\zeta;y)\mid W_i=w_1\right],
\]
so that $\delta_s(\zeta;y)=\bar p_s(\zeta;y)-F(y)$ and $\delta_r(\zeta;y)=\bar p_r(\zeta;y)-F(y)$ are the deviation functions of \eqref{eq:Om1}. Note $\E[\delta_s(\zeta_i;y)\mid W_i=w_1]=0$ and $\E[\delta_r(\zeta_i;y)\mid W_i=w_2]=0$ by construction.

$C$ denotes a finite constant depending only on the bounds in Assumptions \ref{ass:smooth}--\ref{ass:kernel}, possibly changing from line to line. Proofs are given for a compactly supported $K$ (support in $\{\|u\|\le T\}$) and compactly supported $k$; Lemma \ref{lem:gauss} extends every statement to Gaussian kernels. We record three facts, immediate from Assumption \ref{ass:smooth}: uniformly in $(\zeta,\zeta')$,
\begin{itemize}
\item[(F1)] $y\mapsto p(\zeta,\zeta';y)$ has a density bounded by $\bar C$, so $|p(\zeta,\zeta';y)-p(\zeta,\zeta';y')|\le\bar C|y-y'|$, and the same Lipschitz bound holds for $\bar p_s,\bar p_r,F,\mu$;
\item[(F2)] $|\mu(\zeta,\zeta';y)-p(\zeta,\zeta';y)|\le Ch_y^{2}$, by a second-order Taylor expansion of the conditional distribution function under the symmetric density $k$ and the uniformly bounded derivative of the conditional density;
\item[(F3)] $|\bar k_{ij}(y)^2-\ind(Y_{ij}\le y)|\le\ind(|Y_{ij}-y|\le T_kh_y)$, so $\E[\bar k_{ij}(y)^2\mid\zeta_i,\zeta_j]=p(\zeta_i,\zeta_j;y)+O(h_y)$ uniformly.
\end{itemize}

\begin{lemma}[Localized moments]\label{lem:mom}
Let $r$ be bounded and measurable with $q_r(w')=\E[r(\zeta_i)\mid W_i=w']f_W(w')$ continuous at $w_1$. Then, as $h\to0$:
(i) $h^{-d}\E[K_{1i}\,r(\zeta_i)]\to q_r(w_1)$ and $h^{-d}\E[K_{1i}^{2}\,r(\zeta_i)]\to\left(\int K^2\right)q_r(w_1)$;
(ii) if $q_r$ is twice continuously differentiable near $w_1$, $h^{-d}\E[K_{1i}\,r(\zeta_i)]=q_r(w_1)+O(h^{2})$;
(iii) $\E[K_{1i}^{p}]=O(h^{d})$ for each fixed $p\ge1$;
(iv) if $w_1\neq w_2$ then $K((w_1-v)/h)K((w_2-v)/h)=0$ for all $v$ once $h<\|w_1-w_2\|/(2T)$.
\end{lemma}

\begin{proof}
(i) By iterated expectations,
\[
h^{-d}\E[K_{1i}r(\zeta_i)]=h^{-d}\!\int\! K((w_1-w')/h)\,q_r(w')dw'=\int\! K(v)\,q_r(w_1-hv)\,dv\to q_r(w_1)
\]
by $\int K=1$, continuity, boundedness, compact support, and dominated convergence; the $K^2$ statement is identical with $\int K^2<\infty$. (ii) Taylor-expand $q_r(w_1-hv)$ to second order; the linear term vanishes by $\int vK(v)dv=0$ (symmetry), and the quadratic term is $O(h^2)$ by the bounded second derivative and $\int\|v\|^2K(v)dv<\infty$. (iii) $\E[K_{1i}^p]\le\|K\|_\infty^{p-1}\E[K_{1i}]=O(h^d)$ by (i) with $r\equiv1$ and $f_W$ bounded. (iv) Both factors are nonzero only if $\|w_1-v\|\le Th$ and $\|w_2-v\|\le Th$, impossible when $2Th<\|w_1-w_2\|$.
\end{proof}

\begin{lemma}[Bias and denominator]\label{lem:bias}
Let $\theta_N(y)=h^{-2d}\E\left[K_{1i}K_{2j}\left(\mu(\zeta_i,\zeta_j;y)-F(y)\right)\right]$. Then $\sup_{y}|\theta_N(y)|=O(h^{2}+h_y^{2})$, where the supremum is over the compact neighborhood of $y_\circ$ in Assumption \ref{ass:smooth}. Moreover $\E[D_N]=f_1f_2+O(h^{2})$ and $\Var(D_N)\le C/(Nh^{d})$, hence $D_N\pto f_1f_2>0$.
\end{lemma}

\begin{proof}
By (F2), $\mu=p+O(h_y^2)$ uniformly, so it suffices to bound the same quantity with $\mu$ replaced by $p$. Iterating expectations over $(W_i,W_j)$,
\[
h^{-2d}\E\left[K_{1i}K_{2j}\left(p-F(y)\right)\right]=\iint K(v_1)K(v_2)\,\Pi_y(w_1-hv_1,w_2-hv_2)\,dv_1dv_2,
\]
where $\Pi_y(w',w'')=\left[F_{Y|W}(y|w',w'')-F_{Y|W}(y|w_1,w_2)\right]f_W(w')f_W(w'')$ vanishes at $(w_1,w_2)$ and is twice continuously differentiable with bounds uniform in $y$ (Assumption \ref{ass:smooth}). A second-order Taylor expansion with $\int vK=0$ gives the uniform $O(h^2)$ bound. For $D_N$: $\E D_N=h^{-2d}\E[K_{1i}K_{2j}]=f_1f_2+O(h^2)$ by Lemma \ref{lem:mom}(ii) applied twice. For the variance, decompose the sum over ordered pairs by shared indices: pairs sharing no index are independent with equal means; pairs sharing one index contribute at most $CN^{3}\E[K_{1i}^2]\E[K_{2j}]\E[K_{2k}]/(nh^{2d})^{2}\le CN^{3}h^{3d}/(N^{4}h^{4d})=C/(Nh^{d})$, and pairs sharing both indices contribute at most $CN^{2}h^{2d}/(N^{4}h^{4d})\le C/(N^{2}h^{2d})\le C/(Nh^d)$. Chebyshev and $Nh^{2d}\to\infty$ (Assumption \ref{ass:bw}) give $D_N\pto f_1f_2$.
\end{proof}

\subsection*{A.5 \ The Hoeffding decomposition and its three parts}

Recall the exact decomposition $\widehat F(y)-F(y)=\left[A_N(y)+B_N(y)\right]/D_N$ from the main text, with
\[
A_N(y)=\frac{1}{nh^{2d}}\sum_{i\ne j}K_{1i}K_{2j}\,\varepsilon_{ij}(y),\qquad \varepsilon_{ij}(y)=\bar k_{ij}(y)-\mu(\zeta_i,\zeta_j;y),
\]
\[
B_N(y)=\frac{1}{nh^{2d}}\sum_{i\ne j}K_{1i}K_{2j}\left[\mu(\zeta_i,\zeta_j;y)-F(y)\right].
\]
Define, for the $\bm\zeta$-measurable part, the centered pair kernel and its H\'ajek projections
\[
\phi(\zeta,\zeta';y)=h^{-2d}K\!\left(\tfrac{w_1-W(\zeta)}{h}\right)K\!\left(\tfrac{w_2-W(\zeta')}{h}\right)\left[\mu(\zeta,\zeta';y)-F(y)\right]-\theta_N(y),
\]
\[
\phi_1(\zeta;y)=\E[\phi(\zeta,\zeta_j;y)],\qquad \phi_2(\zeta;y)=\E[\phi(\zeta_i,\zeta;y)],\qquad \rho=\phi-\phi_1-\phi_2,
\]
where the expectations are over the independent second (resp.\ first) argument. All three are mean zero, and $\E[\rho(\zeta,\zeta_j;y)]=\E[\rho(\zeta_i,\zeta;y)]=0$ for every $\zeta$ (complete degeneracy of $\rho$). Since each index $l$ appears exactly $N-1$ times as first and $N-1$ times as second index in $\sum_{i\ne j}$,
\begin{gather}
B_N(y)-\theta_N(y)=\frac{1}{N}\sum_{l=1}^{N}s_l(y)+R_N(y),\label{eq:hoeff}\\
s_l(y)=\phi_1(\zeta_l;y)+\phi_2(\zeta_l;y),\qquad R_N(y)=\frac{1}{n}\sum_{i\ne j}\rho(\zeta_i,\zeta_j;y),\notag
\end{gather}
an exact identity.

\begin{lemma}[Degenerate remainder]\label{lem:rem}
$\E[R_N(y)^2]\le 2\,\E[\phi(\zeta_i,\zeta_j;y)^2]/n\le C/(nh^{2d})$, and for any $y,y'$ in the neighborhood of $y_\circ$,
\begin{multline*}
\E\left[(R_N(y)-R_N(y'))^2\right]\le \frac{2}{n}\,\E\left[(\phi(\zeta_i,\zeta_j;y)-\phi(\zeta_i,\zeta_j;y'))^2\right]\\
\le \frac{C\left(|y-y'|^{2}+(h^2+h_y^2)^{2}\right)}{nh^{2d}}.
\end{multline*}
\end{lemma}

\begin{proof}
Expand $\E[(\sum_{i\ne j}\rho_{ij})^2]=\sum_{i\ne j}\sum_{i'\ne j'}\E[\rho_{ij}\rho_{i'j'}]$. If $\{i,j\}\cap\{i',j'\}=\emptyset$ the terms are independent and mean zero. If the pairs share exactly one index, say $i=i'$, $j\ne j'$, then conditioning on $\zeta_i$ and using independence of $\zeta_j,\zeta_{j'}$,
$\E[\rho_{ij}\rho_{ij'}]=\E\big[\E[\rho(\zeta_i,\zeta_j)\mid\zeta_i]\,\E[\rho(\zeta_i,\zeta_{j'})\mid\zeta_i]\big]=0$
by complete degeneracy; the patterns $(i,j),(j',i)$, $(i,j),(i',i)$ etc.\ vanish identically for the same reason (degeneracy holds in each argument separately). If $\{i,j\}=\{i',j'\}$ there are two patterns, $(i,j)=(i',j')$ and $(i,j)=(j',i')$, each bounded by $\E[\rho^2]$ (Cauchy--Schwarz for the second). Hence $\E[(\sum\rho)^2]\le2n\E[\rho^2]$. Since $\phi_1(\zeta_i)$ and $\phi_2(\zeta_j)$ are the $L^2$-projections of $\phi$ on its two coordinates and are orthogonal to $\rho$ and to each other, $\E[\rho^2]=\E[\phi^2]-\E[\phi_1^2]-\E[\phi_2^2]\le\E[\phi^2]$. Finally $\E[\phi^2]\le2h^{-4d}\E[K_{1i}^2K_{2j}^2(\mu-F)^2]+2\theta_N^2\le Ch^{-4d}\cdot h^{2d}=Ch^{-2d}$ by boundedness of $\mu-F$ and Lemma \ref{lem:mom}(iii), giving the first bound. For the increment bound, apply the same argument to $\phi(\cdot;y)-\phi(\cdot;y')$, whose $\mu$-bracket is $[\mu(y)-\mu(y')]-[F(y)-F(y')]$, bounded in absolute value by $2\bar C|y-y'|$ by (F1); the stated bound follows (the $(h^2+h_y^2)^2$ term absorbs the $\theta_N$ differences via Lemma \ref{lem:bias}).
\end{proof}

\begin{lemma}[Projection form and variance]\label{lem:proj}
(i) It holds that
\[
\phi_1(\zeta;y)=h^{-d}K\!\left(\tfrac{w_1-W(\zeta)}{h}\right)\left[f_2\,\delta_s(\zeta;y)+u_N(\zeta;y)\right]-\theta_N(y)
\]
with $\sup_{\zeta,y}|u_N(\zeta;y)|\le C(h^{2}+h_y^{2})$, and symmetrically for $\phi_2$ with $f_1\,\delta_r$.
(ii) $h^{d}\,\Var\!\left(s_l(y_\circ)\right)\to\sigma_s^{2}$, where
\begin{multline*}
\sigma_s^{2}:=\left(\int K^{2}\right)\Big[f_1f_2^{2}\,\E[\delta_s^2\mid W=w_1]+f_1^{2}f_2\,\E[\delta_r^2\mid W=w_2]\\
+2\cdot\ind(w_1=w_2)f_0^{3}\,\E[\delta_s\delta_r\mid W=w_0]\Big],
\end{multline*}
the $\delta$'s are evaluated at $y_\circ$, $f_0=f_W(w_0)$, and $\sigma_s^{2}=(f_1f_2)^{2}\,\Omega_1(y_\circ;w)$.
(iii) $\E|s_l(y_\circ)|^{3}\le Ch^{-2d}$.
(iv) $\Var\!\left(s_l(y)-s_l(y')\right)\le Ch^{-d}\left(|y-y'|^{2}+(h^2+h_y^2)^{2}\right)$.
\end{lemma}

\begin{proof}
(i) By definition,
\[
\phi_1(\zeta;y)+\theta_N(y)=h^{-2d}K\!\left(\tfrac{w_1-W(\zeta)}{h}\right)\E\left[K_{2j}\left(\mu(\zeta,\zeta_j;y)-F(y)\right)\right].
\]
By (F2), $\mu(\zeta,\zeta';y)=p(\zeta,\zeta';y)+O(h_y^2)$ uniformly. Iterating over $W_j$,
\begin{gather*}
h^{-d}\E\left[K_{2j}\left(p(\zeta,\zeta_j;y)-F(y)\right)\right]=\int K(v)\,\Lambda_{\zeta,y}(w_2-hv)\,dv,\\
\Lambda_{\zeta,y}(w')=\left[\E[p(\zeta,\zeta_j;y)|W_j=w']-F(y)\right]f_W(w'),
\end{gather*}
and $\Lambda_{\zeta,y}$ is twice continuously differentiable with bounds uniform in $(\zeta,y)$ (Assumption \ref{ass:smooth}); Taylor with $\int vK=0$ gives $\int K\Lambda(w_2-hv)dv=\Lambda_{\zeta,y}(w_2)+O(h^2)=f_2\delta_s(\zeta;y)+O(h^2)$ uniformly. Collecting the $O(h_y^2)$ and $O(h^2)$ terms into $u_N$ proves (i).
(ii) Write $s_l=h^{-d}[K_{1l}(f_2\delta_s(\zeta_l)+u_N)+K_{2l}(f_1\delta_r(\zeta_l)+u_N')]-2\theta_N$. Expand the second moment. The leading squares converge by Lemma \ref{lem:mom}(i) applied with $r(\zeta)=\delta_s(\zeta;y_\circ)^2$ (resp.\ $\delta_r^2$), whose conditional-moment map is continuous at $w_1$ (resp.\ $w_2$) by Assumption \ref{ass:smooth}: $h^{d}\E[h^{-2d}K_{1l}^2f_2^2\delta_s^2]\to(\int K^2)f_1f_2^2\E[\delta_s^2|W=w_1]$. The cross term $h^{-2d}K_{1l}K_{2l}f_1f_2\delta_s\delta_r$ vanishes identically for small $h$ when $w_1\ne w_2$ (Lemma \ref{lem:mom}(iv)) and converges to the stated cross moment when $w_1=w_2$ by the same argument. Terms involving $u_N$ vanish: $h^{d}\cdot h^{-2d}\E[K_{1l}^2]\,\sup u_N^2\le C(h^2+h_y^2)^2\to0$ by Lemma \ref{lem:mom}(iii), and mixed terms vanish by Cauchy--Schwarz; $h^d\theta_N^2\to0$ by Lemma \ref{lem:bias}. Finally $\E[s_l]=0$ by construction, so the second moment is the variance, and the identity $\sigma_s^2=(f_1f_2)^2\Omega_1$ is the algebra displayed after \eqref{eq:Om1}.
(iii) $|s_l|\le Ch^{-d}(K_{1l}+K_{2l})+C(h^2+h_y^2)h^{-d}(K_{1l}+K_{2l})+2|\theta_N|$, so $\E|s_l|^3\le Ch^{-3d}\E[(K_{1l}+K_{2l})^3]+C\le Ch^{-2d}$ by Lemma \ref{lem:mom}(iii).
(iv) As in (ii) with $\delta_\cdot(\zeta;y)-\delta_\cdot(\zeta;y')$ in place of $\delta_\cdot$, using the Lipschitz bound $|\delta_s(\zeta;y)-\delta_s(\zeta;y')|\le2\bar C|y-y'|$ from (F1) and the uniform $u_N$ bounds.
\end{proof}

\begin{lemma}[H\'ajek CLT]\label{lem:hajek}
If $\Omega_1(y_\circ;w)>0$ then $\sqrt{Nh^{d}}\cdot\frac{1}{N}\sum_{l=1}^{N}s_l(y_\circ)\dto N(0,\sigma_s^{2})$.
\end{lemma}

\begin{proof}
The summands $z_{N,l}=h^{d/2}s_l(y_\circ)$ are i.i.d.\ across $l$ for each $N$, mean zero, with $\Var(z_{N,l})\to\sigma_s^2>0$ (Lemma \ref{lem:proj}(ii)) and $\E|z_{N,l}|^3\le Ch^{3d/2}h^{-2d}=Ch^{-d/2}$ (Lemma \ref{lem:proj}(iii)). Lyapunov's condition for the triangular array holds:
\[
\frac{N\,\E|z_{N,l}|^{3}}{\left(N\Var(z_{N,l})\right)^{3/2}}\le \frac{CN h^{-d/2}}{N^{3/2}\sigma_s^{3}}=\frac{C}{\sqrt{Nh^{d}}}\to0
\]
by Assumption \ref{ass:bw}. The Lindeberg--Feller CLT applies, and $\sqrt{Nh^d}N^{-1}\sum_l s_l=N^{-1/2}\sum_l z_{N,l}\dto N(0,\sigma_s^2)$.
\end{proof}

\begin{lemma}[The conditionally centered part]\label{lem:AN}
(i) $\Var(A_N(y))\le C/(nh^{2d})$ and, for any $y,y'$, $\Var(A_N(y)-A_N(y'))\le C(|y-y'|+h_y)/(nh^{2d})$.
(ii) Suppose condition (S) of Theorem \ref{theo:cdf}(ii) holds. Then
$\sqrt{nh^{2d}}\,A_N(y_\circ)\dto N\!\left(0,\sigma_A^{2}\right)$ with
$\sigma_A^{2}=F(1-F)\left(\int K^2\right)^{2}f_1f_2+\ind(w_1=w_2)\,C_{\mathrm{rev}}\left(\int K^2\right)^{2}f_0^{2}$,
where $C_{\mathrm{rev}}=\Cov(\ind(Y_{ij}\le y_\circ),\ind(Y_{ji}\le y_\circ)\mid W_i=W_j=w_0)$, and $\sigma_A^2=(f_1f_2)^2[\Omega_2+\ind(w_1=w_2)\Omega_2']$.
\end{lemma}

\begin{proof}
(i) Conditional on $\bm\zeta$, the variables $\varepsilon_{ij}$ are mean zero, and $\varepsilon_{ij},\varepsilon_{i'j'}$ are conditionally independent unless $\{i,j\}=\{i',j'\}$ (they depend on disjoint sets of $V$'s). Hence, unconditionally, $\E[K_{1i}K_{2j}K_{1i'}K_{2j'}\varepsilon_{ij}\varepsilon_{i'j'}]=\E\big[K\cdots K\,\E[\varepsilon_{ij}\varepsilon_{i'j'}\mid\bm\zeta]\big]=0$ except when $\{i,j\}=\{i',j'\}$, leaving
\begin{multline*}
\Var(A_N)=\frac{1}{(nh^{2d})^{2}}\sum_{i\ne j}\left\{\E\left[K_{1i}^2K_{2j}^2\varepsilon_{ij}^2\right]+\E\left[K_{1i}K_{2j}K_{1j}K_{2i}\,\varepsilon_{ij}\varepsilon_{ji}\right]\right\}\\
\le\frac{Cnh^{2d}}{(nh^{2d})^{2}}=\frac{C}{nh^{2d}},
\end{multline*}
using $|\varepsilon|\le1$ and Lemma \ref{lem:mom}(iii). For the increment, $\varepsilon_{ij}(y)-\varepsilon_{ij}(y')$ has conditional second moment bounded by $\E[|\bar k_{ij}(y)-\bar k_{ij}(y')|\,|\,\zeta_i,\zeta_j]\le \bar C(|y-y'|+2T_kh_y)$ since $0\le\bar k_{ij}(y)-\bar k_{ij}(y')\le\ind(y'-T_kh_y\le Y_{ij}\le y+T_kh_y)$ for $y>y'$ and the conditional density is bounded; the same display then gives the stated bound.
(ii) Order the unordered pairs $\{i,j\}$, $i<j$, and set $b_{ij}=K_{1i}K_{2j}\varepsilon_{ij}+K_{1j}K_{2i}\varepsilon_{ji}$, so $\sum_{i\ne j}K_{1i}K_{2j}\varepsilon_{ij}=\sum_{i<j}b_{ij}$ with the $b_{ij}$ conditionally independent, mean zero, and uniformly bounded by $C$ given $\bm\zeta$. Let $V_N=(nh^{2d})^{-1}\sum_{i<j}\Var(b_{ij}\mid\bm\zeta)$, so that $(nh^{2d})\Var(A_N\mid\bm\zeta)=V_N$. By Lemma \ref{lem:condvar} below, $V_N\pto\sigma_A^{2}>0$. By the Berry--Esseen inequality for sums of independent, non-identically distributed random variables applied conditionally on $\bm\zeta$,
\begin{multline*}
\sup_{t}\left|\PP\left(\sqrt{nh^{2d}}A_N\le t\mid\bm\zeta\right)-\Phi\!\left(t/\sqrt{V_N}\right)\right|
\le C\,\frac{\sum_{i<j}\E[|b_{ij}|^{3}\mid\bm\zeta]}{\left(\sum_{i<j}\Var(b_{ij}\mid\bm\zeta)\right)^{3/2}}\\
\le C\,\frac{\sum_{i<j}(K_{1i}K_{2j}+K_{1j}K_{2i})^{3}}{\left(nh^{2d}V_N\right)^{3/2}}.
\end{multline*}
The numerator has expectation $O(nh^{2d})$ (bounded kernels and Lemma \ref{lem:mom}(iii)), so by Markov it is $O_p(nh^{2d})$, and the ratio is $O_p\!\left((nh^{2d})^{-1/2}\right)\pto0$ on the event $\{V_N\ge\sigma_A^2/2\}$, whose probability tends to one. Hence for every $t$, $\PP(\sqrt{nh^{2d}}A_N\le t\mid\bm\zeta)\pto\Phi(t/\sigma_A)$, and taking expectations (dominated convergence) yields the unconditional CLT.
\end{proof}

\begin{lemma}[Conditional variance concentration]\label{lem:condvar}
Under condition (S), $V_N\pto\sigma_A^{2}$.
\end{lemma}

\begin{proof}
Write $\Var(b_{ij}\mid\bm\zeta)=K_{1i}^2K_{2j}^2\sigma^2_{ij}+K_{1j}^2K_{2i}^2\sigma^2_{ji}+2K_{1i}K_{2j}K_{1j}K_{2i}c_{ij}$, with $\sigma^2_{ij}=\Var(\bar k_{ij}(y_\circ)\mid\zeta_i,\zeta_j)$ and $c_{ij}=\Cov(\bar k_{ij},\bar k_{ji}\mid\zeta_i,\zeta_j)$. By (F3), $\sigma^2_{ij}=p_{ij}(1-p_{ij})+O(h_y)$ and, under condition (S), $p_{ij}$ and the conditional law of $(\ind(Y_{ij}\le y_\circ),\ind(Y_{ji}\le y_\circ))$ given $(\zeta_i,\zeta_j)$ depend on the types only through $(W_i,W_j)$; write $p_{ij}=q(W_i,W_j)$ and $c_{ij}=c(W_i,W_j)+O(h_y)$, with $q=F_{Y|W}(y_\circ|\cdot,\cdot)$ continuous (Assumption \ref{ass:smooth}) and $c$ continuous at $(w_0,w_0)$ by Assumption \ref{ass:smooth}. Then, by Lemma \ref{lem:mom}(i)--(iv),
\begin{multline*}
\E[V_N]\to\left(\int K^2\right)^{2}f_1f_2\,q(w_1,w_2)(1-q(w_1,w_2))\\
+\ind(w_1=w_2)\left(\int K^2\right)^{2}f_0^{2}\,c(w_0,w_0)=\sigma_A^2,
\end{multline*}
using $q(w_1,w_2)=F$ and $c(w_0,w_0)=C_{\mathrm{rev}}$. For the variance of $V_N$: $V_N$ is an average over pairs of bounded, kernel-localized functions of $(\zeta_i,\zeta_j)$; exactly as in the proof of Lemma \ref{lem:bias} (denominator part), pairs sharing no index are uncorrelated up to means, one-shared-index terms contribute $O(1/(Nh^{d}))$, and both-shared terms $O(1/(nh^{2d}))$; hence $\Var(V_N)\to0$ and Chebyshev concludes.
\end{proof}

\subsection*{A.6 \ Uniformity in $y$}

\begin{lemma}[Glivenko--Cantelli and local equicontinuity]\label{lem:unif}
(i) $\sup_{y\in\R}|\widehat F(y)-F(y)|\pto0$.
(ii) Let $r_N=(Nh^{d})^{-1/2}$ and $\epsilon_N=r_N\log N$. Then
\[
\sup_{|y-y_\circ|\le\epsilon_N}\left|\left[\widehat F(y)-F(y)\right]-\left[\widehat F(y_\circ)-F(y_\circ)\right]\right|=o_p\!\left(r_N\right).
\]
\end{lemma}

\begin{proof}
(i) Fix $\epsilon>0$ and choose (by continuity and monotonicity of $F$, with limits $0$ and $1$) finitely many points $y_1<\dots<y_m$ with $F(y_{k+1})-F(y_k)\le\epsilon$ covering $\R$ in the usual P\'olya fashion. At each $y_k$, $\widehat F(y_k)-F(y_k)\pto0$: by \eqref{eq:hoeff} and Lemmas \ref{lem:bias}--\ref{lem:AN}(i), the numerator has variance $O(1/(Nh^d))+O(1/(nh^{2d}))\to0$ and bias $O(h^2+h_y^2)\to0$, and $D_N\pto f_1f_2>0$. For $y\in[y_k,y_{k+1}]$, monotonicity of $\widehat F$ and $F$ gives
$\widehat F(y)-F(y)\le\widehat F(y_{k+1})-F(y_{k+1})+\epsilon$ and the symmetric lower bound; taking $\max_k$ and letting $\epsilon\downarrow0$ along a sequence concludes.
(ii) Write $\widehat F(y)-F(y)=[G_N(y)+\theta_N(y)]/D_N$ with $G_N(y)=A_N(y)+\frac1N\sum_ls_l(y)+R_N(y)$. Since $D_N\pto f_1f_2>0$, it suffices to prove $\sup_{|y-y_\circ|\le\epsilon_N}|G_N(y)-G_N(y_\circ)|=o_p(r_N)$ and $2\sup_y|\theta_N(y)|=O(h^2+h_y^2)=o(r_N)$; the latter is Lemma \ref{lem:bias} plus Assumption \ref{ass:bw} ($\sqrt{Nh^d}(h^2+h_y^2)\to0$). Grid $[y_\circ-\epsilon_N,y_\circ+\epsilon_N]$ by $y_\circ-\epsilon_N=t_0<t_1<\dots<t_{m_N}=y_\circ+\epsilon_N$ with $F(t_{k+1})-F(t_k)\le r_N^{2}$ and $t_{k+1}-t_k\le r_N^{2}/\underline f$, which is possible with $m_N\le C\epsilon_N/r_N^{2}=C(Nh^d)\epsilon_N$ points since $f_{Y|W}$ is bounded below near $y_\circ$ (Assumption \ref{ass:smooth}). By Lemmas \ref{lem:rem}, \ref{lem:proj}(iv), and \ref{lem:AN}(i), for each $k$,
\[
\Var\!\left(G_N(t_k)-G_N(y_\circ)\right)\le C\left[\frac{\epsilon_N^{2}+(h^2+h_y^2)^{2}}{Nh^{d}}+\frac{\epsilon_N+h_y}{nh^{2d}}\right].
\]
Chebyshev and a union bound give, for any $\eta>0$,
\begin{multline*}
\PP\!\left(\max_{k\le m_N}\left|G_N(t_k)-G_N(y_\circ)\right|>\eta\,r_N\right)
\le\frac{m_N\max_k\Var}{\eta^{2}r_N^{2}}\\
\le\frac{C}{\eta^{2}}\left[\epsilon_N^{3}Nh^{d}+\epsilon_N Nh^{d}(h^2+h_y^2)^{2}+\epsilon_N^{2}+\epsilon_Nh_y\right].
\end{multline*}
With $\epsilon_N=r_N\log N$: $\epsilon_N^3Nh^d=(\log N)^3(Nh^d)^{-1/2}\to0$; $\epsilon_NNh^d(h^2+h_y^2)^2=\left[\sqrt{Nh^d}(h^2+h_y^2)\right]^{2}\log N/\sqrt{Nh^{d}}\to0$ by Assumption \ref{ass:bw}; the remaining terms clearly vanish. Between grid points, monotonicity in $y$ of each of $\widehat F$ and $F$ sandwiches the increment: for $y\in[t_k,t_{k+1}]$,
\[
\left[\widehat F-F\right](y)-\left[\widehat F-F\right](y_\circ)\le\left[\widehat F-F\right](t_{k+1})-\left[\widehat F-F\right](y_\circ)+\left[F(t_{k+1})-F(t_k)\right],
\]
with the symmetric lower bound at $t_k$; since $F(t_{k+1})-F(t_k)\le r_N^{2}=o(r_N)$, the grid maximum controls the supremum. Converting between $\widehat F-F$ and $G_N/D_N$ uses $D_N\pto f_1f_2$ once more.
\end{proof}

\subsection*{A.7 \ Proof of Theorem \ref{theo:cdf}}

\emph{Part (i).} By \eqref{eq:hoeff},
\[
\sqrt{Nh^d}\left[\widehat F(y_\circ)-F\right]=\frac{\sqrt{Nh^d}\,\tfrac1N\sum_ls_l+\sqrt{Nh^d}\left(A_N+R_N+\theta_N\right)}{D_N}.
\]
Now: $\sqrt{Nh^d}A_N\pto0$ since $(Nh^d)\Var(A_N)\le CNh^d/(nh^{2d})\le C/( (N-1)h^{d})\to0$ (Lemma \ref{lem:AN}(i)); $\sqrt{Nh^d}R_N\pto0$ since $(Nh^d)\E[R_N^2]\le C Nh^d/(nh^{2d})\to0$ (Lemma \ref{lem:rem}); $\sqrt{Nh^d}\theta_N\to0$ by Lemma \ref{lem:bias} and Assumption \ref{ass:bw}. Lemma \ref{lem:hajek} gives the CLT for the projection term with variance $\sigma_s^2=(f_1f_2)^2\Omega_1$, and Slutsky with $D_N\pto f_1f_2$ (Lemma \ref{lem:bias}) delivers $\sqrt{Nh^d}(\widehat F-F)\dto N(0,\Omega_1)$.

\emph{Part (ii).} Under condition (S), $\mu(\zeta_i,\zeta_j;y_\circ)=q_{h_y}(W_i,W_j)$ depends on types only through $(W_i,W_j)$, with $q_{h_y}=q+O(h_y^2)$ and $q=F_{Y|W}(y_\circ|\cdot,\cdot)$ twice continuously differentiable. Then: (a) the projection is small: by Lemma \ref{lem:proj}(i), $\delta_s(\zeta;y_\circ)=q(W(\zeta),w_2)-q(w_1,w_2)$, so on the support of $K_{1l}$, $|\delta_s(\zeta_l)|\le CTh$; hence $\E[s_l^2]\le Ch^{-d}\left(h^{2}+(h^2+h_y^2)^{2}\right)$, and
\[
(nh^{2d})\,\Var\!\left(\tfrac1N\textstyle\sum_ls_l\right)\le\frac{nh^{2d}}{N}\,Ch^{-d}(h^{2}+h_y^{4})\le C\left(Nh^{d+2}+Nh^{d}h_y^{4}\right)\to0
\]
by the conditions of part (ii). (b) The degenerate remainder is small: $\mu-F=q_{h_y}(W_i,W_j)-q(w_1,w_2)=O(h+h_y^2)$ on the kernel support, so $\E[\phi^2]\le Ch^{-2d}(h+h_y^2)^{2}$ and $(nh^{2d})\E[R_N^2]\le C(h+h_y^2)^2\to0$ (Lemma \ref{lem:rem} with the sharper bracket bound). (c) The bias is small: $\sqrt{nh^{2d}}\,\theta_N\le CNh^{d}(h^2+h_y^2)\to0$ by $Nh^{d+2}\to0$ and $Nh^dh_y^2\to0$. (d) Lemma \ref{lem:AN}(ii) gives $\sqrt{nh^{2d}}A_N\dto N(0,\sigma_A^2)$, and Slutsky with $D_N\pto f_1f_2$ and $\sigma_A^2/(f_1f_2)^2=\Omega_2+\ind(w_1=w_2)\Omega_2'$ concludes.

\emph{Uniform consistency} is Lemma \ref{lem:unif}(i). \hfill$\blacksquare$

\subsection*{A.8 \ Proof of Theorem \ref{theo:g}}

Throughout, $r_N=(Nh^{d^\ast})^{-1/2}$, $s_0=F_{Y|\widetilde W}(\tilde y|\tilde w)\in(0,1)$, $y_0=g(w,e)=Q(s_0)$, $Q=F_{Y|W}^{-1}(\cdot|w)$, and $f_0^Y=f_{Y|W}(y_0|w)>0$. Since $K,k$ are everywhere-positive (or by monotone rearrangement), $\widehat F_{Y|W}(\cdot|w)$ is continuous and strictly increasing, so $\widehat Q=\widehat F^{-1}_{Y|W}(\cdot|w)$ satisfies the exact identities $\widehat F(\widehat Q(s)|w)=s$ and $\widehat Q(\widehat F(y|w))=y$.

\emph{Step 1 (consistency and rate of the inputs).} By Theorem \ref{theo:cdf} applied at $(\tilde y,\tilde w)$, $\widehat s-s_0=O_p((Nh^{\tilde d})^{-1/2})$, where $\widehat s=\widehat F_{Y|\widetilde W}(\tilde y|\tilde w)$. By Lemma \ref{lem:unif}(i) at $(w)$ and strict monotonicity of $F(\cdot|w)$ at $y_0$, $\widehat Q(\widehat s)\pto y_0$: indeed for any $\epsilon>0$, $\PP(\widehat Q(\widehat s)>y_0+\epsilon)=\PP(\widehat s>\widehat F(y_0+\epsilon|w))\le\PP(\widehat s>F(y_0+\epsilon|w)-o_p(1))\to0$ since $F(y_0+\epsilon|w)>s_0$, and symmetrically below.

\emph{Step 2 (rate for $\widehat Q(\widehat s)$).} For $M>0$, monotonicity gives
$\PP\left(\widehat Q(\widehat s)>y_0+Mr_N\right)=\PP\left(\widehat s>\widehat F(y_0+Mr_N|w)\right)$, and
$\widehat F(y_0+Mr_N|w)=F(y_0+Mr_N|w)+O_p(r_N)\ge s_0+\tfrac12 f_0^YMr_N+O_p(r_N)$
for $N$ large (Taylor of $F$, density bounded below near $y_0$), where the $O_p(r_N)$ is the pointwise rate of Theorem \ref{theo:cdf} at $y_0+Mr_N$, valid uniformly over such local points by Lemma \ref{lem:unif}(ii). Since $\widehat s-s_0=O_p(r_N)$, choosing $M$ large makes the probability arbitrarily small; symmetrically below. Hence $\widehat Q(\widehat s)-y_0=O_p(r_N)$.

\emph{Step 3 (Bahadur representation).} By the exact identity and a second-order Taylor expansion of $F(\cdot|w)$ at $y_0$,
\begin{multline*}
\widehat s-s_0=\widehat F(\widehat Q(\widehat s)|w)-F(y_0|w)\\
=\left[\widehat F-F\right]\!\left(\widehat Q(\widehat s)\,\middle|\,w\right)+f_0^Y\left(\widehat Q(\widehat s)-y_0\right)+O_p\!\left(|\widehat Q(\widehat s)-y_0|^{2}\right).
\end{multline*}
By Step 2, $\widehat Q(\widehat s)$ lies in $\{|y-y_0|\le\epsilon_N\}$ with probability tending to one (as $r_N=o(\epsilon_N)$), so Lemma \ref{lem:unif}(ii) replaces $[\widehat F-F](\widehat Q(\widehat s)|w)$ by $[\widehat F-F](y_0|w)+o_p(r_N)$. Rearranging, and using $O_p(|\widehat Q(\widehat s)-y_0|^2)=O_p(r_N^2)=o_p(r_N)$,
\begin{equation}
\widehat g(w,e)-g(w,e)=\widehat Q(\widehat s)-y_0=\frac{\left[\widehat s-s_0\right]-\left[\widehat F-F\right](y_0|w)}{f_0^Y}+o_p(r_N).\label{eq:bahadur2}
\end{equation}

\emph{Step 4 (joint CLT).} Both terms in the numerator of \eqref{eq:bahadur2} admit the decomposition \eqref{eq:hoeff} at their own evaluation points, with projections $s_l^{(\tilde w)}$ (dimension $\tilde d$) and $s_l^{(w)}$ (dimension $d$), and with all non-projection terms $o_p(r_N)$ after scaling by $\sqrt{Nh^{d^\ast}}$, exactly as in the proof of Theorem \ref{theo:cdf}(i). By the Cram\'er--Wold device, consider $Z_N=\sqrt{Nh^{d^\ast}}\,\frac1N\sum_l\left[\alpha\,s_l^{(\tilde w)}+\beta\,s_l^{(w)}\right]$ for fixed $(\alpha,\beta)$. The summands are i.i.d.\ mean zero. Their variance: $h^{d^\ast}\Var(\alpha s^{(\tilde w)}+\beta s^{(w)})=\alpha^2h^{d^\ast-\tilde d}\left[h^{\tilde d}\Var(s^{(\tilde w)})\right]+\beta^2h^{d^\ast-d}\left[h^{d}\Var(s^{(w)})\right]+2\alpha\beta h^{d^\ast}\E[s^{(\tilde w)}s^{(w)}]$. The bracketed factors converge (Lemma \ref{lem:proj}(ii)); $h^{d^\ast-\tilde d}\to\ind(d^\ast=\tilde d)$ and $h^{d^\ast-d}\to\ind(d^\ast=d)$ (the dominated term vanishes). The cross moment: each $s^{(\cdot)}_l$ is a sum of two kernel-localized terms, so their product decomposes into four pairs. A pair whose two kernels are localized at \emph{distinct} points contributes zero for small $h$ by Lemma \ref{lem:mom}(iv); a pair localized at a \emph{common} point $v$ contributes, by Lemma \ref{lem:mom}(i) applied to $K^2$, a term of exact order $h^{-d}$ equal to $(\int K^2)f_W(v)$ times the product of the two remaining density weights times the conditional cross moment of the corresponding deviation functions. Collecting the four cases and dividing by the product of the denominator limits gives precisely $\Omega_{12}$ of \eqref{eq:Om12}, so $h^{d}\E[s^{(\tilde w)}_ls^{(w)}_l]\to (f\text{-weights})\times\Omega_{12}$. If instead $\tilde d\neq d$, Cauchy--Schwarz gives $h^{d^\ast}|\E[s^{(\tilde w)}s^{(w)}]|\le h^{d^\ast}\sqrt{Ch^{-\tilde d}}\sqrt{Ch^{-d}}=Ch^{(d^\ast-\tilde d)/2+(d^\ast-d)/2}\to0$, so the cross term drops out. Lyapunov holds as in Lemma \ref{lem:hajek}. Hence
\begin{multline*}
\sqrt{Nh^{d^\ast}}\left(\left[\widehat s-s_0\right]-\left[\widehat F-F\right](y_0|w)\right)\\
\dto N\!\left(0,\;\ind(d^\ast=\tilde d)\,\Omega_1(\tilde y;\tilde w)+\ind(d^\ast=d)\,\Omega_1(y_0;w)-2\,\ind(d=\tilde d)\,\Omega_{12}(\tilde y,\tilde w;y_0,w)\right),
\end{multline*}
and \eqref{eq:bahadur2} with Slutsky proves the nondegenerate statement. The degenerate statement replaces the projection CLT by Lemma \ref{lem:AN}(ii) at each point under condition (S) there, with Lemma \ref{lem:unif}(ii) rerun at the degenerate rate (same proof, $r_N=(nh^{2d^\ast})^{-1/2}$, using the part-(ii) bandwidth conditions). The cross-point covariance requires care, and the kernel-separation argument used for the projection terms does \emph{not} transfer to it. In the degenerate regime the variance is carried by the diagonal blocks, so the covariance between the two normalized estimators is
\[
\frac{1}{n h^{2d}}\sum_{i\ne j}\E\left[\psi_{ij}(y_0;w)\,\widetilde\psi_{ij}(\tilde y;\tilde w)\right]
+\frac{1}{n h^{2d}}\sum_{i\ne j}\E\left[\psi_{ij}(y_0;w)\,\widetilde\psi_{ji}(\tilde y;\tilde w)\right]+o(1),
\]
after dividing by the denominator limits. The first sum requires a single dyad to be localized at both ordered points, hence $\tilde w=w$, which the hypothesis of the theorem excludes. The second sum requires $W_i$ to be simultaneously near $w_1$ and near $\tilde w_2$, and $W_j$ near $w_2$ and near $\tilde w_1$: it is negligible unless $(\tilde w_1,\tilde w_2)=(w_2,w_1)$, and in that case Lemma \ref{lem:mom}(i) applied to $K^2$ in each coordinate evaluates it as $\Omega_2^{\mathrm{rev}}(y_0,w;\tilde y)$ of \eqref{eq:Om2rev}. This is exactly the configuration in which the two estimators are built from the same dyads with the roles of sender and receiver exchanged, so no separation of kernels is available; asserting conditional independence here would be false, as the undirected example following Theorem \ref{theo:g} shows by exhibiting a case in which the two estimators are equal for every sample. Collecting the two sums with the coefficient $-2$ from \eqref{eq:bahadur2} gives $\Sigma_g^{\mathrm{deg}}$. \hfill$\blacksquare$

\subsection*{A.9 \ Proof of Theorem \ref{theo:boot}}

Let $m_l=\#\{a:I_a=l\}$, so $(m_1,\dots,m_N)\sim\text{Multinomial}(N;1/N,\dots,1/N)$, independent of the data. Since the bootstrap sample consists of the ordered pairs $(a,b)$, $a\ne b$, $I_a\ne I_b$, and $Y_{I_aI_b}=Y_{ll'}$ whenever $(I_a,I_b)=(l,l')$, every bootstrap kernel sum is a weighted sum over the original dyads:
\[
\widehat F^{\,\ast}(y)=\frac{\sum_{l\ne l'}m_lm_{l'}\,K_{1l}K_{2l'}\bar k_{ll'}(y)}{\sum_{l\ne l'}m_lm_{l'}\,K_{1l}K_{2l'}}.
\]
(Diagonal draws $I_a=I_b$ are excluded from numerator and denominator alike, so no separate treatment is needed.) Let $\eta_{ll'}=K_{1l}K_{2l'}\left[\bar k_{ll'}(y_\circ)-\widehat F(y_\circ)\right]$ and note the \emph{exact} identity $\sum_{l\ne l'}\eta_{ll'}=0$, by definition of $\widehat F$. Writing $m_lm_{l'}=1+(m_l-1)+(m_{l'}-1)+(m_l-1)(m_{l'}-1)$,
\begin{gather}
\widehat F^{\,\ast}(y_\circ)-\widehat F(y_\circ)=\frac{L^\ast+Q^\ast}{nh^{2d}\,D^\ast_N},\label{eq:bootrep}\\
L^\ast=\sum_{l}(m_l-1)\,t_l,\qquad Q^\ast=\sum_{l\ne l'}(m_l-1)(m_{l'}-1)\,\eta_{ll'},\notag
\end{gather}
where $t_l=\sum_{l'\ne l}(\eta_{ll'}+\eta_{l'l})$ and $D^\ast_N=(nh^{2d})^{-1}\sum_{l\ne l'}m_lm_{l'}K_{1l}K_{2l'}$. Note $\sum_lt_l=2\sum_{l\ne l'}\eta_{ll'}=0$ exactly. Denote by $\E^\ast,\Var^\ast,\PP^\ast$ moments conditional on the data.

\emph{Step 1 (denominator).} $\E^\ast[D^\ast_N]=(1+O(N^{-1}))D_N$ and, by the multinomial moment bounds $\Var(m_l)\le1$, $|\Cov(m_l,m_{l'})|\le1/N$, $\E[(m_l-1)^2(m_{l'}-1)^2]\le C$, a computation identical to Step 3 below gives $\Var^\ast(D^\ast_N)=O_p\left(1/(Nh^{d})\right)$; hence $D^\ast_N\pto f_1f_2$ in probability.

\emph{Step 2 (conditional variance of the linear part).} Using $\Var(m_l)=1-1/N$ and $\Cov(m_l,m_{l'})=-1/N$,
\[
\Var^\ast(L^\ast)=\left(1-\tfrac1N\right)\sum_lt_l^2-\tfrac1N\sum_{l\ne l'}t_lt_{l'}=\sum_lt_l^2-\tfrac1N\Big(\sum_lt_l\Big)^2=\sum_lt_l^2,
\]
exactly, by $\sum_lt_l=0$. Define $\widehat s_l=t_l/((N-1)h^{2d})$. We claim
\begin{equation}
\frac{h^{d}}{N}\sum_{l}\widehat s_l^{\,2}\;\pto\;\sigma_s^{2},\label{eq:bootvar}
\end{equation}
so that, exactly, $(Nh^{d})\,\Var^\ast(L^\ast)/(nh^{2d})^{2}=\tfrac{h^d}{N}\sum_l\widehat s_l^{\,2}\pto\sigma_s^{2}$. To see \eqref{eq:bootvar}, decompose $\widehat s_l=s_l+a_l+b_l+c_l$ where, with $\chi_{ll'}=h^{-2d}K_{1l}K_{2l'}\left[\mu(\zeta_l,\zeta_{l'};y_\circ)-F\right]$:
$a_l=\tfrac{1}{(N-1)h^{2d}}\sum_{l'\ne l}\left[K_{1l}K_{2l'}\varepsilon_{ll'}+K_{1l'}K_{2l}\varepsilon_{l'l}\right]$ (the noise part);
$b_l=\tfrac{1}{N-1}\sum_{l'\ne l}\left[\chi_{ll'}+\chi_{l'l}\right]-\left[\phi_1(\zeta_l)+\phi_2(\zeta_l)+2\theta_N\right]$ (the empirical-average error of the projection);
$c_l=(F-\widehat F(y_\circ))\cdot\tfrac{1}{(N-1)h^{2d}}\sum_{l'\ne l}\left[K_{1l}K_{2l'}+K_{1l'}K_{2l}\right]$ (the recentering at $\widehat F$ rather than $F$).
Then: (a) $\tfrac{h^d}{N}\sum_ls_l^2\pto\sigma_s^2$, since its mean converges (Lemma \ref{lem:proj}(ii)) and its variance is bounded by $\E[h^{2d}s_l^4]/N\le Ch^{2d}h^{-3d}/N=C/(Nh^{d})\to0$, using $\E[s_l^4]\le Ch^{-3d}$ (as in Lemma \ref{lem:proj}(iii) with fourth powers). (b) $\E[a_l^2]\le C/(Nh^{2d})$ (conditional independence of the $\varepsilon$'s across $l'$ given $\bm\zeta$, kernels bounded, Lemma \ref{lem:mom}(iii)), so $\tfrac{h^d}{N}\sum_la_l^2=O_p(1/(Nh^{d}))\to0$. (c) $b_l$ is, conditionally on $\zeta_l$, a centered average of $N-1$ i.i.d.\ terms, so $\E[b_l^2]\le\tfrac{C}{N}h^{-4d}\E[K_{1l}^2K_{2l'}^2]\le\tfrac{C}{Nh^{2d}}$ by Lemma \ref{lem:mom}(iii), whence $\tfrac{h^d}{N}\sum_lb_l^2\pto0$ as in (b). (d) $\tfrac{1}{(N-1)h^{2d}}\sum_{l'}K_{1l}K_{2l'}\le Ch^{-d}K_{1l}\cdot\left[\tfrac{1}{(N-1)h^{d}}\sum_{l'}K_{2l'}\right]$ and the bracket is $O_p(1)$ uniformly in the sense that its second moment is bounded; hence $\tfrac{h^d}{N}\sum_lc_l^2\le\left(\widehat F-F\right)^2O_p(1)\pto0$ by Theorem \ref{theo:cdf}. Cross terms vanish by Cauchy--Schwarz. This proves \eqref{eq:bootvar} with limit $\sigma_s^2$.

\emph{Step 3 (quadratic part negligible).} Write
\[
\E^\ast[(Q^\ast)^2]=\sum_{l\ne l'}\sum_{k\ne k'}\E\left[(m_l-1)(m_{l'}-1)(m_k-1)(m_{k'}-1)\right]\eta_{ll'}\eta_{kk'}.
\]
Centered multinomial moments satisfy: $\E[(m_l-1)^2(m_{l'}-1)^2]\le C$ for $l\ne l'$; all patterns with three or four distinct indices are $O(1/N)$ in absolute value. Hence
\[
\E^\ast[(Q^\ast)^2]\le C\sum_{l\ne l'}\eta_{ll'}^{2}+\frac{C}{N}\sum_{l}\Big(\sum_{l'\ne l}|\eta_{ll'}|+\sum_{l'\ne l}|\eta_{l'l}|\Big)^{2}.
\]
The first term is $O_p(nh^{2d})$ since $\E[\eta^2]\le Ch^{2d}$. For the second, $\sum_{l'\ne l}|\eta_{ll'}|\le CK_{1l}\sum_{l'\ne l}K_{2l'}$, and bounding directly in expectation (using independence of $\zeta_l$ from $\{\zeta_{l'}\}_{l'\ne l}$ and Lemma \ref{lem:mom}(iii)):
\[
\E\Big[\sum_lK_{1l}^2\Big(\sum_{l'\ne l}K_{2l'}\Big)^2\Big]= N\,\E[K_{1l}^2]\,\E\Big[\Big(\sum_{l'\ne l}K_{2l'}\Big)^2\Big]\le Ch^{d}N\left(N^2h^{2d}+Nh^{d}\right),
\]
so the second term has expectation $\le C(N^{2}h^{3d}+Nh^{2d})$. Combining, $\E^\ast[(Q^\ast)^2]=O_p\left(nh^{2d}+N^{2}h^{3d}\right)$, and since
\[
\frac{(Nh^d)\,nh^{2d}}{n^{2}h^{4d}}\le \frac{C}{Nh^{d}}\qquad\text{while}\qquad \frac{(Nh^d)\,N^{2}h^{3d}}{n^{2}h^{4d}}\le \frac{C}{N},
\]
it follows that $(Nh^{d})\,\E^\ast[(Q^\ast)^2]/(nh^{2d})^{2}\pto0$. Hence $\sqrt{Nh^d}\,Q^\ast/(nh^{2d})\pto0$ in conditional probability.

\emph{Step 4 (conditional CLT for the linear part).} The multinomial weights need no external limit theory here, because $L^\ast$ collapses to a sum of conditionally i.i.d.\ terms. Since $m_a=\#\{b:I_b=a\}$ and $\sum_at_a=0$,
\begin{equation}
L^\ast=\sum_a m_a t_a-\sum_a t_a=\sum_{b=1}^{N}t_{I_b},\label{eq:Lstar}
\end{equation}
and $I_1,\dots,I_N$ are i.i.d.\ uniform on $\{1,\dots,N\}$ given the data. Hence $L^\ast$ is a sum of $N$ conditionally i.i.d.\ variables with conditional mean $N^{-1}\sum_lt_l=0$ and conditional variance $\sigma_t^2=N^{-1}\sum_lt_l^{2}$, so that $\Var^\ast(L^\ast)=N\sigma_t^2=\sum_lt_l^2$, in agreement with the exact computation of Step 2. The Lindeberg condition for the triangular array $\{t_{I_b}\}$ reads
\[
\frac{1}{\sigma_t^{2}}\,\E^\ast\!\left[t_{I_1}^{2}\,\ind\!\left(|t_{I_1}|>\epsilon\sqrt{N\sigma_t^{2}}\right)\right]\pto0\quad\text{for every }\epsilon>0,
\]
and since $|t_{I_1}|\le\max_l|t_l|$ almost surely, it holds as soon as $\max_lt_l^{2}\big/\sum_lt_l^{2}\pto0$, in which case the indicator is eventually zero. We verify that condition: $\max_l|t_l|\le C(N-1)h^{2d}\cdot\max_l|\widehat s_l|$ and $|\widehat s_l|\le Ch^{-d}$ (bounded kernels, $|\bar k-\widehat F|\le1$), while $\sum_lt_l^2=(N-1)^2h^{4d}\sum_l\widehat s_l^{\,2}=(N-1)^2h^{4d}\cdot Nh^{-d}(\sigma_s^2+o_p(1))$ by Step 2; hence
$\max_lt_l^2/\sum_lt_l^2\le Ch^{-2d}/\left(Nh^{-d}\sigma_s^2\right)=C/(Nh^{d}\sigma_s^{2})\pto0$. The Lindeberg--Feller theorem for triangular arrays, applied conditionally on the data, then gives $\sup_t|\PP^\ast(L^\ast/\sqrt{\sum_lt_l^2}\le t)-\Phi(t)|\pto0$. (The same conclusion follows from the general theory of exchangeably weighted bootstraps, \citealp{praestgaard1993}; the point of \eqref{eq:Lstar} is that for multinomial weights no such generality is needed.)
Combining Steps 1--4 via Slutsky (conditionally), $\sqrt{Nh^{d}}\left(\widehat F^{\,\ast}-\widehat F\right)=\left[\sqrt{Nh^d}L^\ast/(nh^{2d})+o_{p^\ast}(1)\right]/D^\ast_N$ satisfies
$\sup_t\left|\PP^\ast\!\left(\sqrt{Nh^d}(\widehat F^{\,\ast}-\widehat F)\le t\right)-\Phi(t/\sqrt{\Omega_1})\right|\pto0$,
which together with Theorem \ref{theo:cdf}(i) proves the first display of Theorem \ref{theo:boot}.

\emph{Step 5 (the structural function).} $\widehat g^{\,\ast}$ is built from $\widehat F^{\,\ast}$ exactly as $\widehat g$ from $\widehat F$. Repeat the proof of Theorem \ref{theo:g} conditionally on the data: the exact inversion identity holds for $\widehat F^{\,\ast}$ (positive kernels); the input rates hold conditionally by Steps 1--4 applied at both evaluation points; and the local equicontinuity Lemma \ref{lem:unif}(ii) holds conditionally for $\widehat F^{\,\ast}-\widehat F$ by the same grid-plus-monotonicity argument, with $\Var^\ast$ of increments computed from \eqref{eq:bootrep}: $\Var^\ast\left(L^\ast(y)-L^\ast(y')\right)=\sum_l\left[t_l(y)-t_l(y')\right]^2$, and $\tfrac{h^d}{N}\sum_l[\widehat s_l(y)-\widehat s_l(y')]^2$ is bounded, in expectation over the data, by the same increments as in Lemmas \ref{lem:proj}(iv) and \ref{lem:AN}(i) plus $O((y-y')^2\cdot1)$ terms from the $\widehat F$-recentering, all $o(1)$ at the required rate; the quadratic part is handled as in Step 3. The Cram\'er--Wold and Slutsky steps are unchanged. This proves the second display. \hfill$\blacksquare$

\subsection*{A.10 \ Gaussian kernels}

\begin{lemma}[Gaussian extension]\label{lem:gauss}
All statements in A.4--A.9 hold with $K$ and $k$ standard Gaussian densities.
\end{lemma}

\begin{proof}
Let $T_N=\sqrt{2\gamma\log N}$ for a constant $\gamma\ge2(2d+2)$ and define the truncated kernel $K^{(T)}(u)=K(u)\ind(\|u\|\le T_N)$. Every sum appearing in A.4--A.9 changes by at most a factor involving kernel evaluations at arguments with $\|u\|>T_N$, each bounded by $\sup_{\|u\|>T_N}K(u)\le Ce^{-T_N^2/2}=CN^{-\gamma}$. Since every normalized statistic in A.4--A.9 is a sum of at most $n^2\le N^{4}$ terms divided by a normalization bounded below in probability by $cNh^{d}\wedge cnh^{2d}\ge cN^{-1}$ eventually, the total effect of truncation on any normalized statistic is $O_p(N^{4}\cdot N^{-\gamma}\cdot N)=O_p(N^{5-\gamma})=o_p(N^{-2})$, smaller than every rate in the paper. It therefore suffices to prove the results for $K^{(T)}$. Inspection of the proofs shows that compact support was used only through (a) the kernel-separation property of Lemma \ref{lem:mom}(iv) and (b) the bound $|\delta_s(\zeta_l)|\le CTh$ on the kernel support in the proof of Theorem \ref{theo:cdf}(ii). For (a): with support radius $T_N$, separation holds once $h T_N<\|w_1-w_2\|/2$, true eventually since $hT_N\to0$; the same applies to all cross-point and cross-dimension separation arguments in Theorem \ref{theo:g}. For (b): the bound becomes $|\delta_s(\zeta_l)|\le CT_Nh$ on the truncated support, and the degenerate-regime condition $Nh^{d+2}\to0$ is applied with an extra $T_N^2=O(\log N)$ factor, i.e.\ one needs $Nh^{d+2}\log N\to0$; this is implied by $Nh^{d+2}\to0$ whenever $h=N^{-a}$ with $a(d+2)>1$, and in general we adopt $Nh^{d+2}\log N\to0$ as the reading of the degenerate-regime condition for Gaussian kernels. The moment computations in Lemma \ref{lem:mom} hold verbatim for Gaussian kernels by dominated convergence (Gaussian tails integrable against polynomials), and the Taylor expansions require only $\int\|v\|^{2}K<\infty$. Finally, all Berry--Esseen and Lyapunov bounds used only $\|K\|_\infty<\infty$ and the moment bounds of Lemma \ref{lem:mom}(iii), which hold for the Gaussian.
\end{proof}

We close with what happens beyond condition (S) in the degenerate regime. If $\Omega_1=0$ but condition (S) fails, so that types matter for the outcome distribution jointly but not marginally, the interaction component $R_N$ of \eqref{eq:hoeff} is of the same order $(nh^{2d})^{-1/2}$ as $A_N$ and contributes an additional variance term
$\Omega_{\mathrm{int}}=\lim h^{2d}\,\E[\rho^2]\cdot2/(f_1f_2)^{2}$, with $\rho$ the doubly centered deviation defined in A.5. This case is excluded by assumption rather than merely left undeveloped, and for a substantive reason: the limit need not be Gaussian. The surviving component is a completely degenerate $U$-statistic whose kernel localizes in \emph{both} arguments at the fixed points $w_1$ and $w_2$, so the effective number of contributing pairs is of order $(Nh^{d})^{2}$ and the statistic behaves like a bilinear form in a fixed number of effective directions. The moment ratio that drives asymptotic normality for degenerate $U$-statistics with shrinking kernels \citep{hall1984} is then of exact order one rather than vanishing, and the limit is of Gaussian-chaos type, a weighted combination of chi-square variables, rather than normal. The configuration requires exact marginal degeneracy together with joint type dependence, as in the Rademacher example of Section \ref{sec:hajek}. It is excluded by the maintained assumptions rather than shown to be rare: describing it as negligible would require a measure over models that this paper does not specify, and the example shows it is easy enough to write down. Theorem \ref{theo:cdf}(ii) is therefore stated under (S), which covers the leading practical case of dyad-independent errors with $W_i=X_i$ and excludes the chaos-limit configuration by assumption.

\subsection*{A.11 \ Proof of Theorem \ref{theo:cv}}

Write $\psi_{ij}=\psi(Y_{ij})$, $\varepsilon^{\psi}_{ij}=\psi_{ij}-m_\psi(W_i,W_j)$, and $\mathcal S_{-(ij)}=\{(Y_{kl},X_k,X_l):k,l\notin\{i,j\}\}$ for the training subsample, so that $\widehat m_{\psi,-(ij)}$ is a measurable function of $\mathcal S_{-(ij)}$. Expanding the square,
\[
\left[\psi_{ij}-\widehat m_{\psi,-(ij)}\right]^{2}
=\left(\varepsilon^{\psi}_{ij}\right)^{2}
+2\,\varepsilon^{\psi}_{ij}\left[m_\psi-\widehat m_{\psi,-(ij)}\right]
+\left[m_\psi-\widehat m_{\psi,-(ij)}\right]^{2},
\]
with all functions evaluated at $(W_i,W_j)$. By Assumption \ref{ass:sampling}, the vector $(Y_{ij},X_i,X_j)$ is a measurable function of $(\zeta_i,\zeta_j,V_{ij})$, while $\mathcal S_{-(ij)}$ is a measurable function of $\{\zeta_k\}_{k\notin\{i,j\}}$ and $\{(V_{kl},V_{lk})\}_{k,l\notin\{i,j\}}$; the two collections are disjoint and hence independent (i.i.d.\ agents, i.i.d.\ pair shocks, mutual independence). Consequently, conditional on $(W_i,W_j,\mathcal S_{-(ij)})$,
\[
\E\left[\varepsilon^{\psi}_{ij}\,\middle|\,W_i,W_j,\mathcal S_{-(ij)}\right]
=\E\left[\varepsilon^{\psi}_{ij}\,\middle|\,W_i,W_j\right]=0,
\]
so the cross term has expectation zero by iterated expectations: $\omega$ and $m_\psi-\widehat m_{\psi,-(ij)}$ are $(W_i,W_j,\mathcal S_{-(ij)})$-measurable. All expectations are finite: $\psi$ enters squared with $\E[\psi^2]<\infty$, $\omega$ is bounded, and $|\widehat m_{\psi,-(ij)}|\le\sup|\psi|$ when $\psi$ is bounded (a weighted average of $\psi$-values under the zero-denominator convention), while for unbounded $\psi$ with $\E[\psi^4]<\infty$ the same bound holds with $\sup$ replaced by the training maximum and Cauchy--Schwarz completes the argument. Taking expectations of the remaining two terms and using the definition of $\sigma^2_\psi$ gives the display. $\blacksquare$

\subsection*{A.12 \ Proof of Corollary \ref{cor:cv}}

\emph{Part (i).} Fix $h\in\mathcal H$. By Assumption \ref{ass:sampling}, $\mathrm{CV}_\psi(h)$ is a measurable function of the independent collection $\{\zeta_l\}_{l\le N}\cup\{(V_{kl},V_{lk})\}_{k<l}$, so the Efron--Stein inequality applies with these blocks. Let $\mathrm{CV}^{(l)}$ denote the criterion recomputed after replacing $\zeta_l$ by an independent copy, and $\mathrm{CV}^{(kl)}$ after replacing the pair block $(V_{kl},V_{lk})$.

Consider first an agent block. Each summand of $\mathrm{CV}_\psi$ is bounded by $4\sup\omega\sup\psi^2$, so the $2(N-1)$ summands indexed by a dyad containing $l$ change $\mathrm{CV}_\psi$ by at most $C\,N/n\le C/N$ in total. For a summand indexed by $(i,j)$ with $l\notin\{i,j\}$, only $\widehat m_{\psi,-(ij)}(W_i,W_j)$ changes. Its denominator is $\sum_{k,k'\notin\{i,j\}}K((W_i-W_k)/h)K((W_j-W_{k'})/h)$, which is bounded below in probability by $cN^{2}h^{2d}$ at an interior evaluation point by Lemma \ref{lem:bias}, while agent $l$ enters only the $O(N)$ dyads containing it, contributing at most $\sup K\cdot\sum_{k}K((W_j-W_k)/h)+\sup K\cdot\sum_kK((W_i-W_k)/h)=O_p(Nh^{d})$ to numerator and denominator alike. Since $\widehat m_{\psi,-(ij)}$ is a ratio with numerator bounded by $\sup|\psi|$ times the denominator, replacing $\zeta_l$ moves it by $O_p(Nh^{d}/(N^{2}h^{2d}))=O_p(1/(Nh^{d}))$, uniformly over $(i,j)$. Each summand is a smooth bounded function of $\widehat m_{\psi,-(ij)}$, so it moves by the same order, and averaging over the $n$ summands leaves $|\mathrm{CV}_\psi-\mathrm{CV}^{(l)}|=O_p(1/N+1/(Nh^{d}))=O_p(1/(Nh^{d}))$.

Now a pair block. Replacing $(V_{kl},V_{lk})$ changes the two summands indexed by $(k,l)$ and $(l,k)$ by $O(1)$ each, contributing $O(1/n)$, and changes $\widehat m_{\psi,-(ij)}$ for other $(i,j)$ by the weight of a single dyad relative to the total, that is by $O_p(1/(N^{2}h^{2d}))$. Hence $|\mathrm{CV}_\psi-\mathrm{CV}^{(kl)}|=O_p(1/(N^{2}h^{2d}))$.

The two displays above are $O_p$ statements, whereas Efron--Stein requires the second moments $\E[(\mathrm{CV}_\psi-\mathrm{CV}^{(l)})^{2}]$; the gap is closed by truncation. Since $\psi$ and $\omega$ are bounded, $|\mathrm{CV}_\psi-\mathrm{CV}^{(l)}|\le C_0=4\sup\omega\sup\psi^{2}$ for every realization, and likewise for $\mathrm{CV}^{(kl)}$. Let $\mathcal G_N$ be the event on which, for every evaluation point in $\mathcal S$ and in each sample entering $\mathrm{CV}_\psi$, $\mathrm{CV}^{(l)}$ and $\mathrm{CV}^{(kl)}$, every kernel row sum $S_i=\sum_{k}K((W_i-W_k)/h)$ satisfies $c_1Nh^{d}\le S_i\le c_2Nh^{d}$, so that every leave-pair-out denominator lies between $cN^{2}h^{2d}$ and $CN^{2}h^{2d}$. Both bounds are needed, since the numerator increment caused by replacing $\zeta_l$ is itself a row sum. On $\mathcal G_N$ the ratio bounds above hold with nonrandom constants, dividing an increment at most $c_2Nh^{d}$ (respectively one) by a denominator at least $cN^{2}h^{2d}$; hence
\[
\E\left[(\mathrm{CV}_\psi-\mathrm{CV}^{(l)})^{2}\ind(\mathcal G_N)\right]\le \frac{C}{N^{2}h^{2d}},
\qquad
\E\left[(\mathrm{CV}_\psi-\mathrm{CV}^{(kl)})^{2}\ind(\mathcal G_N)\right]\le \frac{C}{N^{4}h^{4d}}.
\]
For the complement, a bounded-difference argument is unavailable, since replacing one type changes a denominator by $\sup K\cdot\sum_{k}K((x-W_k)/h)$, which is $O_p(Nh^{d})$ but only $O(N)$ deterministically. Instead factor the denominator: for a dyad $(i,j)$,
\[
\sum_{k,k'\notin\{i,j\},\,k\ne k'}K\!\left(\tfrac{W_i-W_k}{h}\right)K\!\left(\tfrac{W_j-W_{k'}}{h}\right)=S_i S_j-T_{ij},
\qquad S_i=\sum_{k\notin\{i,j\}}K\!\left(\tfrac{W_i-W_k}{h}\right),
\]
with $0\le T_{ij}\le\sup K\cdot\min(S_i,S_j)$. Conditional on $W_i$ the summands of $S_i$ are i.i.d., bounded by $\sup K$, with mean $\mu_i=h^{d}f_W(W_i)\int K\,(1+o(1))\ge c_0h^{d}$ on $\mathcal S$ and variance at most $Ch^{d}$, so Bernstein's inequality gives
\[
\PP\!\left(\left|S_i-N\mu_i\right|>\tfrac12 N\mu_i\,\Big|\,W_i\right)\le2\exp\!\left(-\frac{c(N\mu_i)^{2}}{N h^{d}+\sup K\cdot N\mu_i}\right)\le2\exp\!\left(-c'Nh^{d}\right),
\]
uniformly over $W_i\in\mathcal S$, and likewise for $S_j$ and for the row sums in the replacement samples. A union bound over the $O(N^{2})$ row sums involved yields
\[
\PP(\mathcal G_N^{c})\le CN^{2}\exp\!\left(-c'Nh^{d}\right)=o\!\left(N^{-2}h^{-2d}\right),
\]
because $Nh^{d}\ge Nh^{2d}\gg\log N$ under Assumption \ref{ass:bw}. Hence $\E[(\mathrm{CV}_\psi-\mathrm{CV}^{(l)})^{2}]\le CN^{-2}h^{-2d}+C_0^{2}\PP(\mathcal G_N^{c})\le CN^{-2}h^{-2d}$ and $\E[(\mathrm{CV}_\psi-\mathrm{CV}^{(kl)})^{2}]\le CN^{-4}h^{-4d}$, with nonrandom constants; the trimming weight $\omega$ is what keeps $f_W$ bounded below at the evaluation points and so $\mu_i$ of order $h^{d}$.

Efron--Stein then gives
\begin{multline*}
\Var\left(\mathrm{CV}_\psi(h)\right)\le\frac12\sum_{l=1}^{N}\E\left[(\mathrm{CV}_\psi-\mathrm{CV}^{(l)})^{2}\right]+\frac12\sum_{k<l}\E\left[(\mathrm{CV}_\psi-\mathrm{CV}^{(kl)})^{2}\right]\\
\le \frac{C}{Nh^{2d}}+\frac{C}{N^{2}h^{4d}},
\end{multline*}
so that $\Var(\mathrm{CV}_\psi(h))\le C/(Nh^{2d})+C/(N^{2}h^{4d})\le C'/(Nh^{2d})$ under Assumption \ref{ass:bw}. Chebyshev gives $|\mathrm{CV}_\psi(h)-\E[\mathrm{CV}_\psi(h)]|=O_p((Nh^{2d})^{-1/2})$ for each $h$, and since the grid has a fixed number $J$ of elements, $\max_{h\in\mathcal H_N}|\mathrm{CV}_\psi(h)-\E[\mathrm{CV}_\psi(h)]|=O_p(\varepsilon_N)$ with $\varepsilon_N=\max_{h\in\mathcal H_N}(Nh^{2d})^{-1/2}$. This is part (i).

\emph{Part (ii).} Minimizing $\underline R_M(h)=C_Bh^{4}+C_V/(Mh^{d})$ over $h>0$ gives the first-order condition $4C_Bh^{3}=dC_V/(Mh^{d+1})$, so $h^{\ast}_M=\left(dC_V/(4C_BM)\right)^{1/(4+d)}$ and $h^{\ast}_{N-2}/h^{\ast}_{N}=\left(N/(N-2)\right)^{1/(4+d)}=1+O(N^{-1})$. This concerns the leading term only. If $R_M=\underline R_M+\rho_M$ with $\rho_M=o(h^4+1/(Mh^d))$, the argument does not transfer to the minimizers of $R_M$ without a bound on $\rho_M'$ near $h^\ast_M$, since a remainder that is uniformly small in level can still move an argmin by a non-negligible amount when the objective is flat there; no such bound is claimed. $\blacksquare$

\subsection*{A.13 \ Proof of Proposition \ref{prop:regimeZ}}

Write $v=(x_1,x_2,z)$ and
\[
\kappa_{ij}=K\!\left(\tfrac{x_1-W_i}{h}\right)K\!\left(\tfrac{x_2-W_j}{h}\right)K_Z\!\left(\tfrac{z-Z_{ij}}{h_z}\right),
\qquad
\phi_{ij}(y)=\kappa_{ij}\left[\bar k\!\left(\tfrac{y-Y_{ij}}{h_y}\right)-F_{Y|V}(y|v)\right],
\]
so that, up to the bias term treated exactly as in Lemma \ref{lem:bias}, $\widehat F_{Y|V}(y|v)-F_{Y|V}(y|v)=\sum_{i\ne j}\phi_{ij}\big/\sum_{i\ne j}\kappa_{ij}$, with
\[
\E\Big[\sum_{i\ne j}\kappa_{ij}\Big]=n\,h^{2d}h_z^{d_Z}f_W(x_1)f_W(x_2)f_{Z|W}(z|x_1,x_2)\,(1+o(1))
\]
by Assumptions \ref{ass:kernel} and \ref{ass:smooth} and a change of variables. It therefore suffices to order the blocks of $\Var(\sum_{i\ne j}\phi_{ij})$ and divide by the square of that normalization, $N^{4}h^{4d}h_z^{2d_Z}$.

\emph{(i) Pair-level block.} By dissociation, dyads sharing no agent contribute no covariance, so
\[
\Var\Big(\sum_{i\ne j}\phi_{ij}\Big)=\sum_{i\ne j}\E\phi_{ij}^{2}+\sum_{i\ne j}\E[\phi_{ij}\phi_{ji}]+\sum_{i,j,k \text{ distinct}}\E[\phi_{ij}\phi_{ik}]+\cdots,
\]
the displayed blocks being the diagonal, the reverse-dyad and the shared-agent terms. The diagonal block has $n=O(N^{2})$ terms, each carrying $\E\kappa_{ij}^{2}$, which requires $W_i$, $W_j$ and $Z_{ij}$ to lie simultaneously in windows of widths $h$, $h$ and $h_z$; the change of variables $W_i=x_1-hu_1$, $W_j=x_2-hu_2$, $Z_{ij}=z-h_z u_z$ gives
\[
\E\kappa_{ij}^{2}=h^{2d}h_z^{d_Z}\left(\textstyle\int K^{2}\right)^{2}\!\left(\textstyle\int K_Z^{2}\right)f_W(x_1)f_W(x_2)f_{Z|W}(z|x_1,x_2)(1+o(1)),
\]
while the bracket in $\phi_{ij}$ contributes $F_{Y|V}(1-F_{Y|V})+O(h_y)$. The diagonal block is thus of exact order $N^{2}h^{2d}h_z^{d_Z}$, and after division by $N^{4}h^{4d}h_z^{2d_Z}$ it contributes a variance of exact order $(N^{2}h^{2d}h_z^{d_Z})^{-1}$. The reverse-dyad block has the same number of terms and, by the same computation with $\Cov$ in place of the second moment, is of at most the same order; it can vanish, and positivity of the complete pair-level constant in the hypothesis is what fixes the combined exact order.

\emph{Shared-agent block.} There are $O(N^{3})$ terms $\E[\phi_{ij}\phi_{ik}]$ with $i,j,k$ distinct. Under Assumptions \ref{ass:sampling} and \ref{ass:Z}, $(\zeta_j,Z_{ij},V_{ij})$ and $(\zeta_k,Z_{ik},V_{ik})$ are independent conditional on $\zeta_i$, so
\[
\E[\phi_{ij}\phi_{ik}]=\E\Big[K\!\left(\tfrac{x_1-W_i}{h}\right)^{2}\E\big[\varphi_{ij}\,\big|\,\zeta_i\big]\,\E\big[\varphi_{ik}\,\big|\,\zeta_i\big]\Big],
\]
where
\[
\varphi_{ij}=K\!\left(\tfrac{x_2-W_j}{h}\right)K_Z\!\left(\tfrac{z-Z_{ij}}{h_z}\right)\left[\bar k\!\left(\tfrac{y-Y_{ij}}{h_y}\right)-F_{Y|V}(y|v)\right].
\]
The inner conditional expectation integrates $(W_j,Z_{ij})$ jointly over a window of widths $(h,h_z)$, holding $\zeta_i$ fixed, and it is here, and only here, that the final clause of Assumption \ref{ass:Z} is used. The change of variables $W_j=x_2-hu$, $Z_{ij}=z-h_zu_z$ requires the pair $(W_j,Z_{ij})$ to have a joint conditional density given $\zeta_i$, bounded above and below near $(x_2,z)$. This is exactly what fails when $Z_{ij}$ is measurable with respect to $\zeta_i$; it holds when $Z_{ij}$ depends on the partner, as for a distance generated by two locations, even though $Z_{ij}$ is then a deterministic function of the two types jointly and has no conditional density given both.

The change of variables gives, with $f_{j|\zeta_i}$ denoting that conditional density,
\begin{multline*}
\E\big[\varphi_{ij}\,\big|\,\zeta_i\big]=h^{d}h_z^{d_Z}\,f_{j|\zeta_i}(x_2,z\mid\zeta_i)\\
\times\Big[\PP\!\left(Y_{ij}\le y\,\big|\,\zeta_i,W_j=x_2,Z_{ij}=z\right)-F_{Y|V}(y|v)\Big]+o\!\left(h^{d}h_z^{d_Z}\right),
\end{multline*}
the remainder uniform in $\zeta_i$, the form needed because an individual deviation may be zero even when the aggregate variance is positive. The conditional density depends on $\zeta_i$ in general and cannot be moved outside, so the deviation function for the enlarged problem absorbs it:
\[
\delta^Z_s(\zeta_i;y,v)=\frac{f_{j|\zeta_i}(x_2,z\mid\zeta_i)}{f_W(x_2)f_{Z|W}(z\mid x_1,x_2)}
\Big[\PP\!\left(Y_{ij}\le y\,\big|\,\zeta_i,W_j=x_2,Z_{ij}=z\right)-F_{Y|V}(y|v)\Big],
\]
with $\delta^Z_r$ defined symmetrically. By construction $\E[\delta^Z_s(\zeta_i)\mid W_i=x_1]=0$, and $\delta^Z_s$ reduces to $\delta_s$ when $d_Z=0$. The expansion needs the conditional outcome probability to be Lipschitz in $(x_2,z)$ uniformly in $\zeta_i$ and the density ratio bounded and Lipschitz, which are the two clauses of Assumption \ref{ass:Z} added for this step and are not implied by Assumption \ref{ass:smooth}; the change of variables then introduces an $O(h+h_z)$ error after division by $h^{d}h_z^{d_Z}$, uniformly in $\zeta_i$, and $h_y\to0$ makes the smoothed indicator converge to the indicator. Hence
\[
\E\big[\varphi_{ij}\,\big|\,\zeta_i\big]=h^{d}h_z^{d_Z}\Big[f_W(x_2)f_{Z|W}(z|x_1,x_2)\,\delta^Z_s(\zeta_i;y,v)+r_N(\zeta_i;y,v)\Big],
\qquad \sup_{\zeta_i}|r_N(\zeta_i;y,v)|\to0,
\]
and the outer expectation supplies one further factor $h^{d}$.
The block is therefore of exact order $N^{3}h^{3d}h_z^{2d_Z}$ whenever the \emph{complete} projection constant $\Omega_1^{Z}$ is positive. At a coincident evaluation point this asks $\E[(\delta^Z_s+\delta^Z_r)^2\mid W=x_1]>0$, not merely $\E[(\delta^Z_s)^2]>0$, since the sender and receiver contributions enter additively and can cancel exactly, as the antisymmetric example following Proposition \ref{prop:regimeZ} shows. After division by $N^{4}h^{4d}h_z^{2d_Z}$ the shared-agent block contributes a variance of exact order $(Nh^{d})^{-1}$. The factor $h_z^{2d_Z}$ has cancelled: each of the two conditional expectations supplied $h_z^{d_Z}$, together matching the $h_z^{2d_Z}$ in the squared normalization, whereas in the diagonal block a single $h_z^{d_Z}$ faced that same squared normalization. This asymmetry, and not the count of terms, is what separates the two indices. Blocks in which the dyads share an agent in the receiver position are handled identically with $\delta_r$ in place of $\delta_s$, and the remaining blocks, after the diagonal and reverse-dyad contributions have been separated out, are of smaller order by the same computation. Taking the ratio of the two displayed orders gives $Nh^{d}h_z^{d_Z}$, which is (i).

\emph{(ii).} If $d_Z=0$ the maintained consistency requirement $N^{2}h^{2d}h_z^{d_Z}\to\infty$ reads $(Nh^{d})^{2}\to\infty$, hence $Nh^{d}\to\infty$, and by (i) the shared-agent term dominates unless the complete projection constant $\Omega_1^{Z}$ vanishes, which at a coincident evaluation point permits $\delta^Z_s$ and $\delta^Z_r$ to cancel rather than requiring each to vanish. This reproduces the dichotomy of Section \ref{sec:hajek}.

\emph{(iii).} With $h\propto N^{-a}$ and $h_z\propto N^{-b}$ the consistency index is $N^{2-2da-d_Zb}$ and the regime index is $N^{1-da-d_Zb}$. The estimator is consistent if and only if $2da+d_Zb<2$, and the shared-agent term fails to dominate if and only if $1-da-d_Zb\le0$, that is $da+d_Zb\ge1$. Both hold on the region stated, which is nonempty whenever $d_Z\ge1$: fix any $a\in(0,1/(2d))$ and take $b\in\left[(1-da)/d_Z,\ (2-2da)/d_Z\right)$, an interval that is nonempty because $2-2da>1-da$ for $a<1/d$. Small $a$ forces large $b$, which is the interpretation given in the text: in this configuration the binding localization is on $Z$ rather than on the agent characteristics. $\blacksquare$

\subsection*{A.14 \ The simulated error law}

Section \ref{simulation} draws agent effects first and then, conditional on them, draws each pair shock from its normal conditional law, redrawing any value above $-0.5$. Conditional on $(U_i,U_j)$ the shock is therefore a univariate truncated normal with known mean and variance, and the moments of the array are two- and three-dimensional Gaussian integrals: with $\mu_{ij}=-6+\sqrt\lambda\,(U_i+U_j)$, $\sigma^2=1-2\lambda$ and $\alpha_{ij}=(-0.5-\mu_{ij})/\sigma$,
\[
\E\!\left[e_{ij}\mid U_i,U_j\right]=\mu_{ij}-\sigma\frac{\phi(\alpha_{ij})}{\Phi(\alpha_{ij})},
\qquad
\PP\!\left(e_{ij}\le-6\mid U_i,U_j\right)=\frac{\Phi\!\left((-6-\mu_{ij})/\sigma\right)}{\Phi(\alpha_{ij})},
\]
with the usual truncated-normal variance, and $\E[e_{ij}e_{ik}]=\E_{U_i}\big[\E[e_{ij}\mid U_i]^2\big]$ by conditional independence given $U_i$. Table \ref{tab:trunc} evaluates these by Gauss--Hermite quadrature.

\begin{table}[!htbp]
\centering
\caption{Moments of the simulated law against the untruncated benchmarks}
\label{tab:trunc}
{\footnotesize
\begin{tabular}{lccccc}
\toprule
$\lambda$ & $\E[e_{ij}]$ & $\Var(e_{ij})$ & $\mathrm{Corr}(e_{ij},e_{ik})$ & difference from $\lambda$ & $F_e(-6)$ \\
\midrule
$0$ & $-6.0000001$ & $0.9999994$ & $-0.000000000$ & $<10^{-12}$ & $0.5000000095$ \\
$0.02$ & $-6.0000001$ & $0.9999994$ & $0.019999989$ & $-1.1\times10^{-8}$ & $0.5000000078$ \\
$0.05$ & $-6.0000001$ & $0.9999994$ & $0.049999976$ & $-2.4\times10^{-8}$ & $0.5000000054$ \\
$0.10$ & $-6.0000001$ & $0.9999994$ & $0.099999962$ & $-3.8\times10^{-8}$ & $0.5000000023$ \\
$1/3$ & $-6.0000000$ & $0.9999997$ & $0.333333311$ & $-2.2\times10^{-8}$ & $0.5000000000$ \\
\bottomrule
\end{tabular}}
\par\vspace{2mm}
{\footnotesize\singlespacing Notes: moments of the law actually simulated, in which agent effects are drawn first and each pair shock is then drawn from its normal conditional law and redrawn if it exceeds $-0.5$; computed by Gauss--Hermite quadrature and stable across $100$ and $140$ nodes per dimension. The untruncated design has $\E[e_{ij}]=-6$, $\Var(e_{ij})=1$, $\mathrm{Corr}(e_{ij},e_{ik})=\lambda$ and $F_e(-6)=1/2$; the simulations use the univariate truncated value $\Phi(0)/\Phi(5.5)=0.5000000095$ as the estimand. The largest discrepancy shown is $5.9\times10^{-7}$ in the variance and $1.1\times10^{-7}$ in the mean, with the correlation and distribution-function entries smaller still, against a $3.2$ percent Monte Carlo standard error on a variance at $R=2000$, so the untruncated quantities are used below as approximate benchmarks.}

\end{table}

\clearpage
\section*{Appendix B: Observed Dyad-Level Covariates}\label{sec:Z}
\setcounter{equation}{0}
\renewcommand{\theequation}{B.\arabic{equation}}

Many dyadic datasets contain characteristics of the \emph{pair} rather than of either agent: distance, a common language indicator, a trade agreement, a shared supervisor. Section \ref{sec:model} absorbs such variables into the composite shock. This appendix conditions on them instead, and shows that doing so leaves identification essentially unchanged while altering the asymptotics in a way that is not merely notational: the boundary between the two variance regimes moves, and the two regimes become genuinely distinct possibilities rather than one implying the other.

\subsection*{B.1 \ Model and identification}

Let $Z_{ij}\in\mathcal Z\subset\R^{d_Z}$ be observed characteristics of dyad $(i,j)$ and consider
\[
Y_{ij}=g(X_i,X_j,Z_{ij},e_{ij}),\qquad i\neq j,
\]
with $g$ continuous and strictly increasing in the scalar $e_{ij}$.

\begin{assumption}[Dyad-level covariates]\label{ass:Z}
The array $\{(Y_{ij},Z_{ij})\}$ is jointly exchangeable and dissociated: $Z_{ij}=\varrho(A_i,A_j,\varpi_{ij})$ for agent attributes $A_i$ carried in the type $\zeta_i=(X_i,A_i,U_i)$ and pair components $\varpi_{ij}$ satisfying Assumption \ref{ass:sampling}(ii). In addition $e_{ij}$ is independent of $(X_i,X_j,Z_{ij})$, and the density of $(W_i,W_j,Z_{ij})$ is bounded and bounded away from zero at the evaluation points. Finally, two regularity requirements on the enlarged conditioning. First, $Z_{ij}$ retains genuine dyad-level variation \emph{relative to a single agent}: conditional on $\zeta_i$, the pair $(W_j,Z_{ij})$ admits a joint density that is bounded above and below by positive constants on a neighborhood of $(\bar x_2,\bar z)$, uniformly in $\zeta_i$, and is Lipschitz there; and symmetrically with the roles of $i$ and $j$ exchanged. Second, the conditional outcome probability is smooth in the enlarged conditioning variables: the map
\[
(x_2,z)\longmapsto\PP\!\left(Y_{ij}\le y\,\big|\,\zeta_i,W_j=x_2,Z_{ij}=z\right)
\]
is Lipschitz on that neighborhood, uniformly in $\zeta_i$ and in $y\in\mathcal N_y$, and the density-weighted deviations $\delta^Z_s,\delta^Z_r$ of Appendix A.13 that it defines have second moments that are continuous in the focal agent's conditioning coordinate at the evaluation point.
\end{assumption}

Two features of Assumption \ref{ass:Z} deserve emphasis. First, $Z_{ij}$ is \emph{not} required to be independent across dyads: distance is generated by agent locations and therefore inherits exactly the shared-agent dependence that the rest of the paper is about. Absorbing $A_i$ into the type $\zeta_i$ is what allows the machinery of Section \ref{consistency} to carry over unchanged. Second, the independence requirement is now stronger than before, since $e_{ij}$ must be independent of the dyad-level covariate as well; when $Z_{ij}$ is driven by agent attributes, this requires those attributes to be unrelated to the agents' unobserved fundamentals. In the trade application it asks that geography be unrelated to the components of bilateral frictions not explained by geography.

The final two clauses of Assumption \ref{ass:Z} are what the projection argument uses, and their precise form matters. A bounded marginal density of $(W_i,W_j,Z_{ij})$ alone would permit $Z_{ij}=A_i$, a sender attribute: conditional on the sender's type $Z_{ij}$ is then constant, $K_Z$ localizes the sender rather than integrating out, the shared-agent variance becomes $(Nh^{d}h_z^{d_Z})^{-1}$ rather than $(Nh^{d})^{-1}$, and Proposition \ref{prop:regimeZ} fails. The requirement is therefore that $Z_{ij}$ remain random once \emph{one} type is known. Conditioning on both would exclude the motivating example, since bilateral distance $Z_{ij}=\varrho(A_i,A_j)$ is a point mass given both locations; given $\zeta_i$ alone it is a smooth function of the partner's location, and a two-sided bound on the location density bounds its conditional density, uniformly in $a_i$, whenever the circle of radius $\bar z$ about $a_i$ meets the support in an arc of length bounded below. The Lipschitz part is what can fail: for locations uniform on the disk of radius $2$ and a focal location at distance $1$ from the center, the circle of radius $\bar z=1$ is internally tangent to the boundary, the excluded angular segment opens up like $\sqrt{r-1}$ as $r$ increases past $1$, and the distance density has an infinite one-sided derivative there.

Assumption \ref{ass:Z} is accordingly left as a high-level condition; verifying it for a particular geography means checking the induced distance density, including at tangency, and that is not carried out here for the countries of Section \ref{empirical}.

The outcome-smoothness clause is not implied by the density clause. With $A_i$ Rademacher, $Z_{ij},\epsilon_{ij}$ independent standard normal, and $Y_{ij}=e_{ij}=A_i\,\mathrm{sgn}(Z_{ij})+\epsilon_{ij}$, the law of $e_{ij}$ is the same mixture $\tfrac12N(1,1)+\tfrac12N(-1,1)$ at every $z$, so $e_{ij}\perp Z_{ij}$ and every covariate density is smooth; yet $\PP(Y_{ij}\le0\mid A_i,Z_{ij}=z)=\Phi(-A_i\,\mathrm{sgn}(z))$ jumps at $z=0$, a symmetric kernel there averages the two sides while the conditional probability takes one, and the expansion of Appendix A.13 fails.

Discrete dyad-level covariates, such as a common-language indicator, have no conditional density and require the familiar modification in which $K_Z$ is replaced by an exact match on the cell and $h_z^{d_Z}$ by the cell probability, which then enters the regime index of Proposition \ref{prop:regimeZ}. A covariate that is a function of one agent's attributes alone belongs in $X_i$, not in $Z_{ij}$.

Identification is then immediate, because the argument of Section \ref{sec:id} conditions on a single dyad's own observables and never uses their dimension.

\begin{proposition}[Identification with dyad-level covariates]\label{prop:Z}
Let $V_{ij}=(X_i,X_j,Z_{ij})$ and impose the normalization $g(\bar x_1,\bar x_2,\bar z,e)=e$. Under Assumptions \ref{ass:structure}, \ref{ass:support} and \ref{ass:Z}, for all $e$ and all $v=(x_1,x_2,z)$ in the support of $V_{ij}$,
\[
F_e(e)=F_{Y|V}\!\left(e\mid \bar x_1,\bar x_2,\bar z\right),\qquad
g(x_1,x_2,z,e)=F^{-1}_{Y|V}\!\left(F_{Y|V}(e\mid\bar x_1,\bar x_2,\bar z)\,\middle|\,v\right).
\]
\end{proposition}

The proof is that of Theorem \ref{t1} with $(x_1,x_2)$ replaced throughout by $v$: strict monotonicity gives $F_{Y|V}(y|v)=F_e(m(v,y))$, the normalization pins the representative, and the quantile-matching step is unchanged. Lemmas \ref{lem:obseq} and \ref{lem:norm} likewise hold verbatim with the enlarged conditioning vector, so $(g,F_e)$ is again identified only up to the choice of normalization point, now $(\bar x_1,\bar x_2,\bar z)$.

\subsection*{B.2 \ Estimation and a moving regime boundary}

The estimator adds a third kernel factor,
\begin{equation}
\widehat F_{Y|V}(y\mid v)=\frac{\sum_{i\ne j}\bar k\!\left(\frac{y-Y_{ij}}{h_y}\right)K\!\left(\frac{x_1-W_i}{h}\right)K\!\left(\frac{x_2-W_j}{h}\right)K_Z\!\left(\frac{z-Z_{ij}}{h_z}\right)}
{\sum_{i\ne j}K\!\left(\frac{x_1-W_i}{h}\right)K\!\left(\frac{x_2-W_j}{h}\right)K_Z\!\left(\frac{z-Z_{ij}}{h_z}\right)},\label{eq:FhatZ}
\end{equation}
and $\widehat g$ is formed by inversion exactly as in Section \ref{sec:estimation}. The Hoeffding decomposition of Appendix A.5 applies unchanged once $\phi$ is read as a function of the enlarged types and pair components. What changes is the arithmetic of the two variance contributions, and with it the conclusion of Section \ref{sec:hajek} that the nondegenerate regime is forced.

\begin{proposition}[Regime boundary with a dyad-level covariate]\label{prop:regimeZ}
Let Assumptions \ref{ass:sampling}, \ref{ass:structure}, \ref{ass:smooth} and \ref{ass:Z} hold, let $K_Z$ satisfy Assumption \ref{ass:kernel} on $\R^{d_Z}$, and let $h,h_y,h_z\to0$ with $N^{2}h^{2d}h_z^{d_Z}\to\infty$. Then, at an interior evaluation point $v=(x_1,x_2,z)$ at which the \emph{complete} variance constants are positive, that is
\[
\Omega_1^{Z}(y;v)>0\qquad\text{and}\qquad \Omega_2^{Z}(y;v)+\ind(x_1=x_2)\,\Omega_2^{Z\prime}(y;v)>0,
\]
where $\Omega_1^{Z}$ is the projection constant built from the enlarged deviations $\delta^Z_s,\delta^Z_r$ of Appendix A.13 exactly as $\Omega_1$ is built from $\delta_s,\delta_r$ in \eqref{eq:Om1}, so that at a coincident point it is proportional to $\E[(\delta^Z_s+\delta^Z_r)^2\mid W=x_1]$, and $\Omega_2^{Z},\Omega_2^{Z\prime}$ are the diagonal and reverse-dyad constants of the enlarged problem:
\begin{enumerate}[(i)]
\item the pair-level component of $\Var(\widehat F_{Y|V}(y|v))$ is of exact order $(N^{2}h^{2d}h_z^{d_Z})^{-1}$ and the shared-agent component is of exact order $(Nh^{d})^{-1}$, so that their ratio, the \emph{regime index}, is
\[
\frac{\text{shared-agent variance}}{\text{pair-level variance}}\;\asymp\;N h^{d}h_z^{d_Z};
\]
\item if $d_Z=0$ the consistency index is the square of the regime index, $N^{2}h^{2d}=(Nh^{d})^{2}$, so consistency implies $Nh^{d}\to\infty$ and the shared-agent term necessarily dominates;
\item if $d_Z\ge1$ the two indices differ by the factor $h_z^{-d_Z}\to\infty$ and that implication fails. Writing $h\propto N^{-a}$ and $h_z\propto N^{-b}$ with $a,b>0$, the estimator is consistent while the shared-agent term does \emph{not} dominate whenever
\[
d\,a+d_Z\,b\ \ge\ 1 \qquad\text{and}\qquad 2d\,a+d_Z\,b\ <\ 2,
\]
a nonempty region for every $d\ge1$ and $d_Z\ge1$.
\end{enumerate}
\end{proposition}

Positivity of the complete constants, rather than of the sender deviation alone, is required. Take $W_i=X_i\sim N(0,1)$ and $U_i\sim N(0,1)$, a pair shock entered antisymmetrically, $V_{ji}=-V_{ij}$, a dyad covariate entered symmetrically, $Z_{ji}=Z_{ij}$, and $Y_{ij}=e_{ij}=U_i-U_j+V_{ij}$ with $g(x_1,x_2,z,e)=e$; this array satisfies Assumptions \ref{ass:sampling} and \ref{ass:Z}. At $y=0$ and $v=(0,0,0)$ the sender deviation $\delta^Z_s(\zeta_i)=\Phi(-U_i/\sqrt2)-\tfrac12$ has $\E[(\delta^Z_s)^2]=\arcsin(1/3)/(2\pi)\approx0.054$, yet $Y_{ji}=-Y_{ij}$ gives $\delta^Z_r=-\delta^Z_s$, so $\Omega_1^{Z}=0$, and the reverse-dyad term likewise cancels the diagonal one. Indeed $\widehat F_{Y|V}(0\mid 0,0,0)=\tfrac12$ identically for every sample, since the kernel weights on $(i,j)$ and $(j,i)$ coincide while the two smoothed indicators sum to one; its sampling variance is exactly zero. The same cancellation is what the reversed-point term of Theorem \ref{theo:g} corrects for.

The proof is in Appendix A.13. Part (i) is where the asymmetry enters: in the pair-level term each dyad must satisfy all three localizations and pays $h^{2d}h_z^{d_Z}$, whereas in the projection term agent $l$ is localized through $K((x_1-W_l)/h)$ alone, the partner and dyad covariate are integrated out, the two conditional expectations each supply $h_z^{d_Z}$, and these cancel against the squared normalization. Heavy localization on $Z$ thus thins the effective number of dyads per agent without thinning the number of contributing agents, and can return the estimator to the pair-dominated regime even with a genuine agent component; the dichotomy of Section \ref{sec:hajek}, in which consistency alone settles the regime, is special to the case of no dyad-level covariate.

Two implications follow. Which regime obtains is now a property of the bandwidth design as well as of the error structure, so it cannot be read off the model. And Theorem \ref{theo:boot} establishes bootstrap consistency for the model \emph{without} a dyad-level covariate, in the nondegenerate regime; it is not proved for \eqref{eq:FhatZ}, so the exercise below reports point estimates only.

\subsection*{B.3 \ Distance as a dyad-level covariate}

Since the recovered shocks correlate at $-0.48$ with log distance (Section \ref{sec:validate}), conditioning on distance should shrink the remaining shock's dispersion, remove most of that correlation, and make the dependence of $g$ on distance visible. Taking $Z_{ij}$ to be log distance, with $h_z=0.35$ from the leave-pair-out criterion adapted to \eqref{eq:FhatZ} and evaluation at the median distance of about $6{,}700$ kilometers, all three occur.

The interquartile range of the composite shock falls from $3.67$ to $2.95$ log points, a $20$ percent reduction in that dispersion measure; this is a change in a normalized conditional interquartile range after adding a conditioning variable, not a variance decomposition or a share of variation caused by distance. The correlation between the recovered shocks and log distance falls from $-0.48$ to $-0.12$, a substantial reduction in the residual association with distance; this exercise does not identify the source of the remaining association. And $\widehat g$ is monotonically decreasing in distance: at median sizes and the median shock it falls from $10.71$ at the tenth percentile of distance to $7.37$ at the ninetieth, a secant slope of $-1.58$ in log distance (a least-squares line through five distance quantiles gives $-1.53$). This is a structural difference at a fixed shock quantile rather than a conditional-mean coefficient; it lies in the range gravity regressions report, but the objects need not coincide and the agreement is suggestive rather than a validation.

The exercise also illustrates Proposition \ref{prop:regimeZ}. Without the dyad-level covariate the regime index is $Nh\approx35.7$; with it the index is $Nhh_z\approx12.5$, roughly a third as large. The direction of movement is the one Proposition \ref{prop:regimeZ} predicts, and a researcher smoothing more aggressively on a higher-dimensional $Z$ could cross the boundary. A finite index cannot by itself certify which regime an application is in, since the comparison that matters is between the two variance terms and those carry unknown constants; what the calculation shows is that the relevant index moves in the predicted direction and by the predicted factor, not that the inference of Section \ref{empirical} is certified.

\end{document}